\documentclass[11pt,a4paper]{article}

\usepackage{jheppub}

\usepackage{amsmath,amssymb}
\usepackage{hyperref}
\usepackage{array}
\usepackage{booktabs}
\usepackage{bbm}
\usepackage{mathrsfs}
\usepackage{tikz}
\usetikzlibrary{arrows.meta,calc}

\newcommand{\Tr}{\operatorname{Tr}}
\newcommand{\PE}{\operatorname{PE}}
\newcommand{\dd}{\mathrm{d}}

\newcommand{\bm}{\textbf{}}
\newcommand{\pp}{\mathrm{pp}}
\newcommand{\eff}{\mathrm{eff}}

\hypersetup{
  unicode=true,
  pdflang={en},
  pdfdisplaydoctitle=true,
  pdftitle={Matrix Theory from Holography II: Beyond Counting},
  pdfauthor={Shota Komatsu, Eunwoo Lee, Chintan Patel},
  pdfsubject={
    Matrix Theory beyond state counting:
    coupling-dependent topological correlators from holography.
    Building on arXiv:2609.13386, this work formulates and tests
    a new triality between a large-charge sector of
    three-dimensional N=4 ADHM theory,
    the Berenstein-Maldacena-Nastase (BMN) matrix model,
    and the discrete light-cone quantization (DLCQ) of M-theory
    on the maximally supersymmetric eleven-dimensional pp-wave.
    Using supersymmetric localization, it matches normalized
    measures and operator insertions in fixed screened
    magnetic-flux sectors, yielding an explicit operator
    dictionary for topological correlators with nontrivial
    BMN-coupling dependence.
    The triple-scaling limit matches the normalized ADHM
    superconformal index to the grand-canonical BMN index,
    with positive magnetic-flux sectors labeling
    fuzzy-sphere vacua.
    For minuscule monopole charges, physical ADHM correlators
    map directly to topological correlators in the trivial
    BMN vacuum.
    The work provides evidence connecting monopole bubbling
    to BMN instantons and tests a mirror map from
    bubbling-completed monopoles to Higgs-branch winding
    operators in necklace quiver theory.
    Finally, it develops a proposed type IIB analogue linking
    four-dimensional orbifold quiver theories,
    DLCQ of type IIB pp-wave string theory,
    and level-one mass-deformed ABJM theory on a spatial
    two-torus, supported by a protected vacuum-counting match
    in the sector of vanishing torsion wrapping charge.
  },
  pdfkeywords={
    Matrix Theory,
    AdS/CFT correspondence,
    M-theory,
    BMN matrix model,
    three-dimensional N=4 ADHM theory,
    large-charge limit,
    triple-scaling limit,
    supersymmetric localization,
    coupling-dependent topological correlation functions,
    protected operator dictionary,
    superconformal index,
    grand-canonical BMN index,
    fuzzy-sphere vacua,
    monopole operators,
    monopole bubbling,
    BMN instantons,
    three-dimensional mirror symmetry,
    Higgs-branch winding operators,
    necklace quiver theory,
    discrete light-cone quantization,
    maximally supersymmetric pp-wave,
    type IIB string theory,
    orbifold quiver gauge theories,
    level-one mass-deformed ABJM theory,
    mABJM vacuum counting
  }
}
\title{Matrix Theory from Holography II: Beyond Counting}

\author[a]{Shota Komatsu,}
\author[b]{Eunwoo Lee,}
\author[c]{Chintan Patel}

\affiliation[a]{Abdus Salam Centre for Theoretical Physics, Imperial College London,
SW7 2AZ London, United Kingdom}
\affiliation[b]{Department of Theoretical Physics, CERN, 1211 Meyrin, Switzerland}
\affiliation[c]{Jadwin Hall, Princeton University,
Princeton, NJ 08540, U.S.A.}

\emailAdd{sk2925@ic.ac.uk}
\emailAdd{eunwoo.lee@cern.ch}
\emailAdd{cp3849@princeton.edu}

\abstract
{Extending our earlier realization of Matrix
Theory for M-theory within AdS/CFT, we propose and test a new triality among a large-charge sector of
three-dimensional $\mathcal N=4$ ADHM theory, the BMN matrix model, and
lightlike-compactified M-theory on the maximally supersymmetric pp-wave. In an appropriate triple-scaling limit, the superconformal index of ADHM theory becomes the grand-canonical index of the BMN model, with 
magnetic-flux sectors corresponding to  fuzzy-sphere vacua.
But the real interest of this new setup is that we can go beyond counting states. Using supersymmetric localization, we show that correlation functions in topological subsectors agree as well, including their nontrivial dependence on the BMN coupling. In each 
magnetic-flux sector, both the localization measure and the operator insertions can be matched precisely with those of the corresponding BMN vacuum. Our results also point to a deeper connection between
monopole bubbling in ADHM theory and instantons in the BMN model. For minuscule monopole
charges, ADHM correlators map directly to topological correlators in
the trivial BMN vacuum. For non-minuscule charges, monopole bubbling produces additional screened sectors, whose relative weights reproduce the classical actions of BMN instantons. Mirror symmetry offers another perspective on this structure, mapping the monopole operators to Higgs-branch ``winding" operators in the mirror necklace theory. Finally, we discuss a type IIB counterpart, connecting a large-charge sector of four-dimensional $\mathbb Z_K$-orbifold quiver theories, the DLCQ of type IIB string theory on the pp-wave, and level-one mass-deformed ABJM theory on $\mathbb R_t\times T^2$. As a first test, we show that the superconformal index of the orbifold theory reproduces the vacuum degeneracy of the mass-deformed ABJM theory.}

\begin{document}
\begin{flushright}
CERN-TH-2026-282
\end{flushright}

\maketitle

\flushbottom

\section{Introduction}
Matrix Theory conjecture \cite{Banks:1996vh} makes a remarkable claim: M-theory in flat spacetime can be described nonperturbatively by a quantum mechanics of matrices. The idea is simple to state, but notoriously difficult to test. The trouble is that extracting the physics of M-theory requires understanding the strongly coupled dynamics of the matrix model, precisely where direct calculations become difficult. How can we get around this? AdS/CFT \cite{Maldacena:1997re, Gubser:1998bc, Witten:1998qj} offers a possible way forward. Over the past three decades, AdS/CFT has developed into a powerful and well-tested framework for studying quantum gravity in asymptotically AdS spacetimes. If Matrix Theory can be recovered as a limit of AdS/CFT, we can turn the logic around and use the tools of holography to probe Matrix Theory from an entirely different perspective.

In our recent work~\cite{Komatsu:2026ano} (see also~\cite{Kovacs:2013una,Du:2026lbd}) we proposed a concrete realization of this idea. We argued that three seemingly different theories become equivalent in a suitable limit:
\begin{enumerate}
    \item The $U(m)$ BMN matrix model with dimensionless coupling $g_{\rm BMN}$.\footnote{Throughout this paper, we follow the convention of \cite{Komatsu:2024vnb} for the BMN coupling.}

    \item The large-charge monopole sector of
    $U(N)_k\times U(N)_{-k}$ ABJM theory, in the triple-scaling limit
    \[
        N,k,J\to\infty,
        \qquad
        g_{\rm BMN}^2=\frac{4\pi^2}{27}\frac{N}{k^2}\ \text{fixed},
        \qquad
        m=\frac{J}{k}\ \text{fixed},
    \]
    where $J$ denotes the R-charge corresponding to the angular momentum along one of the equators
    of the $S^7$ in the bulk dual geometry
    $\mathrm{AdS}_4\times S^7/\mathbb Z_k$.

    \item M-theory on the maximally supersymmetric eleven-dimensional
    pp-wave with a compact lightlike direction, carrying $m$ units
    of momentum along the lightlike circle.
\end{enumerate}
The reason why this works is geometric.
ABJM theory is dual to M-theory on $\mathrm{AdS}_4\times S^7/\mathbb Z_k$~\cite{Maldacena:1997re,Aharony:2008ug}. The relevant states
with large angular momentum $J$ are localized near a null geodesic
at the center of $\mathrm{AdS}_4$, moving along the chosen equator
of $S^7$. Zooming in on the neighborhood of this geodesic gives
the maximally supersymmetric eleven-dimensional pp-wave. This is just the familiar Penrose limit~\cite{Blau:2002dy,Blau:2002mw}.

But there is an important twist. 
The $\mathbb{Z}_k$ orbifold identifies points along the angular direction
containing the chosen geodesic. Under the Penrose limit together
with the large-$k$ scaling, this identification becomes a
compactification of one of the lightlike directions of the
pp-wave. The resulting background therefore defines the discrete
light-cone quantization (DLCQ) of M-theory on the eleven-dimensional
pp-wave, with
\[
    J=mk
\]
identifying $m$ as the number of units of light-cone momentum. We have therefore arrived at the DLCQ of M-theory on the pp-wave, which is conjectured to be described by the $U(m)$ BMN matrix model~\cite{Berenstein:2002jq}. This gives a direct route from AdS/CFT to Matrix Theory: by focusing on states with large angular momentum, one can identify the large-charge
sector of the dual field theory with the matrix quantum mechanics for M-theory.\footnote{Earlier work on the large-angular-momentum ABJM--BMN correspondence can be found in~\cite{Kovacs:2013una}. However, their scaling limit differs from ours: they considered a regime with large $J$ and $k = \mathcal{O}(1)$, leading to a decompactified pp-wave geometry rather than its DLCQ. Related analyses of ABJM--BMN indices appear in~\cite{Chang:2026yeg}. See our previous work \cite{Komatsu:2026ano} for a detailed comparison.} The real value of this connection, however, is not simply that different theories can be related by dualities. It is that we can use AdS/CFT, in particular the dual CFT, to study Matrix Theory itself.

A first nontrivial test of this idea was given in~\cite{Komatsu:2026ano, Du:2026lbd}. In the triple-scaling limit, the ABJM superconformal index was shown to agree with a grand-canonical supersymmetric index of the BMN matrix model, obtained by summing over all nonnegative integer gauge-group ranks.
This grand-canonical structure arises because the ABJM index sums
over the charge $J$, which is related to the BMN rank by $m=J/k$.
This agreement is encouraging, but it only takes us so far. The BMN index is independent of the coupling $g_{\rm BMN}$, so it cannot test the proposed relation between the coupling constants of the two theories. To probe the dynamics more deeply, we need observables that actually depend on the coupling, such as correlation functions. Here we run into a problem. The triple-scaling limit of ABJM theory is not the usual 't Hooft limit. Instead, the 't Hooft coupling grows as
$
\lambda_{\rm ABJM}=\frac{N}{k}\propto\sqrt{N}\to\infty,
$
driving the theory to strong coupling. Thus, precisely the quantities we would like to compute become difficult to access directly.

\begin{figure}[t]
\centering
\resizebox{\linewidth}{!}{%
\begin{tikzpicture}[
    font=\small,
    >={Latex[length=2.3mm,width=1.6mm]},
    box/.style={
        draw,
        line width=0.75pt,
        rounded corners=5pt,
        align=center,
        inner xsep=10pt,
        inner ysep=9pt,
        outer sep=0pt,
        minimum height=1.25cm
    },
    relation/.style={
        <->,
        line width=0.8pt,
        shorten <=3pt,
        shorten >=3pt
    },
    label/.style={
        font=\small,
        inner sep=0pt
    }
]

\node[box,draw=red,minimum width=7.2cm] (gravity) at (0,3.2)
    {\bfseries DLCQ of M-theory on the pp-wave\\[3pt]
     \bfseries with $m$ units of light-cone momentum};

\node[box,draw=blue,minimum width=4.2cm] (abjm) at (-4.6,0)
    {\bfseries Large-charge monopole sector\\
     \bfseries in triple-scaled ADHM};

\node[box,draw=green!60!black,minimum width=4.8cm] (bmn) at (4.6,0)
    {\bfseries $U(m)$ BMN matrix model\\[3pt]
     \bfseries with coupling $g_{\rm BMN}$};

\node[box,draw=purple,minimum width=5.6cm] (necklace) at (0,-3.2)
    {\bfseries Large-charge ``winding'' sector\\[3pt]
     \bfseries in triple-scaled necklace theory};

\coordinate (gravityLeft)
    at ($(gravity.south)+(-2.35cm,0)$);
\coordinate (gravityRight)
    at ($(gravity.south)+(2.35cm,0)$);

\draw[relation]
    (abjm.north) --
    node[label,midway,above left=5pt]
    {\bfseries AdS$_4$/CFT$_3$}
    (gravityLeft);

\draw[relation]
    (bmn.north) --
    node[label,midway,above right=5pt]
    {\bfseries Matrix Theory conjecture}
    (gravityRight);

\draw[relation]
    (abjm.east) -- (bmn.west);

\draw[relation]
    (abjm.south) --
    node[label,midway,below left=6pt]
    {\textbf{Mirror symmetry}}
    (necklace.north west);

\draw[relation]
    (bmn.south) --
    (necklace.north east);

\end{tikzpicture}%
}
\caption{A web of dualities explored in this paper. We extend the BMN triality of \cite{Komatsu:2026ano} to three-dimensional $\mathcal{N}=4$ ADHM theory. Although ABJM and ADHM theories have distinct holographic duals, both backgrounds reduce to the same eleven-dimensional pp-wave geometry in the triple-scaling limit, connecting the large-charge monopole sector of ADHM theory to the BMN matrix model. Mirror symmetry provides an alternative description in terms of $\mathcal{N}=4$ necklace quiver theory, whose large-charge winding sector offers a complementary perspective on this extended triality.}
\label{fig:BMNtriality}
\end{figure}
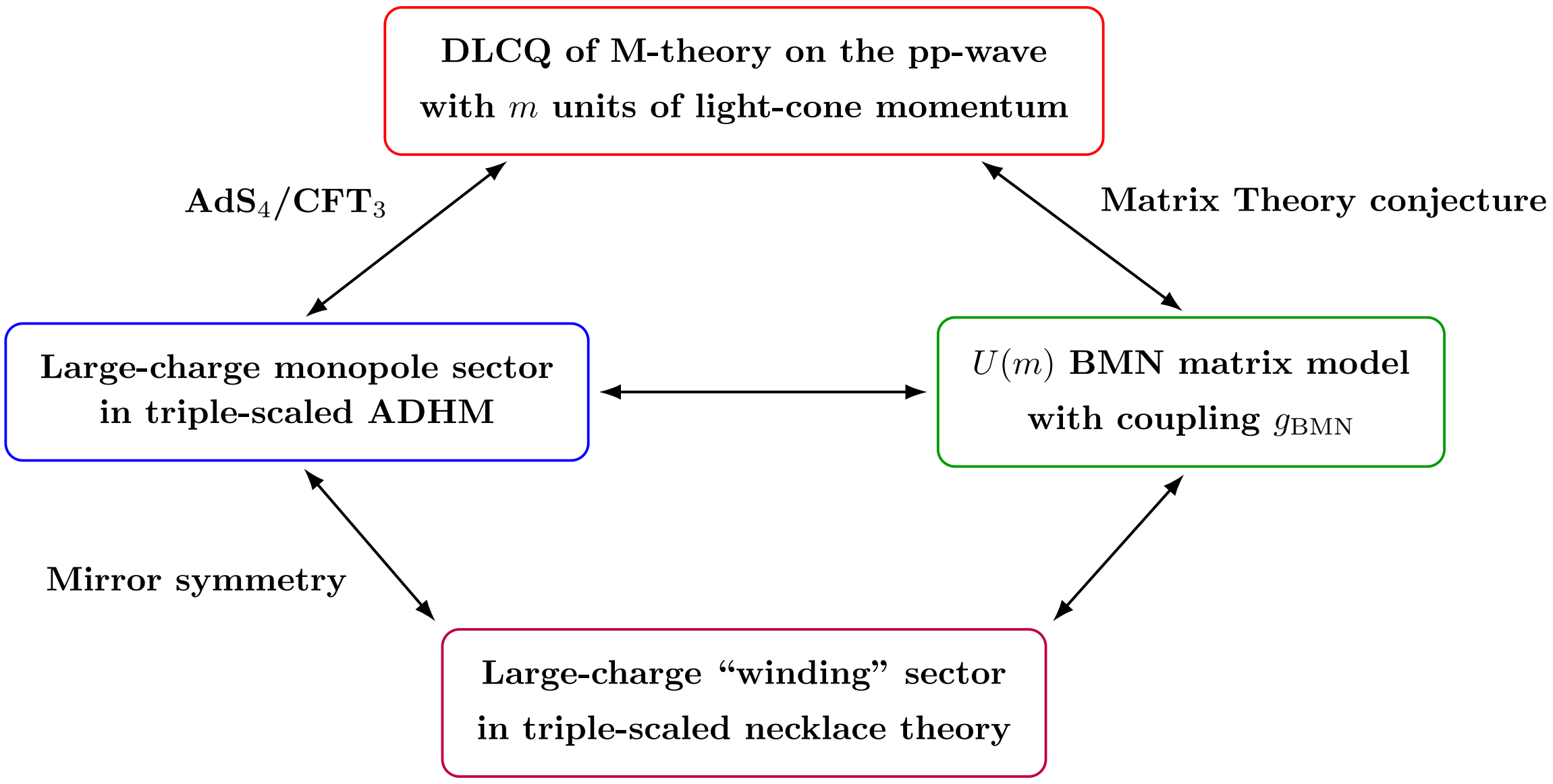

One way to overcome this limitation is to study topological subsectors of
$\mathcal N=4$ SCFTs, whose correlation functions can be computed
exactly by supersymmetric localization \cite{Chester:2014mea,Beem:2016cbd,Dedushenko:2017avn,Dedushenko:2018icp}. For ABJM theory, and more generally
for theories with Chern--Simons terms, the corresponding localization
framework has not yet been developed. We therefore turn to a different theory
that admits, in an appropriate limit, the same holographic description in
terms of the maximally supersymmetric eleven-dimensional pp-wave with a compact
lightlike direction. 
The theory we consider is the three-dimensional $\mathcal N=4$ ADHM theory,
which has a UV description as a $U(N)$ gauge theory with one adjoint
hypermultiplet and $N_f$ fundamental hypermultiplets. We propose an ADHM
counterpart of the BMN triality relating (see Figure \ref{fig:BMNtriality})
\begin{enumerate}
    \item The $U(m)$ BMN matrix model with dimensionless coupling $g_{\rm BMN}$.

    \item The large-charge sector with positive monopole charge of the
    three-dimensional $\mathcal N=4$ $U(N)$ ADHM theory.

    \item M-theory on the maximally supersymmetric eleven-dimensional
    pp-wave with a compact lightlike direction.
\end{enumerate}
The corresponding triple-scaling limit is
\begin{equation}
    N,N_f,J\longrightarrow\infty,
    \qquad
    g_{\rm BMN}^2
    =\frac{4\pi^2}{27}\frac{N}{N_f^2}\ \text{fixed},
    \qquad
    m\equiv\frac{J}{N_f}\ \text{fixed}.
    \label{eq:adhm-triple-scaling-summary}
\end{equation}

As in the ABJM case, the geometric origin of this triality is
the Penrose limit of the bulk dual of the ADHM theory,
M-theory on $\mathrm{AdS}_4\times S^7/\mathbb Z_{N_f}$~\cite{Itzhaki:1998uz,Pelc:1999ms}.
The action of the $\mathbb{Z}_{N_f}$ orbifold on $S^7$ differs from that in ABJM,
as we review in Section~\ref{sec:adhm-review}. Nevertheless, taking the analogous
Penrose limit around an appropriate null geodesic leads to the
same maximally supersymmetric eleven-dimensional pp-wave with
a compact lightlike direction. Here, the angular momentum $J$
is chosen along one of the two equatorial directions that
participate in the $\mathbb Z_{N_f}$ orbifold action.
Analogously to ABJM, this angular momentum is tied to the
monopole charge $m$:
\begin{equation}
    J=N_f m.
\end{equation}
Consequently, the monopole charge $m$ is identified with the
number of units of light-cone momentum and, in the BMN description,
with the rank of the $U(m)$ gauge group.
\medskip

We provide two main pieces of evidence in support of this triality.

\paragraph{Indices.}
First, we establish a match between the superconformal index of
the ADHM theory in the triple-scaling limit and the grand-canonical
supersymmetric index of the BMN matrix model. More precisely,
after dividing out the zero-monopole contribution, we show that
\begin{equation}
    \frac{I_{\rm ADHM}}{I_{\rm ADHM}^{(0)}}
    \longrightarrow
    \sum_{m=0}^{\infty}
    X^m I_{{\rm BMN},\ U(m)}(a,b,c).
    \label{eq:adhm-bmn-grand-summary}
\end{equation}
Here $I_{\rm ADHM}^{(0)}$ denotes the contribution from the sector
in which all magnetic fluxes vanish, while $X$ keeps track of the
positive monopole charge $m$. The transverse excitation fugacities
$(a,b,c)$ and their ADHM counterparts are specified in Section~\ref{adhmindlim}.
The same calculation identifies positive magnetic-flux
configurations with fuzzy-sphere vacua: a flux of magnitude $s$
corresponds to an irreducible $SU(2)$ block of dimension $s$,
and a partition of the total monopole charge $m$ determines a
vacuum of the $U(m)$ BMN model~\cite{Berenstein:2002jq,Dasgupta:2002hx}. Together with our previous result \cite{Komatsu:2026ano}, this also identifies these indices with the index on the eleven-dimensional pp-wave background.

\paragraph{Correlators.} As a second piece of evidence, we compare Coulomb-branch
correlators of the ADHM theory with correlators in a topological subsector of the
BMN matrix model. The Coulomb-branch operators consist of monopole
operators dressed by polynomials in the vector-multiplet scalars,
together with scalar polynomials of zero magnetic charge.
Suitable position-dependent combinations of these operators,
inserted along a great circle of $S^3$, define a topological sector~\cite{Chester:2014mea,Beem:2016cbd}
whose correlators can be computed using supersymmetric
localization~\cite{Dedushenko:2017avn,Dedushenko:2018icp, Gaiotto:2020vqj,Hatsuda:2021oxa}.
We consider monopole insertions of opposite charges,
together with protected scalar insertions and dressings.
On the BMN side, the corresponding protected scalar $\Phi$ is a
linear combination of one of the $SO(3)$ scalars and a
time-dependent combination of two of the $SO(6)$ scalars,
annihilated by the localizing supercharge
of~\cite{Asano:2012zt}. The localization supercharges themselves can also be
matched directly: we choose the ADHM supercharge so that the corresponding
bulk Killing spinor reduces, in the Penrose limit, to the supercharge used in
BMN localization.

The comparison of localized correlators proceeds as follows.
In both theories, localization reduces the protected correlators
to matrix integrals with specified measures and operator insertions.
On the ADHM side, the integral initially involves $N$ eigenvalues.
In a fixed magnetic-flux sector $\mathbf v$,
we integrate out the eigenvalues carrying zero flux.
In the triple-scaling limit, the resulting effective measure
matches the BMN localization measure:
\begin{equation}
    d\mu_{\rm ADHM,\,\mathbf v}^{\rm eff}
    =
    \mathscr A_{\mathbf v}
    \left.
    d\mu_{\rm BMN}^{L(\mathbf v)}
    \right|_{q_i=2\sigma_i}
    \bigl[1+o(1)\bigr].
    \label{eq:intro-adhm-bmn-measure}
\end{equation}
Here $L(\mathbf v)$ denotes the fuzzy-sphere vacuum associated
with $\mathbf v$, while $\sigma_i$ and $q_i$ are the charged
ADHM eigenvalues and the corresponding BMN localization variables.
The number of charged eigenvalues equals the number of irreducible
fuzzy-sphere blocks, whose dimensions sum to the BMN rank $m$.
The factor $\mathscr A_{\mathbf v}$ is independent of these
eigenvalues and therefore cancels from correlators normalized
within that sector.

The operator insertions match as well. Let $\Phi$ denote the
protected ADHM scalar, in the dimensionless conventions used below.
The dressing operation $\mathcal{D}_n$ associated with a BMN trace
$\operatorname{Tr}_m\phi^n$ turns out to be
\begin{equation}
    \mathcal D_n\mathcal O
    :=[P_n(\Phi),\mathcal O]_\star,
    \qquad
    P_n(\Phi)
    :=\frac{2^n}{n+1}
      \operatorname{Tr}
      B_{n+1}\!\left(\frac12+i\Phi\right),
    \label{eq:intro-adhm-bmn-dressing}
\end{equation}
where $B_{n+1}$ is a Bernoulli polynomial and $\star$ denotes
the ordered product in the protected Coulomb-branch operator
algebra. Acting on a monopole operator, this commutator reproduces
the localized BMN trace insertion; repeated actions reproduce
products of traces.

\paragraph{Monopole bubbling.}
To apply this dictionary to physical monopole correlators, one must also take
{\it monopole bubbling} into account. A monopole operator specifies a singular
magnetic charge $\mathbf b$, but smooth dynamical monopoles can partially
screen it, leaving a smaller asymptotic magnetic flux $\mathbf v$. The full
correlator therefore receives contributions from all allowed screened sectors
$\mathbf v$. 

The unscreened sector, $\mathbf v=\mathbf b$, is particularly simple. In this
sector, the dictionary above gives the following relation between normalized
correlators:
\begin{equation}
    \frac{
        \left\langle
        \bigl(\mathcal D_{n_k}\cdots\mathcal D_{n_1}
        \mathcal M_{\mathbf b}\bigr)
        \mathcal M_{-\mathbf b}
        \right\rangle_{\rm ADHM}^{(\mathbf{v}=\mathbf b)}
    }{
        \left\langle
        \mathcal M_{\mathbf b}\mathcal M_{-\mathbf b}
        \right\rangle_{\rm ADHM}^{(\mathbf{v}=\mathbf{b})}
    }
    \longrightarrow
    \left\langle
        \prod_{\ell=1}^{k}
        \operatorname{Tr}_m\!\left(\phi^{n_\ell}\right)
    \right\rangle_{\rm BMN,\,\mathbf b},
    \qquad m=\sum_i b_i .
    \label{eq:intro-adhm-bmn-correlators}
\end{equation}
Here the superscript $(\mathbf{v})$ denotes restriction to the
magnetic-flux sector $\mathbf v$, the arrow denotes the
triple-scaling limit, and the BMN expectation value is normalized and evaluated 
in the vacuum labelled by the
partition $\mathbf b$.
For minuscule charges $\mathbf b=(1^m,0^{N-m})$, monopole bubbling is absent. In this case,
\eqref{eq:intro-adhm-bmn-correlators} relates the full normalized ADHM
correlator directly to the corresponding BMN correlator in the trivial fuzzy-sphere 
vacuum, without any need to project onto a particular screened sector.



For non-minuscule monopole charges, bubbling is present, and the left-hand side
of \eqref{eq:intro-adhm-bmn-correlators} represents only the unscreened
contribution rather than the full physical ADHM correlator. The latter requires
summing over the allowed screened sectors $\mathbf v$. Their relative weights
are controlled  by the factors $\mathscr A_{\mathbf v}$ appearing in
\eqref{eq:intro-adhm-bmn-measure}. We find
\begin{equation}
    -\frac12
    \log\frac{\mathscr A_{\mathbf v}}{\mathscr A_{\mathbf b}}
    =
    S_{\rm cl}(\mathbf b\to\mathbf v)+\text{subleading terms},
    \label{eq:intro-bubbling-instanton}
\end{equation}
where $S_{\rm cl}(\mathbf b\to\mathbf v)$ is the classical action of a BMN
instanton interpolating between the vacua associated with the unscreened and
screened partitions
\cite{Yee:2003ge,Lin:2006tr}.\footnote{In
\cite{Asano:2012zt}, BMN instanton contributions were neglected under the
assumption that they are exponentially suppressed in the 't~Hooft limit. We
find that the remaining eigenvalue integrals can in some cases compete
with this classical exponential suppression, suggesting that the instanton
sector may deserve a more careful treatment. See Subsection \ref{subsec:adhm-bmn-bubbling-scope} for more details.}
This points to a natural correspondence between monopole bubbling in ADHM
theory and instanton transitions between BMN vacua, although a complete
identification remains open. It also motivates extension of the BMN localization
analysis of \cite{Asano:2012zt} to include instanton sectors explicitly.
\paragraph{Mirror description.}We also study the correspondence using the mirror dual of the ADHM theory, an
$\mathcal N=4$ necklace quiver theory~\cite{Intriligator:1996ex,Hanany:1996ie,Kapustin:2010xq}. Mirror symmetry maps
ADHM Coulomb-branch operators to Higgs-branch operators in the
necklace theory, providing a complementary description in
terms of ordinary gauge-invariant polynomials.
We propose and test a map between bubbling-completed monopole
operators and operators built
from products of bifundamental scalars ``winding" around the necklace, together
with their protected scalar dressings. This description allows
selected correlators to be studied without separately computing
the monopole-bubbling coefficients. We test the proposed map
through analytic and numerical calculations at finite rank,
while leaving the general operator dictionary for future work.

\begin{figure}[t]
\centering
\resizebox{\linewidth}{!}{%
\begin{tikzpicture}[
    font=\small,
    >={Latex[length=2.3mm,width=1.6mm]},
    box/.style={
        draw,
        line width=0.75pt,
        rounded corners=5pt,
        align=center,
        inner xsep=10pt,
        inner ysep=9pt,
        outer sep=0pt,
        minimum height=1.25cm
    },
    relation/.style={
        <->,
        line width=0.8pt,
        shorten <=3pt,
        shorten >=3pt
    },
    label/.style={
        font=\small,
        inner sep=0pt
    }
]

\node[box,draw=red,minimum width=7.2cm] (gravity) at (0,3.2)
    {\bf DLCQ of IIB string theory on the pp-wave\\[3pt] \bf
     with $m$ units of light-cone momentum};

\node[box,draw=blue,minimum width=4.2cm] (abjm) at (-4.6,0)
    {\bf Large-charge ``winding" sector\\ \bf  in $\mathbb{Z}_K$-orbifold quiver theory};

\node[box,draw=green!60!black,minimum width=4.8cm] (bmn) at (4.6,0)
    {\bf Mass-deformed $U(m)_1\times U(m)_{-1}$ \\[3pt]
     \bf ABJM on $\mathbb{R}_t\times T^2$};

\coordinate (gravityLeft)
    at ($(gravity.south)+(-2.35cm,0)$);
\coordinate (gravityRight)
    at ($(gravity.south)+(2.35cm,0)$);

\draw[relation]
    (abjm.north) --
    node[label,midway,above left=5pt] {\bf AdS$_5$/CFT$_4$}
    (gravityLeft);

\draw[relation]
    (bmn.north) --
    node[label,midway,above right=5pt] {\bf Matrix Theory conjecture}
    (gravityRight);

\draw[relation]
    (abjm.east) --
    node[label,midway,below=6pt] {\huge \bf ?}
    (bmn.west);

\node[align=center,anchor=north,inner sep=0pt]
    at ($(abjm.south)+(0,-0.25cm)$)
    {\footnotesize $ N,g_{\rm YM},K,J\to\infty$\\[3pt]
     $g_{s}= \frac{g_{\rm YM}^2}{4\pi K}$,\quad $\mu_{\rm ABJM}^2 A_{T^2}= \pi\frac{K}{N}$,\quad  $m=\frac{J}{K}$ fixed};

\node[align=center,anchor=north,inner sep=0pt]
    at ($(bmn.south)+(0,-0.25cm)$)
    {$\mu_{\rm ABJM}$: mass, \quad  $A_{T^2}$: area of $T^2$\\[3pt] $\tau_{T^2}=C_0+\frac{i}{g_s}$: torus complex structure};

\end{tikzpicture}%
}
\caption{Type IIB analogue of the BMN triality. A large-charge ``winding'' sector of $\mathbb Z_K$ orbifold quiver theories is related to type IIB string theory on the maximally supersymmetric pp-wave, whose nonperturbative description is conjectured to be mass-deformed ABJM theory on $\mathbb R_t\times T^2$. The relevant limit of the quiver theory sends $N$, $g_{\rm YM}$, $K$ and $J$ to infinity with $g_{\rm YM}^2/K$, $K/N$, and $J/K$ fixed. These combinations map, respectively, to the type IIB string coupling $g_s$, the torus area $A_{T^2}$ in units measured by the mass deformation, and the rank of the mass-deformed ABJM gauge group $m$.}
\label{fig:typeIIBtriality}
\end{figure}
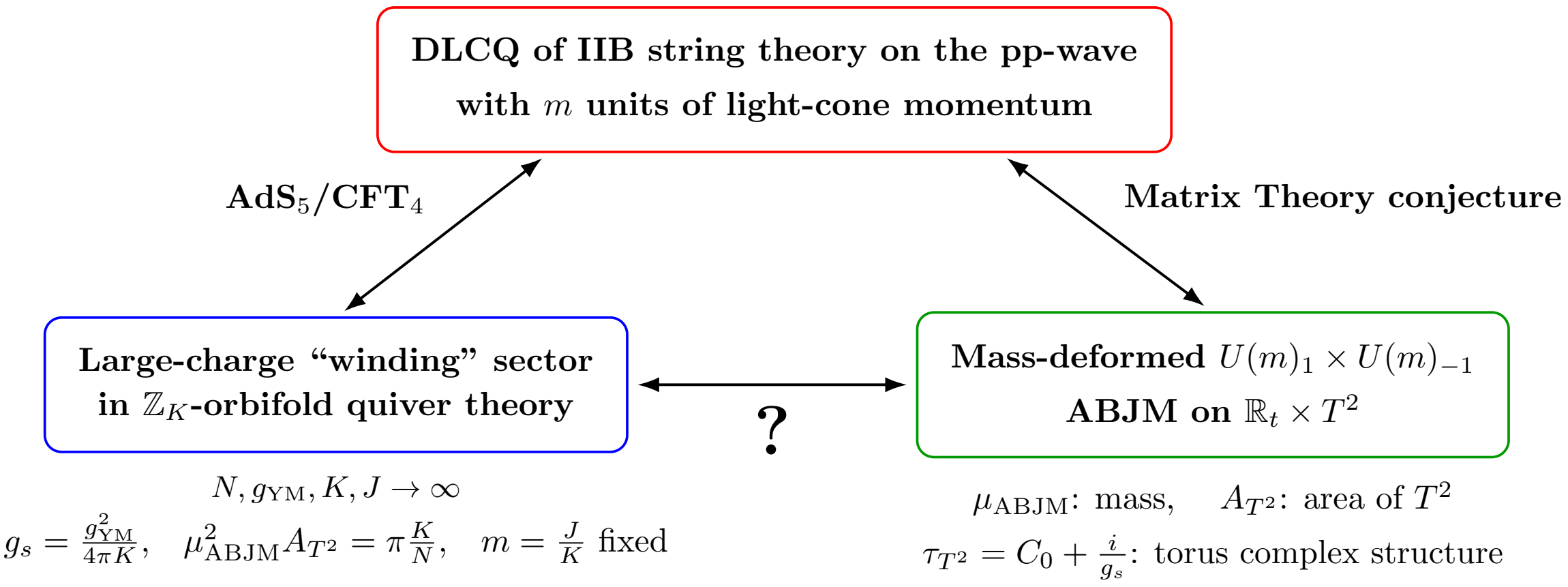

\paragraph{Type IIB analogue.}Finally, we propose a type IIB analogue of the triality, building on the work
of Mukhi, Rangamani, and Verlinde \cite{Mukhi:2002ck}. The starting point is a
four-dimensional $\mathcal N=1$ or $\mathcal N=2$ quiver theory obtained as a
$\mathbb Z_K$ orbifold of $\mathcal N=4$ super Yang--Mills. We conjecture a
triality relating a large-charge sector of this quiver theory, the DLCQ of the
maximally supersymmetric type IIB pp-wave, and level-one mass-deformed ABJM
theory on $\mathbb R_t\times T^2$. 

The relevant limit is somewhat unusual. We send the rank $N$ and gauge coupling $g_{\rm YM}$ of each quiver node, the number of nodes $K$, and the charge $J$ to infinity, while keeping the three combinations $g_{\rm YM}^2/K$, $K/N$, and $J/K$ fixed. These combinations have a natural interpretation in the mass-deformed ABJM description:
\begin{align}
g_s=\frac{g_{\rm YM}^2}{4\pi K},\qquad
\mu_{\rm ABJM}^2 A_{T^2}=\pi\frac{K}{N},\qquad
m=\frac{J}{K},.
\end{align}
Here $g_s$ is the type IIB string coupling, which fixes the complex structure of the torus through $\tau_{T^2}=C_0+i/g_s$. The parameter $A_{T^2}$ is the torus area, $\mu_{\rm ABJM}$ is the mass parameter, equal to the pp-wave mass, and $m$ is the rank of the mass-deformed ABJM gauge group.

There is an important subtlety that makes the type IIB story slightly different from its M-theory counterpart. The orbifold geometry contains torsion three-cycles, which can be wrapped by D3-branes. In the quiver theory, these wrapped branes are described by baryonic operators. The interesting point is that these operators can survive the scaling limit, even though the corresponding wrapped branes are not visible in the local pp-wave geometry. To make a clean comparison, we therefore have to restrict to the sector with vanishing torsion wrapping charge.
With this restriction, we find evidence for the triality. The protected counting in the quiver theory reproduces the vacuum degeneracies of level-one mass-deformed ABJM theory.  Whether the two descriptions also agree at the level of excitations and their dynamics remains an interesting question for the future.

\paragraph{Structure of the paper.}The rest of the paper is organized as follows.
In Section~\ref{sec:adhm-review}, we review the ADHM theory,
its Coulomb branch, and its superconformal index. In Section~\ref{sec:ADHMtriality}, we study the Penrose limit of $AdS_4\times S^7/\mathbb{Z}_{N_f}$ and derive the parameter dictionary for the triality. We then establish the index match and
the correspondence between magnetic-flux sectors and BMN vacua.
Section~\ref{sec:adhm-bmn-correlators} develops the localization
comparison of topological correlators, constructs the operator dictionary,
and examines the relation between monopole bubbling and
BMN instantons.
Section~\ref{sec:necklace-bmn-correlators} studies the mirror
necklace description and tests the proposed operator map.
Section~\ref{sec:iib-ppwave-mabjm} presents the type IIB proposal
and its protected counting tests.
We conclude in Section~\ref{sec:discussion} with a discussion
of the results and open questions.
Appendices~\ref{app:susy_matching}
and~\ref{app:generic-positive-flux-measure} contain the detailed
supercharge comparison and the derivation of the effective
localization measure, respectively.
Appendix~\ref{num test} presents numerical tests of the proposed mirror map for neutral insertions.

\section{A brief review of the ADHM theory}
\label{sec:adhm-review}

\subsection{Brane construction of ADHM theory and its holographic dual}

The ADHM theory arises as the worldvolume theory of D2-branes in a D2--D6 brane system~\cite{Itzhaki:1998uz,Pelc:1999ms}. Consider $N$ D2-branes extended along the directions $012$, together with $N_f$ D6-branes extended along $0123456$.
The open strings with both endpoints on the D2-branes give rise to a $U(N)$ vector multiplet and an adjoint hypermultiplet. The vector multiplet contains the $U(N)$ gauge field together with three real adjoint scalars, corresponding to transverse fluctuations of the D2-branes along the directions $789$. The adjoint hypermultiplet contains four real, or equivalently two complex, adjoint scalars, associated with motion of the D2-branes along the directions $3456$.

In addition, open strings stretching between the D2- and D6-branes give rise to $N_f$ fundamental hypermultiplets of the $U(N)$ gauge theory. Altogether, this brane configuration and hence the worldvolume theory preserves $3d$ $\mathcal N=4$ supersymmetry. The gauge coupling of this theory is dimensionful, and the theory becomes strongly coupled in the infrared. It is believed to flow to an interacting $3d$ $\mathcal N=4$ superconformal fixed point, which we also refer to as the ADHM theory. 

The brane configuration preserves the geometric symmetry
\[
SO(4)_{3456}\times SO(3)_{789}
\simeq
SU(2)_H\times SU(2)_F\times SU(2)_C .
\]
The $SO(4)_{3456}$ factor acts on the four scalars of the adjoint
hypermultiplet and decomposes as $SU(2)_H\times SU(2)_F$, while
$SO(3)_{789}\simeq SU(2)_C$\footnote{This identification is somewhat schematic. As we will see below, after uplifting to M-theory the $SU(2)_C$ symmetry acts nontrivially on the full Taub--NUT geometry, including the M-theory circle, and not only on the $789$ directions.} rotates the three real scalars in the
vector multiplet. The factors $SU(2)_H\times SU(2)_C$ constitute the
$3d$ $\mathcal N=4$ R-symmetry, whereas $SU(2)_F$ is a flavor symmetry.

In addition, the theory has a topological $U(1)_T$ symmetry, with
conserved current
\[
j_T^\mu=\frac{1}{4\pi}\epsilon^{\mu\nu\rho}
\operatorname{Tr}F_{\nu\rho},
\qquad
Q_T=\frac{1}{2\pi}\int_{S^2}\operatorname{Tr}F
=\sum_{i=1}^N m_i.
\]
Here $F=\frac12F_{\mu\nu}\,dx^\mu\wedge dx^\nu$, and
$\boldsymbol m=(m_1,\ldots,m_N)$ denotes the GNO magnetic charge~\cite{Goddard:1976qe}.
In the M-theory uplift, $U(1)_T$ is associated with translations along
the Taub--NUT circle, so that the isometry acting on the Coulomb-branch
directions is $SU(2)_C\times U(1)_T$.

The matter content of the ADHM theory is listed in table \ref{tab:adhm-field-content}.
\begin{table}[th]
\centering
\renewcommand{\arraystretch}{1.25}
\setlength{\tabcolsep}{6pt}
\begin{tabular}{
    >{\centering\arraybackslash}p{2.6cm}
    >{\centering\arraybackslash}p{2.2cm}
    >{\centering\arraybackslash}p{1.6cm}
    >{\centering\arraybackslash}p{1.6cm}
    >{\centering\arraybackslash}p{1.6cm}
    >{\centering\arraybackslash}p{1.6cm}
    >{\centering\arraybackslash}p{1.4cm}
}
\toprule
Field
& $U(N)_{\mathrm{gauge}}$
& $SU(2)_H$
& $SU(2)_C$
& $SU(2)_F$
& $SU(N_f)$
& $U(1)_T$
\\
\midrule

Gauge field $A_\mu$
& $\mathrm{adj}$
& $\mathbf{1}$
& $\mathbf{1}$
& $\mathbf{1}$
& $\mathbf{1}$
& $0$
\\

Vector scalars $\phi_{\dot a\dot b}$
& $\mathrm{adj}$
& $\mathbf{1}$
& $\mathbf{3}$
& $\mathbf{1}$
& $\mathbf{1}$
& $0$
\\

Gauginos $\lambda_{a\dot a}$
& $\mathrm{adj}$
& $\mathbf{2}$
& $\mathbf{2}$
& $\mathbf{1}$
& $\mathbf{1}$
& $0$
\\
\midrule

Adjoint-hyper scalars $B_{aA}$
& $\mathrm{adj}$
& $\mathbf{2}$
& $\mathbf{1}$
& $\mathbf{2}$
& $\mathbf{1}$
& $0$
\\

Adjoint-hyper fermions $\psi_{\dot a A}$
& $\mathrm{adj}$
& $\mathbf{1}$
& $\mathbf{2}$
& $\mathbf{2}$
& $\mathbf{1}$
& $0$
\\
\midrule

Fundamental-hyper scalars $q_a{}^{i}$
& $\mathbf{N}\oplus\overline{\mathbf{N}}$
& $\mathbf{2}$
& $\mathbf{1}$
& $\mathbf{1}$
& $\mathbf{N_f}\oplus\overline{\mathbf{N_f}}$
& $0$
\\

Fundamental-hyper fermions $\psi_{\dot a}{}^{i}$
& $\mathbf{N}\oplus\overline{\mathbf{N}}$
& $\mathbf{1}$
& $\mathbf{2}$
& $\mathbf{1}$
& $\mathbf{N_f}\oplus\overline{\mathbf{N_f}}$
& $0$
\\
\midrule

Monopole operator $V_{\boldsymbol m}$
& gauge invariant
& $\mathbf{1}$
& spin $j_C$
& $\mathbf{1}$
& $\mathbf{1}$
& $\displaystyle \sum_{r=1}^{N}m_r$
\\

\bottomrule
\end{tabular}
\caption{
Field content of the $3d$ $\mathcal N=4$ ADHM theory with gauge group
$U(N)$, one adjoint hypermultiplet, and $N_f$ fundamental
hypermultiplets. For $SU(2)$'s, the numbers  in the table are  dimensions of the representation, e.g.~\textbf{1} denotes a singlet.  For the bare monopole,
$j_C=\frac{N_f}{2}\sum_{r=1}^N|m_r|$. 
}
\label{tab:adhm-field-content}
\end{table}

Let us now briefly review the holographic dual of the infrared CFT~\cite{Itzhaki:1998uz,Pelc:1999ms}. Upon uplifting the D2--D6 system to M-theory, the D2-branes become M2-branes, while the D6-branes are uplifted to a Kaluza--Klein monopole geometry of the form
\begin{equation}
    \mathbb R^{1,6}\times \mathcal M_{N_f},
\end{equation}
where $\mathcal M_{N_f}$ is the four-dimensional Taub--NUT space associated with $N_f$ coincident KK monopoles. Near the core of the KK monopoles, the Taub--NUT space approaches the orbifold
\begin{equation}
    \mathcal M_{N_f}\longrightarrow \mathbb R^4/\mathbb Z_{N_f}.
\end{equation}
The M2-branes therefore probe the transverse geometry
\begin{equation}
    \mathbb R^4\times \frac{\mathbb R^4}{\mathbb Z_{N_f}}
    \simeq
    \frac{\mathbb R^8}{\mathbb Z_{N_f}},
\end{equation}
where the orbifold acts nontrivially only on one of the two $\mathbb R^4$ factors.

Taking the infrared limit of the worldvolume theory corresponds, on the gravity side, to taking the near-horizon limit of the M2-branes located at the orbifold singularity. The standard near-horizon geometry of the M2-branes is therefore
\begin{equation}
    AdS_4\times S^7/\mathbb Z_{N_f}.
\end{equation}
The $\mathbb Z_{N_f}$ quotient acts only on four real, or equivalently two complex, coordinates of the eight-dimensional transverse space in which the $S^7$ is embedded. In terms of complex coordinates on $\mathbb C^4$, the action may be written, up to a relabeling of coordinates, as
\begin{equation}
    (Y_1,Y_2;Z_1,Z_2)
    \longrightarrow
    (Y_1,Y_2;\omega Z_1,\omega^{-1}Z_2),
    \qquad
    \omega=e^{2\pi i/N_f}.
\end{equation}

\subsection{Moduli space of ADHM theory}
\label{subsec:adhm-monopole-charge}

The moduli space of the ADHM theory has Higgs, Coulomb, and mixed branches. 

The Higgs branch is obtained by giving expectation values to the scalar fields in the hypermultiplets, subject to the corresponding F-term and D-term constraints and quotienting by the gauge symmetry. From the perspective of the D6-brane worldvolume theory, this branch is identified with the moduli space of instantons~\cite{Atiyah:1978ri,Douglas:1996sw}\footnote{Recall that codimension 4 branes appear as instantons in the worldvolume of higher-dimensional branes.}. The geometric isometries acting on the Higgs-branch directions are
\[
SO(4)_{3456}\simeq SU(2)_H\times SU(2)_F .
\]
In this paper, however, we will not be focusing much on the Higgs branch.

The Coulomb branch has a simple geometric interpretation in the M-theory
description. For a single M2-brane, it parametrizes the motion of the M2-brane
in the Taub--NUT directions. Near the core of the KK monopoles, the Taub--NUT
space approaches $\mathbb C^2/\mathbb Z_{N_f}$, and hence the Coulomb branch
of the abelian theory is
\begin{equation}
    \mathcal M_C^{U(1)}
    \simeq
    \frac{\mathbb C^2}{\mathbb Z_{N_f}}.
\end{equation}
For the $U(N)$ theory, describing $N$ indistinguishable M2-branes, this
generalizes to~\cite{Cremonesi:2013lqa,Bullimore:2015lsa}
\begin{equation}
    \mathcal M_C
    \simeq
    \operatorname{Sym}^N\left(
        \frac{\mathbb C^2}{\mathbb Z_{N_f}}
    \right).
\end{equation}

From the field-theory point of view, the Coulomb branch is parametrized
semiclassically by expectation values of the three real scalars in the vector
multiplet together with the dual photons. Choosing a complex structure, two
of the three real scalars are combined into a complex scalar $\varphi$, while
we denote the remaining real scalar by $\sigma$. At a generic point on the
Coulomb branch the gauge group is broken to its maximal torus, so that the
low-energy gauge fields are abelian. In three dimensions, an abelian gauge
field can be dualized to a compact scalar $\gamma$, conventionally called the
dual photon. Schematically,
\begin{equation}
    d\gamma \propto \frac{1}{g_{\rm YM}^2}\,*F ,
\end{equation}
where the precise normalization depends on the convention for the periodicity
of $\gamma$.

For a single $U(1)$ factor, it is useful to introduce the semiclassical
complex coordinate
\begin{equation}
    Y=\frac{2\pi}{g_{\rm YM}^2}\,\sigma+i\gamma,
\end{equation}
where we choose the normalization $\gamma\sim\gamma+2\pi$. The monopole
operators of magnetic charge $\pm1$ behave semiclassically as~\cite{Borokhov:2002cg,Bullimore:2015lsa}
\begin{equation}
    V_+\sim e^Y,
    \qquad
    V_-\sim e^{-Y}.
\end{equation}
Thus, at a generic point on the Coulomb branch, one may use
$(\varphi,Y)$, or equivalently $(\varphi,V_+)$ on a suitable patch, as
complex coordinates.

Quantum effects modify the global relation among the Coulomb-branch
operators. In the massless theory one finds~\cite{Intriligator:1996ex,Cremonesi:2013lqa,Bullimore:2015lsa}\footnote{This relation arises from quantum fluctuations of the charged
hypermultiplets in the monopole background. In the OPE of monopole
operators of equal and opposite magnetic charge, the corresponding
one-loop determinant produces one factor of $\varphi$ for each
fundamental hypermultiplet.}
\begin{equation}
    V_+V_-=\varphi^{N_f}.
\end{equation}

 The topology of the resulting space can be seen by introducing covering-space
coordinates $Z_1,Z_2$ through
\begin{equation}\label{eq:defVpmZ12}
    V_+=Z_1^{N_f},
    \qquad
    V_-=Z_2^{N_f},
    \qquad
    \varphi=Z_1Z_2,
\end{equation}
the relation $V_+V_-=\varphi^{N_f}$ is automatically satisfied.
Moreover, the transformation
\begin{equation}
    (Z_1,Z_2)
    \longrightarrow
    \left(
        e^{2\pi i/N_f}Z_1,
        e^{-2\pi i/N_f}Z_2
    \right)
\end{equation}
leaves $V_\pm$ and $\varphi$ invariant. Hence
\begin{equation}
    \mathcal M_C^{U(1)}
    \simeq
    \frac{\mathbb C^2}{\mathbb Z_{N_f}}.
\end{equation}

The pair $(Z_1,Z_2^*)$ transforms as a doublet under $SU(2)_C$.
With the conventional normalization in which the monopole operators
$V_\pm$ have topological charges $\pm1$, the $U(1)_T$ action on the
covering-space coordinates is
\begin{equation}
    Z_1\longrightarrow e^{i\alpha/N_f}Z_1,
    \qquad
    Z_2\longrightarrow e^{-i\alpha/N_f}Z_2,
\end{equation}
so that
\begin{equation}
    V_\pm\longrightarrow e^{\pm i\alpha}V_\pm,
    \qquad
    \varphi\longrightarrow\varphi .
\end{equation}
The scaling dimension of a bare monopole is~\cite{Cremonesi:2013lqa}
\begin{equation}
    \Delta(V_{\boldsymbol m})=j_C
    =\frac{N_f}{2}\sum_{i=1}^N|m_i|.
    \label{eq:adhm-monopole-dimension}
\end{equation}
In particular, a bare monopole with $m_i\geq0$ and $Q_T=m$ has
$\Delta=N_fm/2$.
If $J_1$ and $J_2$ generate phase rotations of $Z_1$ and $Z_2$,
respectively, the orbifold charge is
\begin{equation}
    J_{\rm orb}=J_1-J_2=N_fQ_T.
    \label{eq:adhm-orbifold-charge}
\end{equation}

\subsection{The ADHM superconformal index}
\label{subsec:adhm-index-review}

The superconformal index is defined in radial quantization by~\cite{Bhattacharya:2008zy}
\begin{equation}
    \mathcal I_{N,N_f}
    =\Tr_{\mathcal H(S^2)}
      \left[(-1)^F e^{-\beta\{Q,S\}}
      \prod_A \xi_A^{\mathcal Q_A}\right],
    \label{eq:adhm-index-definition}
\end{equation}
where all charges $\mathcal Q_A$ commute with the chosen supercharge.
We use the weight
\[
    p^{(\Delta+j_3)/2}\,t^{2(R_H-R_C)}\,s^{2F_3}\,r^{Q_T},
\]
where $j_3$ is the spatial-rotation Cartan and $R_H$, $R_C$, and $F_3$
are the Cartans of $SU(2)_H$, $SU(2)_C$, and $SU(2)_F$, normalized
so that a doublet has weights $\pm\frac12$.
The contributing states satisfy $\Delta=j_3+R_H+R_C$.
It is convenient to introduce four covering fugacities
\begin{equation}
    x=p^{1/4}ts,
    \qquad
    y=p^{1/4}ts^{-1},
    \qquad
    z=p^{1/4}t^{-1}r^{1/N_f},
    \qquad
    w=p^{1/4}t^{-1}r^{-1/N_f},
    \label{eq:adhm-cover-fugacities}
\end{equation}
so that
\begin{equation}
    u\equiv zw=p^{1/2}t^{-2},
    \qquad
    p=xyu=xyzw.
    \label{eq:adhm-p-u-def}
\end{equation}
The positive and negative unit monopoles then have weights
\begin{equation}
    \mathcal X_+=z^{N_f}=p^{N_f/4}t^{-N_f}r,
    \qquad
    \mathcal X_-=w^{N_f}=p^{N_f/4}t^{-N_f}r^{-1}.
    \label{eq:adhm-longitudinal-fugacities}
\end{equation}
This is the index counterpart of $J_{\rm orb}=\pm N_f$ for unit monopoles.

Let $f$ denote the combined vector-plus-adjoint-hypermultiplet
single-letter index, and let $h$ denote the fundamental
hypermultiplet letter.  They take the factorized form
\begin{align}
    1-f(x,y,u)
    &=\frac{(1-x)(1-y)(1-u)}{1-p},
    \label{eq:adhm-one-minus-letter-f}\\
    h(x,y,u)
    &=\sqrt{xy}\,\frac{1-u}{1-p}.
    \label{eq:adhm-letter-h}
\end{align}
At winding number $n$ we use
\begin{equation}
    f_n=f(x^n,y^n,u^n),
    \qquad
    h_n=h(x^n,y^n,u^n).
\end{equation}
With $SU(N_f)$ flavor fugacities $\boldsymbol a$ satisfying
$\prod_{\alpha=1}^{N_f}a_\alpha=1$, the fundamental sources are
\begin{equation}
    v_n=h_n\,
        \chi_{\mathbf{N_f}}(\boldsymbol a^n),
    \qquad
    \bar v_n=h_n\,
        \chi_{\overline{\mathbf{N_f}}}(\boldsymbol a^n).
    \label{eq:adhm-flavor-sources}
\end{equation}
In the flavor-unrefined limit, $v_n=\bar v_n=N_fh_n$.

To display the monopole dependence, suppose the GNO charge $s\in\mathbb Z$
appears with multiplicity $n_s\geq0$.  The flux breaks the gauge group as
\begin{equation}
    U(N)\longrightarrow\prod_{s\in\mathbb Z}U(n_s),
    \qquad
    \sum_s n_s=N.
    \label{eq:adhm-flux-breaking}
\end{equation}
For $U_s\in U(n_s)$ define
\begin{equation}
    \chi_{s,n}=\Tr U_s^n,
    \qquad
    \bar\chi_{s,n}=\Tr U_s^{-n}.
\end{equation}
The vector-plus-adjoint kernel between two flux blocks is
\begin{equation}
    \mathcal F_{s,t}^{(n)}
    =\delta_{s,t}
     +p^{\frac n2|s-t|}(f_n-1).
    \label{eq:adhm-direct-flux-kernel}
\end{equation}
The Kronecker term ensures that
$\mathcal F_{s,s}^{(n)}=f_n$; the normalized Haar measure of each
unbroken $U(n_s)$ factor is kept separately in $[dU_s]$.

Suppressing only a flux-independent normalization, the localization
formula~\cite{Kim:2009wb,Imamura:2011su}, applied to the ADHM matter
content~\cite{Gang:2011xp}, can be organized as
\begin{align}
    \mathcal I_{N,N_f}
    ={}&\sum_{\substack{\{n_s\geq0\}\\\sum_s n_s=N}}
       \mathcal X_+^{\sum_{s>0}s n_s}
       \mathcal X_-^{\sum_{s<0}|s|n_s}
       \int\prod_s[dU_s]
    \nonumber\\
    &\times\exp\!\left\{
      \sum_{n=1}^{\infty}\frac1n
      \left[
        \sum_{s,t}\mathcal F_{s,t}^{(n)}
             \chi_{s,n}\bar\chi_{t,n}
       +\sum_s p^{\frac n2|s|}
          \bigl(v_n\chi_{s,n}+\bar v_n\bar\chi_{s,n}\bigr)
      \right]\right\}.
    \label{eq:adhm-general-index}
\end{align}
The factor $p^{n|s-t|/2}$ is the angular-momentum cost of a field
connecting two different fluxes, while $p^{n|s|/2}$ is the corresponding
cost for a fundamental field attached to the $s$-node.

\section{BMN triality for ADHM theory}\label{sec:ADHMtriality}
In this section, following \cite{Komatsu:2026ano}, we study the limit in which
$AdS_4\times S^7/\mathbb Z_{N_f}$ reduces to the maximally supersymmetric
eleven-dimensional pp-wave with a compact lightlike direction. This is a
combined Penrose and large-$N_f$ limit. Translating it to the ADHM theory
yields the parameter dictionary relating the ADHM and BMN descriptions. We
then test the resulting triality using the superconformal index and find that,
in the same limit, the ADHM index reduces to the grand-canonical index of the
BMN matrix model.
\subsection{BMN limit of the ADHM theory}
\label{sec:adhm-bmn-limit}

We now derive the limit in which the ADHM theory describes the DLCQ of
the eleven-dimensional pp-wave.  We first obtain the scaling from the
bulk geometry and then show that the same limit of the ADHM index gives
the grand-canonical BMN Witten index.


Write the transverse space as
\begin{equation}
    \mathbb C^4=\mathbb C^2_H\oplus\mathbb C^2_C
\end{equation}
and introduce complex coordinates $(Y_1,Y_2;Z_1,Z_2)$.  The orbifold
acts only on the second factor,
\begin{equation}
    (Y_1,Y_2;Z_1,Z_2)
    \sim
    (Y_1,Y_2;e^{2\pi i/N_f}Z_1,e^{-2\pi i/N_f}Z_2).
    \label{eq:adhm-s7-orbifold-action}
\end{equation}
As is clear from this orbifold action, the Coulomb branch coordinates $Z_{1,2}$ in \eqref{eq:defVpmZ12} correspond to $Z_{1,2}$ here and the global $U(1)$ charges $J_{1,2}$ \eqref{eq:adhm-orbifold-charge} of the ADHM theory are dual to the angular momenta corresponding to phase rotations of the complex coordinates $Z_{1,2}$.

We take the Penrose limit around the null geodesic at the center of
$AdS_4$ that moves along
\begin{equation}
    Z_1=e^{i\phi},
    \qquad
    Y_1=Y_2=Z_2=0.
    \label{eq:adhm-penrose-geodesic}
\end{equation}
Along this orbit the quotient acts as
\begin{equation}
    \phi\sim\phi+\frac{2\pi}{N_f}.
    \label{eq:adhm-geodesic-identification}
\end{equation}
This choice is essential: a geodesic lying entirely in the unorbifolded
$\mathbb C^2_H$ would give the local pp-wave but would not generate the
desired compact null direction.

Let $R$ denote the radius of $S^7$; the $AdS_4$ radius is $R/2$.\footnote{The
notation of Appendix~\ref{app:susy_matching} is related to ours by
$R=2L$, $\phi=\psi$, and
$(Y_1,Y_2;Z_1,Z_2)=(z_3,z_4;z_1,z_2)$, with orbifold order $k=N_f$.
We use dimensionless global $AdS_4$ time $T$; the physical boundary time
in the appendix is $t=rT$, with $r$ the boundary-sphere radius.
The appendix uses $\mu=3$. The lightcone coordinates in the Appendix $x^\pm_{\rm App}$ are related to the lightcone coordinates here \eqref{eq:adhm-canonical-lightcone} by  $x^+=(3/\mu)x^+_{\rm App}$ and
$x^-=(\mu/3)x^-_{\rm App}$.}  Flux
quantization gives \cite{Itzhaki:1998uz,Pelc:1999ms}
\begin{equation}
    R^6=32\pi^2NN_f\,\ell_p^6.
    \label{eq:adhm-radius-flux}
\end{equation}
The metric and four-form flux may be written as
\begin{align}
    ds^2
    &=\frac{R^2}{4}ds^2_{AdS_4}+R^2ds^2_{S^7/\mathbb Z_{N_f}},
    \\
    F_4&=\frac{3R^3}{8}\,\epsilon_{AdS_4}.
    \label{eq:adhm-ads-background}
\end{align}
Near the geodesic, introduce transverse coordinates of order $R^0$ and
lightcone coordinates $U$ and $V$ defined by
\begin{equation}
    T=U+\frac{4V}{R^2},
    \qquad
    \phi=\frac{U}{2}-\frac{2V}{R^2},
    \label{eq:adhm-penrose-coordinates}
\end{equation}
where $T$ is the dimensionless global AdS$_4$ time coordinate.  
Taking $R\to\infty$ gives
\begin{align}
    ds^2
    &=-4\,dU\,dV
      -\left(r^2+\frac{y^2}{4}\right)dU^2
      +d\boldsymbol r^2+d\boldsymbol y^2,
    \label{eq:adhm-ppwave-intermediate}\\
    F_4
    &=3\,dU\wedge dr_1\wedge dr_2\wedge dr_3.
    \label{eq:adhm-ppwave-flux}
\end{align}
This is the maximally supersymmetric eleven-dimensional pp-wave~\cite{Kowalski-Glikman:1984qtj,Figueroa-OFarrill:2001hal}.  The
opposite rotation of $Z_2$ in \eqref{eq:adhm-s7-orbifold-action} becomes
infinitesimal in the scaling limit, so no residual transverse orbifold
survives locally.

At fixed $T$, the quotient shifts $V$ by
\begin{equation}
    \Delta V=\frac{\pi R^2}{2N_f}
    \label{eq:adhm-V-period}
\end{equation}
up to an irrelevant orientation sign.  Passing to the conventional BMN
coordinates
\begin{equation}
    x^+=\frac{3U}{\mu},
    \qquad
    x^-=\frac{2\mu V}{3},
    \label{eq:adhm-canonical-lightcone}
\end{equation}
one obtains
\begin{equation}
    x^-\sim x^-+2\pi R_- ,
    \qquad
    R_-=\frac{\mu R^2}{6N_f}
       =\frac{\mu\ell_p^2}{6}(32\pi^2)^{1/3}
        \left(\frac{N}{N_f^2}\right)^{1/3}.
    \label{eq:adhm-null-radius}
\end{equation}
Therefore a finite null radius requires
\begin{equation}
    N,N_f\longrightarrow\infty,
    \qquad
    \frac{N}{N_f^2}=\lambda_{pp}\quad\text{fixed}.
    \label{eq:adhm-double-scaling-geometry}
\end{equation}
In these conventions, the dimensionless BMN coupling is
\begin{equation}
    g_{\rm BMN}^2
    =\frac{R_-^3}{\mu^3\ell_p^6}
    =\frac{4\pi^2}{27}\frac{N}{N_f^2}=\frac{4\pi^2}{27}\lambda_{pp}.
    \label{eq:adhm-bmn-coupling-map}
\end{equation}

The dimensionless lightcone Hamiltonian is the near-BPS combination
\begin{equation}
    H_{\rm lc}=\Delta-\frac{J_1}{2}.
    \label{eq:adhm-lightcone-Hamiltonian}
\end{equation}
The generator of translations in $x^+$ is
$i\partial_{x^+}=(\mu/3)H_{\rm lc}$.
Orbifold invariance quantizes $J_{\rm orb}=J_1-J_2$ in multiples of
$N_f$.  Thus the integer longitudinal momentum in the Penrose limit is
\begin{equation}
    m=\frac{J_{\rm orb}}{N_f}=Q_T,
    \label{eq:adhm-dlcq-momentum}
\end{equation}
For positive bare-monopole backgrounds, $J_2=0$ and $J_1=N_fm$.
Finite transverse excitations have $J_2=O(1)$, so that
$J_1/N_f=m+J_2/N_f\to m$ in the limit.
The sector with fixed $m$ is described by the $U(m)$ BMN matrix
model.  

Combining \eqref{eq:adhm-double-scaling-geometry} with
\eqref{eq:adhm-dlcq-momentum} reproduces precisely the triple-scaling limit
\eqref{eq:adhm-triple-scaling-summary}. Together with the BMN conjecture,
which identifies the DLCQ of M-theory on the pp-wave with the BMN matrix
model, this yields the triality proposed in the Introduction.

\subsection{BMN limit of the ADHM superconformal index}\label{adhmindlim}

We next analyze the superconformal index \eqref{eq:adhm-general-index} of ADHM theory in the scaling
limit
\begin{equation}
    N,N_f\longrightarrow\infty,
    \qquad
    \lambda_{pp}\equiv\frac{N}{N_f^2}\quad\text{fixed}\,,
    \label{eq:adhm-double-scaling-index}
\end{equation}
in order to test the triality with the BMN matrix model. 
The BMN comparison concerns sectors with finite total monopole charge.
Their nonzero-flux blocks have finite rank, while the zero-flux block
contains $N-O(1)$ eigenvalues.\footnote{
A different ADHM--BMN index limit was studied in~\cite{Duary:2026tub}:
the positive GNO charges grow with their differences and multiplicities
fixed, at fixed $N_f$. Here the nonzero charges and their multiplicities
remain finite while $N$ and $N_f$ grow.
} We first determine the latter's saddle,
keeping the $N_f$-dependence of the fundamental sources explicit, and
then integrate it out.

\paragraph{The neutral large-\texorpdfstring{$N$}{N} saddle}
We first set all GNO charges to zero and introduce the normalized
holonomy moments
\begin{equation}
    \rho_n=\frac1N\Tr U^n,
    \qquad
    \rho_{-n}=\frac1N\Tr U^{-n}.
\end{equation}
The Haar measure and the zero-flux terms of
\eqref{eq:adhm-general-index} give the large-$N$ action as
\begin{equation}
    S^{(0)}
    =\sum_{n=1}^{\infty}\frac1n
      \left[
        -N^2(1-f_n)\rho_n\rho_{-n}
        +Nv_n\rho_n+N\bar v_n\rho_{-n}
      \right].
    \label{eq:adhm-neutral-action}
\end{equation}
Its saddle is
\begin{equation}
    \rho_n^\star=\frac{\bar v_n}{N(1-f_n)},
    \qquad
    \rho_{-n}^\star=\frac{v_n}{N(1-f_n)}.
    \label{eq:adhm-neutral-saddle}
\end{equation}
In the flavor-unrefined limit,
\begin{equation}
    \rho_n^\star
    =\frac{h_n}
      {\sqrt{\lambda_{pp}N}\,(1-f_n)}
    =O(N^{-1/2}).
    \label{eq:adhm-neutral-shift-scaling}
\end{equation}
The limiting density is therefore uniform, but the first displacement
cannot be discarded: its on-shell action is of order $N_f^2=O(N)$.

Substitution of the saddle gives
\begin{equation}
    \log\mathcal I_{\rm saddle}^{(0)}
    =\sum_{n=1}^{\infty}\frac1n
      \frac{v_n\bar v_n}{1-f_n}.
    \label{eq:adhm-neutral-onshell-general}
\end{equation}
For unrefined flavor fugacities,
\begin{equation}
    \log\mathcal I_{\rm saddle}^{(0)}
    =N_f^2\sum_{n=1}^{\infty}\frac1n
      \frac{(xy)^n(1-u^n)}
      {(1-p^n)(1-x^n)(1-y^n)}.
    \label{eq:adhm-neutral-onshell-unrefined}
\end{equation}
The Gaussian determinant is
\begin{equation}
    \mathcal I_{\rm 1-loop}^{(0)}
    =\prod_{n=1}^{\infty}\frac{1}{1-f_n}
    =\prod_{n=1}^{\infty}
      \frac{1-p^n}{(1-x^n)(1-y^n)(1-u^n)}.
    \label{eq:adhm-neutral-one-loop}
\end{equation}
Equivalently, the full neutral factor is
\begin{equation}
    \mathcal I_\infty^{(0)}
    =\PE\!\left[
      i_{\rm tr}
      +g\,\chi_{\mathbf{N_f}}(\boldsymbol a)
          \chi_{\overline{\mathbf{N_f}}}(\boldsymbol a)
    \right],
    \label{eq:adhm-neutral-full}
\end{equation}
where
\begin{align}
    i_{\rm tr}
    &=\frac{x}{1-x}+\frac{y}{1-y}+\frac{u}{1-u}
      -\frac{p}{1-p},
    \\
    g
    &=\frac{xy(1-u)}{(1-p)(1-x)(1-y)}.
    \label{eq:adhm-neutral-g}
\end{align}
The neutral contribution describes the zero-longitudinal-momentum sea and must be divided out
before comparing the finite-$m$ sectors with the BMN matrix model.

\paragraph{Integrating out the zero-flux sea}
Next, we consider a non-zero flux configuration with
\begin{equation}
    n_0=N-O(1),
    \qquad
    \sum_{s\ne0}n_s=O(1).
    \label{eq:adhm-finite-flux-regime}
\end{equation}
Thus the zero-flux block is macroscopic, while all nonzero-flux blocks
have finite rank.  For a fixed winding number $n$, define
\begin{equation}
    T_n=\chi_{0,n},
    \qquad
    \bar T_n=\bar\chi_{0,n},
    \qquad
    c_{s,n}=p^{\frac n2|s|}(f_n-1)
    \quad(s\ne0).
    \label{eq:adhm-T-c-def}
\end{equation}
The part of the exponent involving the zero node is
\begin{equation}
    S_{0,n}
    =\frac1n\left[
       -(1-f_n)T_n\bar T_n
       +A_nT_n+B_n\bar T_n
    \right],
    \label{eq:adhm-zero-node-action}
\end{equation}
with
\begin{align}
    A_n&=v_n+\sum_{s\ne0}c_{s,n}\bar\chi_{s,n},
    \\
    B_n&=\bar v_n+\sum_{s\ne0}c_{s,n}\chi_{s,n}.
    \label{eq:adhm-zero-node-sources}
\end{align}
At large $n_0$, $T_n$ and $\bar T_n$ are Gaussian variables.  Completing
the square gives
\begin{align}
    &-(1-f_n)T_n\bar T_n+A_nT_n+B_n\bar T_n
    \nonumber\\
    &\qquad=-(1-f_n)
      \left(T_n-\frac{B_n}{1-f_n}\right)
      \left(\bar T_n-\frac{A_n}{1-f_n}\right)
      +\frac{A_nB_n}{1-f_n}.
    \label{eq:adhm-completing-square}
\end{align}
The determinant reproduces the neutral factor
$(1-f_n)^{-1}$, while the induced exponent is
\begin{align}
    \frac{A_nB_n}{1-f_n}
    ={}&\frac{v_n\bar v_n}{1-f_n}
    \nonumber\\
    &+\sum_{s\ne0}\frac{c_{s,n}}{1-f_n}
      \bigl(v_n\chi_{s,n}+\bar v_n\bar\chi_{s,n}\bigr)
    \nonumber\\
    &+\sum_{s,t\ne0}
      \frac{c_{s,n}c_{t,n}}{1-f_n}
      \bar\chi_{s,n}\chi_{t,n}.
    \label{eq:adhm-induced-exponent}
\end{align}
The first line is the neutral on-shell contribution.  The second line
exactly cancels the direct fundamental determinant in \eqref{eq:adhm-general-index}:
\begin{align}
   &p^{n|m|/2}
      \bigl(v_n\chi_{m,n}+\bar v_n\bar\chi_{m,n}\bigr)
   +\frac{c_{m,n}}{1-f_n}
      \bigl(v_n\chi_{m,n}+\bar v_n\bar\chi_{m,n}\bigr)
   =0.
   \label{eq:fundamental-cancellation}
\end{align}
The last line of \eqref{eq:adhm-induced-exponent}, combined with the
direct nonzero-node determinant, produces the effective kernel
\begin{equation}
    f_{s,t}^{\rm eff}(n)
    =\delta_{s,t}
     +\left[
       p^{\frac n2|s-t|}
       -p^{\frac n2(|s|+|t|)}
      \right](f_n-1),
    \qquad s,t\ne0.
    \label{eq:adhm-effective-kernel}
\end{equation}
If $st<0$, then $|s-t|=|s|+|t|$ and the effective interaction vanishes.
Consequently the large-$N$ index factorizes as~\cite{Gang:2011xp,Imamura:2011uj}
\begin{equation}
    \mathcal I_\infty
    =\mathcal I_\infty^{(0)}
     \mathcal I_\infty^{(+)}
     \mathcal I_\infty^{(-)}.
    \label{eq:adhm-large-N-factorization}
\end{equation}

\paragraph{The BMN index from the positive-monopole sector}

We now restrict to positive fluxes $s,t>0$.  Using
\eqref{eq:adhm-one-minus-letter-f}, the effective kernel becomes
\begin{align}
    f_{s,t}^{\rm eff}(n)
    &=\delta_{s,t}
      -\frac{
       p^{\frac n2|s-t|}-p^{\frac n2(s+t)}}{1-p^n}
       (1-x^n)(1-y^n)(1-u^n)
    \nonumber\\
    &=\delta_{s,t}
      -\sum_{j=\frac12|s-t|}^{\frac12(s+t)-1}
       p^{nj}(1-x^n)(1-y^n)(1-u^n).
    \label{eq:adhm-effective-kernel-geometric-sum}
\end{align}
This is exactly the BMN single-letter index for fluctuations between fuzzy-sphere blocks of dimensions $s$ and $t$~\cite{Chang:2024lkw}, based on the fluctuation spectrum of~\cite{Dasgupta:2002hx,Kim:2002if}, under the identification
\begin{equation}
    a=x,\qquad b=y,\qquad c=u,\qquad abc=p.
    \label{eq:adhm-bmn-fugacity-map}
\end{equation}

It follows that the positive-monopole contribution is
\begin{align}
    \mathcal I_\infty^{(+)}(X;x,y,u)
    ={}&\sum_{\{n_s\}_{s>0}}
      X^{\sum_s s n_s}
      \int\prod_{s>0}[dU(n_s)]
    \nonumber\\[-0.2em]
    &\times\exp\!\left[
      \sum_{n=1}^{\infty}\frac1n
      \sum_{s,t>0}
      \iota_{s,t}(x,y,u;n)
      \Tr U_s^{-n}\Tr U_t^n
    \right]
    \nonumber\\
    ={}&\sum_{m=0}^{\infty}X^m
      \mathcal I_{{\rm BMN},U(m)}(x,y,u),
    \label{eq:adhm-positive-is-bmn}
\end{align}
where
\begin{equation}
    X=\mathcal X_+=z^{N_f}
    \label{eq:adhm-X-positive}
\end{equation}
counts positive topological charge, equivalently units of DLCQ
momentum. The equality of the effective letter kernels, and hence
the positive-sector index in \eqref{eq:adhm-positive-is-bmn},
already follows after integrating out the large-$N$ zero-flux
sea. It does not require taking $N_f$ large. Unlike in ABJM \cite{Komatsu:2026ano},
there are no two sets of GNO charges whose equality must be
selected by an additional limit.


The negative-flux blocks have the same effective kernel, with fugacity $Y=\mathcal X_-=w^{N_f}$. Thus, before taking the large-$N_f$ limit, the normalized index factorizes as
\begin{equation}
    \frac{\mathcal I_\infty}{\mathcal I_\infty^{(0)}}
    =
    \mathcal I_\infty^{(+)}(X;x,y,u)\,
    \mathcal I_\infty^{(-)}(Y;x,y,u),
    \qquad
    Y=w^{N_f}=\frac{u^{N_f}}{X},
    \quad u=zw.
    \label{eq:adhm-two-orientation-index}
\end{equation}

The large-$N_f$ limit removes these additional contributions.
Keeping $X=z^{N_f}$ and $u=zw$ fixed with $|u|<1$ gives
\begin{equation}
    Y=\frac{u^{N_f}}{X}\longrightarrow0.
    \label{eq:adhm-negative-suppression}
\end{equation}
The negative-flux factor then tends to one, leaving the
positive-sector BMN index. In the correlated limit required for
a finite bulk null radius, we obtain
\begin{equation}
    \lim_{\substack{N,N_f\to\infty\\
                     N/N_f^2=\lambda_{\rm pp}\\
                     z^{N_f}=X,\;zw=u}}
    \frac{\mathcal I_{N,N_f}(x,y,z,w)}
         {\mathcal I_\infty^{(0)}(x,y,u)}
    =\sum_{m=0}^{\infty}X^m
      \mathcal I_{{\rm BMN},U(m)}(a,b,c).
    \label{eq:adhm-bmn-final-index-relation}
\end{equation}


\paragraph{Half-BPS check and the vacuum dictionary}
The vacuum correspondence is already visible in the half-BPS limit.
A purely positive GNO sector is specified by a partition
\begin{equation}
    m=\sum_{s>0}s n_s.
\end{equation}
Because the ADHM theory has no Chern--Simons Gauss-law charge, the bare
monopole does not require the $O(N_fm)$ matter dressing that appears in
ABJM theory.  Its large dimension is instead the monopole zero-point
energy.  The same partition labels a
BMN vacuum, and the residual gauge group agrees on the two sides.

Turning off the transverse excitation fugacities sets $x,y,$ and $u\to0$.
Equation \eqref{eq:adhm-bmn-final-index-relation} reduces to
\begin{equation}
    \PE\!\left[\frac{X}{1-X}\right]
    =\prod_{n=1}^{\infty}\frac{1}{1-X^n}
    =\sum_{m=0}^{\infty}p(m)X^m,
    \label{eq:adhm-half-bps-partitions}
\end{equation}
where $p(m)$ is the number of integer partitions of $m$.  This equals
the number of classical vacua of the $U(m)$ BMN matrix model~\cite{Berenstein:2002jq,Dasgupta:2002hx} and gives
the simplest check of the ADHM--BMN identification.

\section{Topological correlators in the BMN limit}
\label{sec:adhm-bmn-correlators}

The match between the superconformal index of ADHM theory and the
grand-canonical index of BMN quantum mechanics suggests a correspondence
between BPS monopole sectors in ADHM theory and fuzzy-sphere vacua in
BMN quantum mechanics. In this section, we sharpen this correspondence
by comparing the protected correlators computed by the two localization
constructions, together with the supercharges and operator insertions
that define them.

We first review the BMN localization of \cite{Asano:2017nxw} in
Subsection~\ref{subsec:asano-supercharge} and the ADHM Coulomb-branch
localization of \cite{Dedushenko:2017avn,Dedushenko:2018icp} in Subsection~\ref{subsec:adhm-localization}.
Subsection~\ref{subsec:adhm-bmn-supercharge-match} then matches the localizing supercharges. 
In Subsection~\ref{subsec:adhm-bmn-fixed-backgrounds}, we show that the triple-scaling limit reduces the ADHM measure in a fixed positive magnetic-flux sector to the BMN measure in the
corresponding fuzzy-sphere vacuum.
Subsection~\ref{subsec:adhm-bmn-operator-dictionary} establishes the operator dictionary for comparing protected correlators in these fixed sectors.
Finally, we discuss monopole bubbling in Subsection~\ref{subsec:adhm-bmn-bubbling-scope}, which allows a single monopole
operator to contribute to several screened flux sectors, and explore
its possible connection to BMN instantons.

\subsection{Review of BMN localization}
\label{subsec:asano-supercharge}

We review the localization results of \cite{Asano:2012zt,Asano:2017nxw} in this
subsection. In \cite{Asano:2017nxw}, a ``double'' Wick rotation is performed which
makes one of the $SO(6)$ scalar fields imaginary and the worldline time
Euclidean. The labels for the $SO(3)$ scalars are $Y_{2,3,4}$ and those
for the $SO(6)$ scalars are $Y_{5,6,7,8,9,10}$. In their notation,
$Y_{10}$ is Wick rotated by defining $Y_0\equiv iY_{10}$. We relate the
worldline Lorentzian and Euclidean times by $t_A=-i\tau_A$, and the
worldline gauge field is denoted by $A_{\tau_A}\equiv Y_1$.

The supersymmetry parameter used for localization has the following
components as a sixteen-component BMN spinor, corresponding to the
dynamical sector of the eleven-dimensional target-space supersymmetry,
in an appropriate basis (see Appendix A of \cite{Asano:2012zt} for the
explicit choice of basis):\footnote{
The first exponent in \eqref{eq:asano-killing-spinor} has the opposite
sign to that printed in equation (3.2) of \cite{Asano:2012zt}.
Rewriting their $\epsilon_+$ solution in equation (2.8) in compact
form, using their gamma-matrix conventions, gives the sign in
\eqref{eq:asano-killing-spinor}. This suggests a typographical sign
error in their equation (3.2).
}
\begin{equation}
    \epsilon_{\rm A}(\tau_A)
    =
    e^{-\frac{\tau_A}{2}\Gamma^{09}}
    e^{-\frac{\pi}{4}\Gamma^{49}}\chi,
    \qquad
    \chi=
    \begin{pmatrix}
        \eta_1\\
        0\\
        0\\
        0
    \end{pmatrix},
    \qquad
    \eta_1=(1,0,0,0)^{\mathsf T}.
    \label{eq:asano-killing-spinor}
\end{equation}
In the explicit gamma-matrix basis used in \cite{Asano:2017nxw}, the
action of some of the gamma-matrix combinations on $\chi$ and
$\chi'=-\Gamma^{49}\chi$ is as follows:
\begin{equation}
    \Gamma^{49}\chi=-\chi',
    \qquad
    \Gamma^{49}\chi'=\chi,
    \qquad
    \Gamma^{09}\chi=-\chi,
    \qquad
    \Gamma^{09}\chi'=+\chi'.
\end{equation}
Therefore,
\begin{equation}
    \epsilon_{\rm A}(\tau_A)
    =
    \frac{1}{\sqrt{2}}
    \left(
        e^{+\tau_A/2}\chi
        +
        e^{-\tau_A/2}\chi'
    \right).
    \label{eq:asano-spinor-two-components}
\end{equation}
Under
$Spin(3)\times Spin(6)\simeq SU(2)\times SU(4)$, the two components
belong to $(\boldsymbol{2},\boldsymbol{4})$ and
$(\boldsymbol{2},\overline{\boldsymbol{4}})$, respectively, with an
appropriate choice of chirality convention after returning to
Lorentzian signature.\footnote{
This is because they are related by $\Gamma^{49}$, which changes the
chirality of the $SO(6)$ part of the spinor and hence exchanges the
conjugate representations $\boldsymbol{4}$ and
$\overline{\boldsymbol{4}}$ of $SU(4)\simeq Spin(6)$.
} 

In \cite{Asano:2017nxw}, correlation functions of products of functions
of a scalar annihilated by the chosen supercharge are considered.
With the spinor convention above, this scalar takes the form
\begin{equation}
    \Phi_{\rm coh}
    =
    Y_0\cosh\tau_A-Y_4+Y_9\sinh\tau_A,
    \qquad
    \delta_s\Phi_{\rm coh}=0.
    \label{eq:asano-cohomological-scalar}
\end{equation}
After undoing the analytic continuation, the corresponding
Lorentzian protected operator, with the same scalar labels, is
\begin{equation}
    \phi(t_A)
    =
    -Y_4(t_A)
    +
    i\left(
        Y_{10}(t_A)\cos t_A
        +
        Y_9(t_A)\sin t_A
    \right).
    \label{eq:asano-protected-scalar}
\end{equation}
Four BMN supercharges annihilate the corresponding protected scalar;
the spinor \eqref{eq:asano-killing-spinor} selects one of them for
localization.

After including the BRST transformation $\delta_B$, the square of
the combined supercharge $Q=\delta_s+\delta_B$ is a linear combination
of a diagonal $SO(3)\times SO(6)$ rotation (the first term below)
and a gauge transformation (the second term below):
\begin{equation}
    Q^2
    =
    -i\,\delta_{U(1)}
    +
    \delta_{\rm gauge}(a_0),
    \qquad
    \delta_{\rm gauge}(a_0)X:=[X,a_0].
    \label{eq:asano-q-square-review}
\end{equation}

The path integral is performed by imposing fuzzy-sphere boundary
conditions in the far Euclidean past and future. In the fuzzy-sphere
vacua, the $SO(3)$ scalars take a block diagonal form in terms of
generators of $SU(2)$,
\begin{equation}
    Y_{i+1}=-2L_i,
    \qquad
    i=1,2,3.
\end{equation}
On the localization locus, the scalar in
\eqref{eq:asano-cohomological-scalar} takes the following form in terms
of constant Hermitian matrices $M$:
\begin{equation}
    [M,L_i]=0,
    \qquad
    \phi_{\rm loc}
    :=
    \left.\Phi_{\rm coh}\right|_{\rm loc}
    =
    2L_3+iM.
    \label{eq:asano-localization-locus}
\end{equation}
Here $Y_{10}=M/\cosh\tau_A$ and $Y_9=0$ on the locus. The final
correlator involves an integral over these matrices $M$.

\paragraph{Localized matrix integral.}
Write the irreducible decomposition of the fixed vacuum and the
commuting modulus as
\begin{equation}
\begin{aligned}
    L_i
    &=
    \bigoplus_{s=1}^{\Lambda}
    \left(L_i^{[n_s]}\otimes\mathbf{1}_{N_s}\right),
    &
    M
    &=
    \bigoplus_{s=1}^{\Lambda}
    \left(\mathbf{1}_{n_s}\otimes M_s\right),
    \\
    M_s
    &=
    \operatorname{diag}(q_{s,1},\ldots,q_{s,N_s}),
    &
    m
    &=
    \sum_{s=1}^{\Lambda}n_sN_s,
    \qquad
    \nu=\sum_{s=1}^{\Lambda}N_s.
\end{aligned}
\label{eq:pwmm-review-decomposition}
\end{equation}
where $L_i^{[n_s]}$ denotes the irreducible $SU(2)$ representation of
dimension $n_s$. Up to an eigenvalue-independent normalization,
localization gives the measure
\begin{equation}
\begin{aligned}
    d\mu_{\rm BMN}^{L}(M)
    &=
    \frac{1}{\prod_{s=1}^{\Lambda}N_s!}
    \prod_{s=1}^{\Lambda}\prod_{a=1}^{N_s}dq_{s,a}\,
    \exp\left[
        -\frac{2}{g_{\rm loc}^{\,2}}
        \sum_{s=1}^{\Lambda}n_s
        \sum_{a=1}^{N_s}q_{s,a}^{\,2}
    \right]
    Z_{\rm 1\text{-}loop}^{L}(M),
    \\[1mm]
    Z_{\rm 1\text{-}loop}^{L}(M)
    &=
    \prod_{s,t=1}^{\Lambda}
    \prod_{J=|n_s-n_t|/2}^{(n_s+n_t)/2-1}
    \prod_{a=1}^{N_s}
    \prod_{b=1}^{N_t}
    {}'
    \left[
        \frac{
            \bigl((2J+2)^2+q_{st,ab}^{\,2}\bigr)
            \bigl((2J)^2+q_{st,ab}^{\,2}\bigr)
        }{
            \bigl((2J+1)^2+q_{st,ab}^{\,2}\bigr)^2
        }
    \right]^{1/2}.
\end{aligned}
\label{eq:pwmm-fixed-vacuum-measure}
\end{equation}
where $q_{st,ab}:=q_{s,a}-q_{t,b}$. The prime omits the vanishing
$(2J)^2+q_{st,ab}^{\,2}$ factor when $s=t$, $J=0$, and $a=b$.
The products over $(s,t)$ and $(a,b)$ are unordered; the square root
therefore gives one full pair factor for every ordered pair of
eigenvalues. The Weyl factor in
\eqref{eq:pwmm-fixed-vacuum-measure} may instead be absorbed into the
normalization and is displayed only to fix our convention.
Also note that the dimensionless coupling $g_{loc}$ above is related to the dimensionless coupling we defined in \eqref{eq:adhm-bmn-coupling-map} by\footnote{This can be seen as follows: using equations (3.10) and (2.4) in \cite{Asano:2017nxw}, we can deduce that $g_{loc}^2= \frac{54}{\pi^2}\frac{R_-^3}{\mu^3 l_p^6}$, comparing this with our \eqref{eq:adhm-bmn-coupling-map}, we get \eqref{mapcoup}. }
\begin{equation}\label{mapcoup}
    g^2_{loc}= \frac{54}{\pi^2}g^2_{BMN}
\end{equation}
Consequently, for arbitrary smooth functions $f_\alpha$ for which
the matrix integral converges,
\begin{equation}
    \left\langle
        \prod_{\alpha=1}^{p}
        \operatorname{Tr}f_\alpha\bigl(\phi(t_\alpha)\bigr)
    \right\rangle_{\rm BMN,L}
    =
    \frac{
        \displaystyle
        \int d\mu_{\rm BMN}^{L}(M)\,
        \prod_{\alpha=1}^{p}
        \operatorname{Tr}f_\alpha(2L_3+iM)
    }{
        \displaystyle
        \int d\mu_{\rm BMN}^{L}(M)
    }.
    \label{eq:pwmm-review-correlators}
\end{equation}
The right-hand side is independent of the insertion times.
Equations \eqref{eq:pwmm-fixed-vacuum-measure} and
\eqref{eq:pwmm-review-correlators} hold at finite matrix rank and finite
$g_{\rm loc}$ within the perturbative localization sector defined by
the boundary condition at the chosen vacuum.

It is important to note that possible instanton or kink
configurations interpolating between distinct fuzzy-sphere vacua
are not included \cite{Asano:2012zt,Asano:2017nxw}. They may be neglected whenever they are absent
or parametrically suppressed; otherwise the fixed-vacuum matrix
integral is only one contribution to the full answer.

\subsection{Review of ADHM Coulomb-branch localization}
\label{subsec:adhm-localization}

In this subsection, we review the protected ADHM Coulomb-branch correlators, mostly following the computations in \cite{Dedushenko:2017avn,Dedushenko:2018icp}. The correlation functions of interest are the twisted Coulomb-branch correlators. The operator insertions involve dressed monopole operators and polynomials in the vector-multiplet scalars $\phi_{\dot a\dot b}$ shown in Table~\ref{tab:adhm-field-content}. The insertions are placed on a line in $\mathbb{R}^3$ or a great circle in $S^3$.\footnote{These two descriptions are related by a conformal map from $\mathbb{R}^3$ to $S^3$.} The correlators are topological in the sense that they do not depend on the positions of the insertions on the great circle, but only on their relative ordering. The operator insertions lie in the cohomology of a particular supercharge $\mathcal Q^C_\beta$, which we describe below.

Let us
write the Poincar\'e and conformal supersymmetry generators of the superconformal algebra of the ADHM theory,
$\mathfrak{osp}(4|4)$ as
\begin{equation}
    Q_{\alpha a\dot a},
    \qquad
    S_{\alpha a\dot a},
\end{equation}
where $\alpha$ is an $SU(2)_{\rm sp}$ spinor index,
$a$ is an $SU(2)_H$ index, and $\dot a$ is an $SU(2)_C$ index.  The
Coulomb-branch localization supercharge is a linear combination of the Poincare and conformal supercharges and is written as:
\begin{equation}
    \mathcal Q_\beta^C
    =\mathcal Q_1^C+\beta\mathcal Q_2^C,
    \qquad \beta\neq0.
    \label{eq:adhm-Q-beta-review}
\end{equation}
Both $\mathcal Q_1^C$ and $\mathcal Q_2^C$ are linear combinations of $Q$ and $S$. When written as a generator for local supersymmetry on $\mathbb{R}^3$, the supercharge above takes the following form:\footnote{Here
$
s_{\alpha a\dot a}
=
2\,\eta_{\alpha a\dot a}
$
in terms of the conformal-supersymmetry parameter $\eta$ of \cite{Dedushenko:2017avn}, and $\gamma_i^{(3)}$ are the three dimensional gamma matrices. These spinors are obtained after taking a conformal map from $S^3$ to $\mathbb R^3$ of spinors listed in equation 3.1 of \cite{Dedushenko:2017avn}}
\begin{equation}
\xi_{\alpha a\dot a}(x)
=
q_{\alpha a\dot a}
+
\frac{x^i}{2}
(\gamma_i^{(3)})_\alpha{}^\beta
s_{\beta a\dot a}
\label{eq:DFPY_flat_spinor}
\end{equation}
The explicit components of the spinors $q$ and $s$ are given in
\eqref{eq:DFPY_q} and \eqref{eq:DFPY_s}. The equivariant square of
the supercharge at the superconformal fixed point is
\begin{equation}
    \left(\mathcal Q_\beta^C\right)^2
    =
    {4i\beta}\left(P_\tau+R_H\right),
    \label{eq:adhm-Q-square-review}
\end{equation}

We take $\beta=1$.
Here $P_\tau$ generates an isometry whose fixed locus is a great
circle on $S^3$ or equivalently, a line on $\mathbb R^3$, and $R_H$ is the Cartan generator of $SU(2)_H$.
Operators annihilated by $\mathcal Q_\beta^C$ are therefore also
annihilated by $P_\tau+R_H$. The local Coulomb-branch operators are
Lorentz scalars and $SU(2)_H$ singlets, so this bosonic generator
annihilates them on the fixed circle. To obtain operators annihilated
by $\mathcal Q_\beta^C$ itself, we use the position-dependent
R-symmetry polarizations described below.

Denoting the angular coordinate along the circle by $\varphi$,
the Coulomb-branch operators are transported by the
$\mathcal Q_\beta^C$-exact twisted translation
\begin{equation}
    \widehat P_\varphi^C=P_\varphi+R_C,
\end{equation}
where $R_C$ is the Cartan generator of $SU(2)_C$.
Their correlators are consequently independent of the insertion
positions as long as their ordering is preserved.

As we stated above, the Coulomb-branch cohomology contains dressed monopole operators,
as well as gauge-invariant scalar polynomials in the sector of
zero magnetic charge. The dressing is constructed from the
following position-dependent combination of the vector-multiplet
scalars:
\begin{equation}
\begin{aligned}
    \Phi(\varphi)
    &=
    \left.
    v^{\dot a}(\varphi)v^{\dot b}(\varphi)
    \phi_{\dot a\dot b}
    \right|_{\text{circle}},
    \qquad
    v^{\dot a}(\varphi)
    =
    \frac{1}{\sqrt{2}}
    \begin{pmatrix}
        e^{i\varphi/2}\\
        e^{-i\varphi/2}
    \end{pmatrix},
    \\[1mm]
    &=
    \left.
    \left(
        \frac12 e^{i\varphi}\phi_{\dot1\dot1}
        +\phi_{\dot1\dot2}
        +\frac12 e^{-i\varphi}\phi_{\dot2\dot2}
    \right)
    \right|_{\text{circle}}.
\end{aligned}
\label{eq:adhm-twisted-dressing-scalar}
\end{equation}
Here $\phi_{\dot a\dot b}$ are the vector multiplet adjoint scalars. For a monopole of GNO
charge $b$, the dressing polynomial must be invariant under the
subgroup $G_b\subset G$ that preserves $b$; the resulting dressed
operator is then completed to a gauge-invariant insertion.

Solving the localization equations in a fixed magnetic-flux sector
labelled by $\vec b$, the twisted scalar $\Phi$ takes the form
\begin{equation}
    \Phi_N
    =
    \operatorname{diag}\left(\vec\sigma+\frac{i}{2}\vec b\right),
    \qquad
    \Phi_S
    =
    \operatorname{diag}\left(\vec\sigma-\frac{i}{2}\vec b\right),
    \label{eq:adhm-Phi-N-S}
\end{equation}
where $\vec\sigma$ is a constant real $N$-component vector. These expressions describe the discontinuity of the
localized scalar across a monopole insertion: its value depends on
which side of the insertion it is evaluated on along the great
circle.\footnote{
For two antipodal monopole insertions with equal and opposite
charges, the great circle is divided into two arcs. We label the
corresponding scalar values by $N$ and $S$
}

One subtlety is that the magnetic charge labelling a localization
sector need not coincide with the GNO charge of the bare monopole
operator inserted on the great circle. This is due to a phenomenon
called monopole bubbling, explained in detail in
\cite{Dedushenko:2018icp}. A monopole operator with charge $\vec b$
can receive contributions from sectors with smaller screened
charges $\vec m\prec\vec b$, where $\prec$ denotes the dominance
ordering of GNO charges, rather than a componentwise inequality.
For $U(N)$, screening preserves the total magnetic charge,
\begin{equation}
    \sum_{i=1}^{N}m_i=\sum_{i=1}^{N}b_i.
\end{equation}
In such a sector, $\vec b$ in
\eqref{eq:adhm-Phi-N-S} is replaced by $\vec m$. For minuscule GNO
charges, however, there are no lower-charge bubbling
contributions. Minuscule monopole charges are of the form: $\vec m=(1,1,\ldots 1,0,0,\ldots 0)$. As per our correspondence, these map to the trivial vacua of BMN quantum mechanics under the triality.

The full correlator involving two monopole operators with equal
and opposite non-minuscule charges $\pm\vec b$, together with
gauge-invariant polynomial insertions of neutral operators, takes
the schematic form
\begin{equation}
    \mathcal C_{ b}[P]
    =
    \frac{1}{Z_{S^3}}
    \sum_{\vec m\preceq\vec b}
    \int_{\mathbb R^N} d^N\sigma\,
     Z_{m}(\vec\sigma)\,
    \mathcal B_{ b, m}(\sigma)\,
    P\!\left(
        \Phi_N(\sigma, m),
        \Phi_S(\sigma, m)
    \right).
    \label{eq:adhm-correlator-bubbling-schematic}
\end{equation}
Here $ Z_{m}$ denotes the fixed-flux measure density,
including the one-loop determinants and Weyl factors, while
$\mathcal B_{b,m}$ collects the bubbling factors associated
with both monopole insertions, with the appropriate shifted
arguments and Weyl sums understood. The sum includes the
unscreened sector $\vec m=\vec b$, for which
$\mathcal B_{b,b}=1$ in this notation.
Each term on the right-hand side involves an integral over the
localization moduli $\vec\sigma$. The polynomial $P$ denotes the
product of neutral insertions, with each scalar replaced by
\eqref{eq:adhm-Phi-N-S} using the screened charge $\vec m$.
The choice between $\Phi_N$ and $\Phi_S$ is determined by the arc
on which the corresponding operator is inserted, and hence by its
ordering relative to the monopole insertions.

\paragraph{Fixed-flux matrix integral.}
For later comparison, it is useful to write the fixed-flux measure
density $Z_{\vec m}(\vec\sigma)$ explicitly.\footnote{
Since we are working with superconformal ADHM theory, the mass
and FI parameters are set to zero.
}
Define
\begin{align}
    h_n(x)
    &:=\frac{1}{2\pi}
       \left|\Gamma\!\left(\frac{n+1}{2}+ix\right)\right|^2,
    \nonumber\\
    \mathcal F_n(x)
    &:=\left(x^2+\frac{n^2}{4}\right)
       \left[
       \frac{
       \Gamma\!\left(\frac{n+1}{2}+ix\right)
       \Gamma\!\left(\frac{n+1}{2}-ix\right)}{
       \Gamma\!\left(1+\frac n2+ix\right)
       \Gamma\!\left(1+\frac n2-ix\right)}
       \right]^2.
    \label{eq:adhm-review-Fn}
\end{align}
For a fixed screened GNO charge $\vec m=(m_1,\ldots,m_N)$,
one fundamental hypermultiplet contributes
$\prod_i h_{|m_i|}(\sigma_i)$ to the one-loop determinant.
The adjoint hypermultiplet and vector multiplet contributions
combine into $\mathcal F_{|m_i-m_j|}(\sigma_i-\sigma_j)$ for each
pair of eigenvalues, up to eigenvalue-independent factors.

Thus, for ADHM theory with gauge group $U(N)$, one adjoint
hypermultiplet, and $N_f$ fundamental hypermultiplets, the
fixed-flux measure density is
\begin{equation}
    Z_{\vec m}(\vec\sigma)
    =
    \frac{\mathcal N_{\vec m}}{|W_{\vec m}|}
    \prod_{i=1}^{N}h_{|m_i|}(\sigma_i)^{N_f}
    \prod_{i<j}
    \mathcal F_{|m_i-m_j|}(\sigma_i-\sigma_j).
    \label{eq:adhm-fixed-flux-measure-review}
\end{equation}
Here $W_{\vec m}$ is the Weyl subgroup preserving $\vec m$, and
$\mathcal N_{\vec m}$ collects the eigenvalue-independent factors.
The corresponding fixed-flux integral, before including bubbling
factors and neutral operator insertions, is
\begin{equation}
    \int_{\mathbb R^N}d^N\sigma\,
    Z_{\vec m}(\vec\sigma).
\end{equation}

\paragraph{Choice of R-symmetry polarization.}

The preceding formulas use the original localization conventions of
\cite{Dedushenko:2017avn,Dedushenko:2018icp}.
For the Penrose-limit comparison, we use an equivalent choice of
supercharge,
\begin{equation}
    \widetilde{\mathcal Q}
    =\frac{1}{\sqrt2}U_C\mathcal Q^C_{\beta=1}U_C^{-1},
\end{equation}
where the $SU(2)_C$ transformation $U_C$ is specified in
Appendix~\ref{app:Penrose_spinor}.
The common factor $1/\sqrt2$ fixes the normalization of the charge.
Since $U_C$ commutes with $P_\tau$ and $R_H$,
\begin{equation}
    \widetilde{\mathcal Q}^{\,2}=2i(P_\tau+R_H).
\end{equation}
Equation~\eqref{eq:adhm-Q-square-review} continues to refer to the
original charge.

At the undeformed superconformal fixed point, $SU(2)_C$ is an exact
symmetry. We transform every insertion by
$\widetilde{\mathcal O}=U_C\mathcal O U_C^{-1}$ and transform the
monopole boundary conditions, their counterterms, and the localization
deformation accordingly. The auxiliary-field and background data of
the deformation are transformed as well. The transformation of
monopole polarizations and boundary conditions is described in
equations (2.48)--(2.50) of \cite{Dedushenko:2017avn}.
These transformations preserve the correlator:
\begin{equation}
    \left\langle\prod_i\widetilde{\mathcal O}_i\right\rangle
    =\left\langle\prod_i\mathcal O_i\right\rangle.
\end{equation}
The product includes the monopole insertions that define the
correlator. Any boundary states used to represent it are transformed
as well.

We label each rotated localization configuration by the same
parameters $(\vec\sigma,\vec b)$ as its original configuration.
On the rotated locus, the transformed dressing scalar has the values
\begin{equation}
    \widetilde\Phi_N
    =\operatorname{diag}\left(\vec\sigma+\frac{i}{2}\vec b\right),
    \qquad
    \widetilde\Phi_S
    =\operatorname{diag}\left(\vec\sigma-\frac{i}{2}\vec b\right).
\end{equation}
We therefore use the same integration variables and insertion polynomials for both choices of polarization.


\subsection{Matching the localizing supercharges}
\label{subsec:adhm-bmn-supercharge-match}

We compare the BMN localization spinor with the Penrose limit of
the aligned ADHM localization supercharge $\widetilde{\mathcal Q}$.
This charge is related to the original ADHM charge by the $SU(2)_C$
transformation described above. Its whole Penrose limit lies in the
dynamical supersymmetry sector. The detailed construction is given
in Appendix~\ref{app:susy_matching}.

After the conformal map to the cylinder and continuation to
Lorentzian signature, the aligned boundary spinor is
\begin{equation}
    \widetilde\xi_{\rm cyl}(T,n)
    =e^{-iT/2}\widetilde q
    +e^{+iT/2}(n^i\sigma_i)\widetilde s,
    \qquad n^in^i=1.
    \label{eq:scmatch-boundary-spinor}
\end{equation}
Here $T$ is dimensionless cylinder time. The rotation and the chosen
overall normalization give
\begin{equation}
    \widetilde q^{(\dot1)}=0,
    \qquad \widetilde q^{(\dot2)}=q^{(\dot2)},
    \qquad \widetilde s^{(\dot1)}=s^{(\dot1)},
    \qquad \widetilde s^{(\dot2)}=0.
\end{equation}
The coefficients $q$ and $s$ on the right-hand side are the original
ones in \eqref{eq:DFPY_q} and \eqref{eq:DFPY_s}.

The boundary matching fixes the constant bulk spinor as in
\eqref{eq:ADHM_aligned_bulk_constant}.
We evaluate its bulk Killing spinor on the null geodesic
$z_1=e^{i\psi}$ at the center of $AdS_4$.
Along this geodesic, $T=2\psi=x^+$.
The resulting spinor $\widetilde\epsilon_\gamma$ obeys
\begin{equation}
    \Pi_-\widetilde\epsilon_\gamma=0,
    \qquad \Pi_\pm=\frac12(1\pm\gamma_0\gamma_7).
\end{equation}
The Lorentz boost required by the Penrose limit therefore acts as
\begin{equation}
    S_\Omega\widetilde\epsilon_\gamma
    =\Omega^{-1/2}\widetilde\epsilon_\gamma.
\end{equation}
After the dynamical normalization, the limiting spinor is
\begin{equation}
    \epsilon^{\rm PW}_{\rm aligned}(x^+)
    =\lim_{\Omega\to\infty}
      \Omega^{1/2}S_\Omega\widetilde\epsilon_\gamma(x^+)
    =e^{-ix^+/4}\chi_-+e^{+ix^+/4}\chi_+.
    \label{eq:scmatch-penrose-spinor}
\end{equation}
The coefficients $\chi_\pm$ are defined in \eqref{eq:chi_pm_def}.
They coincide with the dynamical coefficients of the original spinor,
because of our normalization of the aligned charge. The comparison
below concerns the whole limiting supercharge.

Recall that $x^+$ acts as time in the BMN quantum mechanics.
Using the sign convention specified with
\eqref{eq:asano-killing-spinor}, the localizing spinor of
\cite{Asano:2012zt} has a similar decomposition into two
time-dependent components,
\begin{equation}
\begin{aligned}
    \epsilon_{\rm A}(\tau_{\rm A})
    &=
    e^{-\frac{\tau_{\rm A}}2\Gamma^{09}}
    e^{-\frac\pi4\Gamma^{49}}\chi
    \\
    &=
    \frac1{\sqrt2}
    \left(
        e^{+\tau_{\rm A}/2}\chi
        -
        e^{-\tau_{\rm A}/2}\Gamma^{49}\chi
    \right)
    \\
    &=
    \frac1{\sqrt2}
    \left(
        e^{+\tau_{\rm A}/2}\chi
        +
        e^{-\tau_{\rm A}/2}\chi'
    \right)\\
    &=\frac1{\sqrt2}
    \left(
        e^{+i t_{\rm A}/2}\chi
        +
        e^{-i t_{\rm A}/2}\chi'
    \right),
    \qquad
    \chi'=-\Gamma^{49}\chi .
\end{aligned}
\label{eq:scmatch-asano-spinor}
\end{equation}
Here $\tau_{\rm A}=it_A$ is Euclidean BMN time, and is related to $x^+$ by $t_A=x^+/2$\footnote{This is due to different normalization of the coordinates in \cite{Asano:2012zt} from our conventions in Appendix \ref{app:susy_matching}.}, and the constant spinor
obeys $\Gamma^{09}\chi=-\chi$ and $\Gamma^{5678}\chi=\chi$.
Since $\{\Gamma^{09},\Gamma^{49}\}=0$, the second component obeys
$\Gamma^{09}\chi'=+\chi'$.

The spatial rotations of the ADHM theory become the rotations of the
three BMN transverse directions inherited from $AdS_4$. The remaining
six transverse directions admit an embedding of $SU(2)_H\times U(1)_C$
into $\mathrm{Spin}(6)$. We compare quantum numbers under the common
subgroup of the spin cover of the BMN rotation group,
\begin{equation}
\begin{aligned}
    G_{\rm common}
    &=
    SU(2)_{\rm space}\times SU(2)_H\times U(1)_C,
    \\
    \epsilon^{\rm PW}_{\rm aligned}
    &\in
    (\mathbf2,\mathbf2)_{+1/2}
    \oplus
    (\mathbf2,\mathbf2)_{-1/2}.
\end{aligned}
\label{eq:scmatch-common-representations}
\end{equation}
The two frequency components carry opposite $U(1)_C$ weights;
their individual signs depend on the common orientation chosen
for the $U(1)_C$ generator.\footnote{
The BPS monopole background selects a direction in $SU(2)_C$
and preserves the corresponding $U(1)_C$ subgroup. Its R-symmetry
polarization is transformed together with the localization charge.
}
The BMN chirality condition $\Gamma^{5678}\chi=\chi$ selects the
same $SU(2)_H$ doublet, while $\chi'=-\Gamma^{49}\chi$ carries the
opposite $U(1)_C$ weight and the same $SU(2)_H$ representation.
Thus the two pairs have the same representation content under
$G_{\rm common}$.

The two constant spinors also have a definite relative polarization.
After expressing the two bulk sectors in the common spatial spinor
basis described in Appendix~\ref{app:susy_matching}, their coefficient
tensors obey
\begin{equation}
    q^{(\dot2)}_{\alpha a}
    =
    -i(\sigma_3)_\alpha{}^\beta
    s^{(\dot1)}_{\beta a}.
    \label{eq:scmatch-boundary-pairing}
\end{equation}

On the BMN side, we check that the action of $\Gamma^{49}$ when restricted to the $SU(2)$ components of the spinors $\chi$ and $\chi'$, also imposes the same relation between their $SU(2)$ indices as in \eqref{eq:scmatch-boundary-pairing} given that we make a particular identification among the $SO(3)$ cartesian frames in pp-wave geometry and the frames chosen at the center of AdS.

\subsection{Matching the fixed-sector measures}
\label{subsec:adhm-bmn-fixed-backgrounds}

In this subsection, we take the triple-scaling limit of the ADHM
measure \eqref{eq:adhm-fixed-flux-measure-review} in a fixed screened
GNO sector and compare it with the BMN measure
\eqref{eq:pwmm-fixed-vacuum-measure}. The detailed computation is
presented in Appendix~\ref{app:generic-positive-flux-measure}.
Here we summarize the main steps.

Consider the following choice of positive screened GNO charges:
\begin{equation}
    \vec m
    =
    \bigl(n_1^{N_1},n_2^{N_2},\ldots,n_\Lambda^{N_\Lambda},
    0^{N-\nu}\bigr),
    \qquad
    \nu=\sum_{s=1}^{\Lambda}N_s,
    \qquad
    m=\sum_{s=1}^{\Lambda}n_sN_s.
    \label{eq:adhm-general-positive-background}
\end{equation}
Here the distinct positive charges $n_s$ occur with multiplicities
$N_s$. We keep all $n_s$ and $N_s$ fixed in the double-scaling limit.
As explained in the index computation, the corresponding BMN
gauge-group rank is $m$, and the fuzzy-sphere vacuum is
\begin{equation}
    L_i(\vec m)
    =
    \bigoplus_{s=1}^{\Lambda}
    \left(L_i^{[n_s]}\otimes\mathbf 1_{N_s}\right),
    \label{eq:adhm-general-bmn-vacuum}
\end{equation}
where $L_i^{[n_s]}$ denotes the irreducible $SU(2)$ representation
of dimension $n_s$.

On the BMN side, localization around this vacuum restricts the
commuting matrix $M$ to
\begin{equation}
    M
    =
    \bigoplus_{s=1}^{\Lambda}
    \left(\mathbf 1_{n_s}\otimes M_s\right),
    \qquad
    M_s=\operatorname{diag}(q_{s,1},\ldots,q_{s,N_s}).
    \label{eq:adhm-general-bmn-modulus}
\end{equation}
On the ADHM side, we denote the eigenvalues carrying positive flux
$n_s$ by $\sigma_{s,a}$, with $a=1,\ldots,N_s$, and the zero-flux
eigenvalues by $\rho_I$, with $I=1,\ldots,N-\nu$.
We integrate out the latter and compare the resulting effective
measure for $\sigma_{s,a}$ with the BMN measure for $q_{s,a}$.

As explained in detail in Appendix \ref{app:generic-positive-flux-measure}, we use the large-$N$ saddle-point approximation, treating the
zero-flux eigenvalues as a continuous distribution with density
\begin{equation}
    \varrho(x)
    =
    \sum_{I=1}^{N-\nu}\delta(x-\rho_I).
\end{equation}
The parameters are scaled so that
\begin{equation}
    \lambda_{\rm pp}=\frac{N}{N_f^2}
\end{equation}
is fixed as $N,N_f\to\infty$.
\label{eq:adhm-correlator-double-scaling}
Since the number of nonzero-flux eigenvalues is fixed, their
contribution to the leading saddle-point equation is subleading.
The bulk part of the zero-flux density is approximately triangular,
\begin{equation}
    \varrho_0(x)\simeq 2N_f(A-|x|),
    \qquad
    A^2\simeq\frac{N}{2N_f},
\end{equation}
away from the central and endpoint regions. Thus the endpoints of
its support move far from the origin in the double-scaling limit.
The treatment of the finite support and the corresponding
corrections to the density is discussed in
Appendix~\ref{app:generic-positive-flux-measure}.

We first consider the interaction between the nonzero-flux
eigenvalues. This receives a direct contribution from the factor
$\mathcal F_{|n_s-n_t|}$ in
\eqref{eq:adhm-fixed-flux-measure-review}, as well as an induced
contribution from integrating out fluctuations of the zero-flux eigenvalues around the large $N$ saddle $\varrho_0$. 
Expanding the density around its leading saddle and integrating
out its quadratic fluctuations gives, to leading order in this
approximation,
\begin{equation}
    \mathcal K_{n_s,n_t}^{\rm ADHM,eff}(x)
    =
    \frac{\mathcal F_{|n_s-n_t|}(x)}{\mathcal F_{n_s+n_t}(x)},
    \qquad
    x=\sigma_{s,a}-\sigma_{t,b}.
\end{equation}
Here the factor $1/\mathcal{F}_{n_s+n_t}$ arose from integrating out the zero-flux eigenvalues, see Appendix \ref{appsub:2body} for details. 
Using the gamma-function recurrence relations, this ratio can be
written as
\begin{equation}
    \frac{\mathcal F_{|n_s-n_t|}(x)}{\mathcal F_{n_s+n_t}(x)}
    =
    \prod_{J=|n_s-n_t|/2}^{(n_s+n_t)/2-1}
    \frac{
        \bigl((2J+2)^2+4x^2\bigr)
        \bigl((2J)^2+4x^2\bigr)
    }{
        \bigl((2J+1)^2+4x^2\bigr)^2
    }.
\end{equation}
This is the same product that appears in the BMN one-loop
determinant $Z^L_{1-loop}$ in second line of \eqref{eq:pwmm-fixed-vacuum-measure} for an unordered pair of eigenvalues belonging to
blocks of dimensions $n_s$ and $n_t$, with the identification
$q_{s,a}=2\sigma_{s,a}$, where $q$ are eigenvalues of the matrix $M$ that appears in \eqref{eq:pwmm-fixed-vacuum-measure}.

We next consider the single-eigenvalue potential. 
 As we explain in detail in Appendix \ref{app:generic-positive-flux-measure}, we show that the effective potential for  charge-$n_s$ eigenvalue
takes the form:
\begin{equation}
    V_{n_s}^{\rm eff}(\sigma)-V_{n_s}^{\rm eff}(0)
    =
    \frac{n_s}{\lambda_{\rm pp}}\sigma^2+o(1),
    \label{eq:adhm-general-one-body-potential}
\end{equation}
where the potential is defined to enter the measure as
$e^{-V^{\rm eff}}$.

Combining the pairwise interaction with the single-eigenvalue potential,
we find that the ADHM effective measure reproduces the BMN fixed-vacuum
measure under the identification $q_{s,a}=2\sigma_{s,a}$.
Matching the Gaussian terms in the first line of
\eqref{eq:pwmm-fixed-vacuum-measure} requires
\begin{equation}
    g_{\rm loc}^2 = 8\lambda_{\rm pp}
    = 8\frac{N}{N_f^2}.
    \label{eq:adhm-bmn-localization-dictionary}
\end{equation}
This is precisely the coupling map obtained by combining the localization
coupling \eqref{mapcoup} with the BMN-ADHM dictionary implied by the triality
\eqref{eq:adhm-bmn-coupling-map}.
Thus, within the saddle-point approximation described above,
\begin{equation}
    d\mu_{\rm ADHM,\vec m}^{\rm eff}
    =
    \mathscr A_{\vec m}(\lambda_{\rm pp})\,
    \left.
    d\mu_{\rm BMN}^{L(\vec m)}
    \right|_{q_{s,a}=2\sigma_{s,a}}
    [1+o(1)].
    \label{eq:adhm-general-measure-match}
\end{equation}
The prefactor $\mathscr A_{\vec m}$ is independent of the charged
eigenvalues and therefore cancels in normalized correlators at fixed
screened background.

\subsection{Operator dictionary and topological correlators}
\label{subsec:adhm-bmn-operator-dictionary}
So far, we have shown that the localization measure for each screened-charge contribution on the right-hand side of \eqref{eq:adhm-fixed-flux-measure-review} matches that of the corresponding BMN fuzzy-sphere vacuum. These contributions are labelled by the screened GNO charges, rather than the bare charges of the inserted monopole operators. Because of the monopole bubbling, recovering the full ADHM correlator
requires weighting each sector by its monopole-bubbling coefficient
$\mathcal B_{b,m}(\sigma)$ and summing over the allowed screened charges,
as in \eqref{eq:adhm-correlator-bubbling-schematic}. 

To avoid this complication, in this subsection, we focus on minuscule GNO charges, for which bubbling is absent altogether and there is no $\mathcal B_{b,m}(\sigma)$, and match the operator insertions in the ADHM and BMN localization formulas. 
The same dictionary
also applies to the unscreened sector of non-minuscule charges, where the bubbling coefficient is trivial. Its extension to sectors with
nontrivial bubbling coefficients remains an open problem.

Ignoring the bubbling coefficients, let us consider the form of the localizing ADHM scalar $\Phi$ in a given screened background charge as given in \eqref{eq:adhm-Phi-N-S}. The localizing BMN scalar $\phi$ takes  the form given in \eqref{eq:asano-localization-locus}. We want to see if the trace of powers of the localization scalar in BMN $\phi$ can be writtes in terms of the insertions of $\Phi$ on ADHM side. 

Consider the insertion of $\Tr(\phi^n)$ on the BMN side in a given
vacuum, which on the localization contour becomes
$\Tr(2L_3+iM)^n$, where $L_3$ is an $SU(2)$ generator in the
reducible representation corresponding to the fuzzy sphere vacuum.
The trace can be written as a sum over traces in the irreducible
$SU(2)$ blocks defined in \eqref{eq:pwmm-review-decomposition}.

As per our correspondence in
subsection~\ref{subsec:adhm-bmn-fixed-backgrounds}, each irreducible
block of size $n_s$ corresponds to a flux eigenvalue with flux $n_s$.
Depending on the position of the scalar insertion, the corresponding
eigenvalue of the scalar $\Phi$ takes the following form, using
\eqref{eq:adhm-Phi-N-S}:
\begin{equation}
    \Phi_{s,a}^{N/S}
    =\sigma_{s,a}\pm\frac{i}{2}n_s.
\end{equation}
Since, within each irreducible block, the matrix $M$ is proportional
to the identity with eigenvalue $q_{s,a}=2\sigma_{s,a}$, we can write
the trace restricted to this block as
\begin{equation}
    \Tr_{n_s}(2L_3+iM)^n
    =
    \sum_{j=-(n_s-1)/2}^{(n_s-1)/2}
    \left(2j+2i\sigma_{s,a}\right)^n,
    \label{tracephi}
\end{equation}
where the sum runs in unit steps.

Using the Bernoulli polynomial identity
\begin{equation}
    \sum_{\ell=0}^{s-1}(x+\ell)^n
    =
    \frac{B_{n+1}(x+s)-B_{n+1}(x)}{n+1},
    \label{eq:adhm-Bernoulli-finite-difference}
\end{equation}
with $s=n_s$ and
$x=\frac{1-n_s}{2}+i\sigma_{s,a}$, the sum on the RHS of
\eqref{tracephi} becomes
\begin{equation}
    \Tr_{n_s}(2L_3+iM)^n
    =
    \frac{2^n}{n+1}
    \left[
    B_{n+1}\left(\frac12+i\Phi^S_{s,a}\right)
    -
    B_{n+1}\left(\frac12+i\Phi^N_{s,a}\right)
    \right].
\end{equation}
This implies that the complete trace on the BMN side can be written as:
\begin{equation}\label{crrmat}
    \Tr(2L_3+iM)^n
    =
    \frac{2^n}{n+1}
    \Tr\left[
    B_{n+1}\left(\frac12+i\Phi^S\right)
    -
    B_{n+1}\left(\frac12+i\Phi^N\right)
    \right]. 
\end{equation}
Note that, even though the matrix $\Phi$ is $N\times N$ and hence contains more terms in the trace, the eigenvalues with zero flux charge do not contribute since the solutions for $\Phi$ are same in both north and south arcs for these eigenvalues and the two terms above cancel.

Therefore, we see that the trace of powers of the BMN localizing
scalar maps to the trace of the difference of the above Bernoulli
polynomials evaluated on the south and north arcs. This difference
might appear not to correspond to a local operator insertion, since
it involves contributions from both arcs. However, it can be
interpreted through the insertion of a star commutator,
\begin{equation}
    [P(\Phi),\mathcal M_{\vec m}]_\star
    =
    P(\Phi)\star\mathcal M_{\vec m}
    -
    \mathcal M_{\vec m}\star P(\Phi),\label{eq:dressedmonopole}
\end{equation}
where $P_n(\Phi):=
    \frac{2^n}{n+1}
    \Tr B_{n+1}\left(\frac12+i\Phi\right)$ and $\star$ denotes the ordered product in the protected
one-dimensional Coulomb-branch operator algebra
\cite{Dedushenko:2017avn,Dedushenko:2018icp}.  $\mathcal M_{\vec m}$ denotes the dressed BPS monopole operator.
We use the star-product convention of Eq.~\eqref{starorder}.
In the North-pole shift-operator representation this gives $\widehat{A\star B}_N=\widehat B_N\widehat A_N$.
Therefore, the RHS of
\eqref{crrmat} can be reproduced, sector by sector, by replacing
the monopole insertion with the star-product commutator\footnote{Similar operators expressed in terms of the Bernoulli polynomials appeared in the study of quantized Coulomb brach algebra in \cite{Hatsuda:2021oxa, Gaiotto:2020vqj}. It would be interesting to understand the precise connection.}
\begin{equation}
    \frac{2^n}{n+1}
    \left[
        \Tr B_{n+1}\left(\frac12+i\Phi\right),
        \mathcal M_{\vec m}
    \right]_\star.
\end{equation}

The same construction applies to insertions of product of multiple traces on
the BMN side. Define 
\begin{align}
\mathcal D_n\mathcal O
    :=
    [P_n(\Phi),\mathcal O]_\star\,.
\end{align}
The order in this commutator is chosen so that, in a fixed screened
sector $\vec m$, it generates the difference
\begin{equation}
    P_n(\Phi_{\vec m}^S)
    -
    P_n(\Phi_{\vec m}^N).
\end{equation}
For a product of $k$ traces, we therefore consider the nested
commutator
\begin{equation}
    \mathcal O_{\vec b;n_1,\ldots,n_k}
    :=
    \mathcal D_{n_k}\cdots\mathcal D_{n_1}\mathcal M_{\vec b}
    =
    \bigl[
        P_{n_k},
        \bigl[P_{n_{k-1}},\ldots,
        [P_{n_1},\mathcal M_{\vec b}]_\star
        \ldots\bigr]_\star
    \bigr]_\star.
\end{equation}

In each screened sector, this corresponds to inserting the product of BMN traces. 
Since the operators $P_n(\Phi)$ are commuting scalar
polynomials in the shift-operator representation
\cite{Dedushenko:2018icp}, their adjoint actions commute,
\begin{equation}
    [\mathcal D_n,\mathcal D_p]=0.
\end{equation}
Thus the nested commutator is independent of the order of the
$n_\ell$, consistently with the product of BMN trace insertions.

The resulting dictionary of insertions is
\begin{equation}
    \prod_{\ell=1}^{k}\Tr_m\bigl(\phi^{n_\ell}\bigr)
    \longleftrightarrow
    \mathcal D_{n_k}\cdots\mathcal D_{n_1}
    \mathcal M_{\vec b}.
\end{equation}
Here the right-hand side denotes a replacement of one monopole
insertion in the ADHM correlator, with the other insertions kept
fixed. 
Therefore, if we only look at the minuscule GNO charges, we can identify the following correlators exactly:
\begin{equation}
\lim_{\substack{N,N_f\to\infty\\
                     N/N_f^2=\lambda_{\rm pp}}}
\frac{
\left\langle
(\mathcal D_{n_k}\cdots\mathcal D_{n_1}\mathcal M_{\vec b})
\mathcal M_{-\vec b}
\right\rangle_{\mathrm{ADHM}}
}{
\left\langle
\mathcal M_{\vec b}\mathcal M_{-\vec b}
\right\rangle_{\mathrm{ADHM}}
}
=
\left\langle
\prod_{\ell=1}^k\operatorname{Tr}_m(\phi^{n_\ell})
\right\rangle_{\mathrm{BMN,trivial}} .
\end{equation}
For non-minuscule GNO charges, the nested commutator generates the same factors
sector by sector on RHS of \eqref{eq:adhm-correlator-bubbling-schematic}, while retaining the corresponding bubbling coefficients.

\subsection{Monopole bubbling and BMN instantons}
\label{subsec:adhm-bmn-bubbling-scope}
The previous subsections compared a fixed ADHM magnetic-flux
sector with a fixed BMN vacuum. A physical monopole operator
can receive contributions from several such sectors. For a
non-minuscule monopole, the magnetic charge imposed at the
insertion can be partly screened, so the flux measured away
from the insertion need not equal the original charge. This
screening is called monopole bubbling.

\paragraph{BMN instanton action from bubbling.} Let $B$ be the charge imposed at the insertion and let
$v\preceq B$ label a possible screened charge. The sector
$v=B$ is the unscreened/unbubbled sector. In the hemisphere description
of \cite{Dedushenko:2018icp}, the state created by the monopole
operator is a sum over outgoing fluxes. Each term is a
fixed-flux hemisphere wavefunction multiplied by its bubbling
coefficient. Gluing the two hemispheres pairs contributions
with the same outgoing flux. Schematically,
\begin{equation}
 \langle\mathcal M_B^\dagger\mathcal M_B\rangle
 =\frac{1}{Z_{S^3}}\sum_{v\preceq B}W_{B\to v},
 \qquad
 W_{B\to v}
 =\int d\mu_{\mathrm{ADHM},v}\,
       Z^-_{B\to v}Z^+_{B\to v}.
 \label{eq:adhm-bubbling-schematic}
\end{equation}
Here $d\mu_{\mathrm{ADHM},v}$ is the fixed-flux measure
introduced earlier, and $Z^\pm_{B\to v}$ are the bubbling
coefficients on the two hemispheres, including the appropriate
Weyl sums. In the unbubbled sector, $Z^\pm_{B\to B}=1$.
To decide whether a screened contribution can be neglected,
we compare $W_{B\to v}/W_{B\to B}$, the relative weights of
two sectors in the same monopole two-point function.

To express this comparison in BMN variables, we integrate out
the neutral ADHM eigenvalues as in
\eqref{eq:adhm-general-measure-match}. The factor $\mathscr A_p$ appearing as the neutral-saddle normalization cancels from
normalized correlators within one sector, but contributes to
the relative weights of different sectors as
\begin{equation}
 \frac{W_{B\to v}}{W_{B\to B}}
 \simeq
 \frac{\mathscr A_v}{\mathscr A_B}
 \frac{\displaystyle
       \int d\mu_{\mathrm{BMN},v}\,
              Z^-_{B\to v}Z^+_{B\to v}}
      {\displaystyle\int d\mu_{\mathrm{BMN},B}}.
 \label{eq:adhm-bubbling-bmn-factorization}
\end{equation}
The BMN integrals here are fixed-vacuum
integrals, with the remaining normalization factors inherited
from the ADHM measure. The bubbling factors are understood
after the neutral-sea reduction and are written in the charged
variables $q=2\sigma$.
The ratio $\mathscr A_v/\mathscr A_B$ turns out to reproduce
the classical BMN instanton action. For fixed partitions of
the same total charge, the neutral-saddle computation gives
\begin{equation}
 -\frac12\log\frac{\mathscr A_v}{\mathscr A_B}
 =S_{\mathrm{cl}}+o(1),
 \label{adhm-bmn-instanton}
\end{equation}
where $S_{\mathrm{cl}}$ is the classical action of a BMN
instanton connecting the associated vacua
\cite{Yee:2003ge,Lin:2006tr} (See Appendix \ref{app:neutral-sea-normalization} for derivation). 

This agreement strongly suggests a connection between monopole bubbling and
BMN instantons, although a complete identification is currently out of reach. First, the
ratio \eqref{eq:adhm-bubbling-bmn-factorization} contains additional factors,
notably $Z^-_{B\to v}Z^+_{B\to v}$, which must also be accounted for.
Second, existing BMN localization analyses do not include instanton sectors.
Our results therefore motivate extending BMN localization to include
instantons, which would allow a more complete test of the proposed
correspondence.

\paragraph{Suppression of bubbling sectors.}Setting aside the correspondence between bubbling and instantons, let us discuss the physical consequence of \eqref{eq:adhm-bubbling-bmn-factorization}. Taken alone, the factor $\frac{\mathcal{A}_v}{\mathcal{A}_B}=e^{-2S_{\mathrm{cl}}}$ suggests
strong suppression when the classical action is large. This was indeed assumed in the analysis in \cite{Asano:2012zt}.
 However, as we discuss below, the remaining integral over
nonzero-flux eigenvalues can either partially compensate for this
suppression or enhance it. Thus the relative weight of a screened sector
cannot be determined from the classical action alone. 

In what follows, we examine this competition
in the elementary channel
\begin{equation}
 B=(2^k),
 \qquad
 v=(2^{k-1},1,1),
 \label{eq:bubbling-elementary-channel}
\end{equation}
which compares two nearby M5-like vacua. The unscreened partition $B$
describes two coincident spherical M5-branes of equal size, while the
screened partition $v$ describes two M5-branes of slightly different sizes.
We first take the ADHM scaling limit at fixed $k$ and integrate out the
zero-flux eigenvalues. We then analyze the charged-eigenvalue saddle in the
large-$k$ limit, keeping the 't~Hooft coupling
$t=kg_{\rm loc}^2$ fixed.
\begin{itemize}
\item \textbf{The zero-flux eigenvalue contribution.}
Replacing one charge-two eigenvalue by two charge-one
eigenvalues changes their interaction with the zero-flux eigenvalues
and reduces the number of neutral eigenvalues by one.
Both effects must be included in the relative normalization.
For the channel~\eqref{eq:bubbling-elementary-channel},
the calculation in Appendix~\ref{app:neutral-sea-normalization}
gives $S_{\mathrm{cl}}=2k/t$ and hence
\begin{equation}
 \log\frac{\mathscr A_v}{\mathscr A_B}
 =-\frac{4k}{t}+o(1).
 \label{eq:bubbling-sea-relative-constant}
\end{equation}
The neutral contribution alone would therefore suppress
the screened sector by $e^{-4k/t}$.

\item \textbf{The charged contribution.}
We now evaluate the remaining charged integral, before
including the bubbling coefficients. At small $t$, the
quadratic potential confines the charged eigenvalues to a
narrow region of width $O(\sqrt t)$. The pair interactions
can therefore be expanded at small eigenvalue separation.
The resulting powers of $t$ give a contribution of order
$k|\log t|$ to the logarithm of the relative weight.
This is smaller in magnitude than the neutral term $-4k/t$.
Thus, at weak 't Hooft coupling, the classical instanton
action controls the leading suppression of the reduced
weight.

To follow the change at larger $t$, we solve the charged
saddle numerically. The table lists the leading contributions
to the logarithm of the relative weight $\frac{W_{B\to v}}{W_{B\to B}}$ \eqref{eq:adhm-bubbling-bmn-factorization}, divided by $k$,
before including the bubbling coefficients.
\begin{center}
\begin{tabular}{c|r|r|r}
 $t$ & Zero-flux eigenvalues & Charged eigenvalues & Combined\\
 \hline
 $0.1$  & $-40$    & $3.15309$    & $-36.84691$\\
 $1$    & $-4$     & $1.28007$    & $-2.71993$\\
 $10$   & $-0.4$   & $0.296571$   & $-0.103429$\\
 $100$  & $-0.04$  & $0.0390701$  & $-9.30\times10^{-4}$\\
 $1000$ & $-0.004$ & $0.00399947$ & $-5.30\times10^{-7}$
\end{tabular}
\end{center}
The charged contribution enhances the screened sector and
nearly cancels the neutral suppression at large $t$.
At fixed $t$, the negative combined entries still imply exponential suppression
as $k\to\infty$, but the exponent can be much smaller than
the classical instanton estimate. In a simultaneous
large-$k$, large-$t$ limit, the remaining exponent and the
subleading terms must both be controlled before deciding
whether bubbling can be neglected.

The cancellation also depends on the background. If the
initial partition contains a macroscopic number of charge-one
entries, as in $B=(2^k,1^{k'})$ with $k'$ proportional to
$k$, the charged saddle instead gives additional suppression
at large $t$ for the same elementary splitting. Thus the
relative weight depends on the charged eigenvalue distribution
in the chosen vacuum, as well as on the classical instanton
action.

\item \textbf{The bubbling coefficients.}
Finally, the sector weight in
\eqref{eq:adhm-bubbling-schematic} includes the bubbling
coefficients. Their effect is to multiply the screened
integral by the expectation value of $Z^-_{B\to v}Z^+_{B\to v}$
in that sector. For example, define $\mathcal M_B$ by the
one-dimensional Coulomb-branch product of two minuscule
charge-$(1^k)$ operators. In this operator convention, the
elementary bubbling coefficient is \cite{Dedushenko:2018icp}
\begin{equation}
 Z^{\mathrm{prod}}_{B\to v}(q_1,q_2)
 =2-\frac{2}{(q_1-q_2)^2+4},
 \label{eq:bubbling-elementary-coefficient}
\end{equation}
where $q_1,q_2$ are the two charge-one eigenvalues.
On the real contour, the squared coefficient entering the
conjugate hemisphere pairing lies between $9/4$ and $4$.
It therefore changes the logarithm of the sector weight
only by $O(1)$ and leaves the leading order-$k$ exponent
unchanged at fixed $t$. This conclusion applies to the
elementary channel and the operator choice specified here.
We leave the analysis of bubbling coefficients at greater
bubbling depth for future work.
\end{itemize}
\section{The mirror necklace theory}
\label{sec:necklace-bmn-correlators}

In this section, we discuss the mirror dual of ADHM
theory~\cite{Intriligator:1996ex,Hanany:1996ie}, referred to as the
necklace theory. The mirror of rank-$N$ ADHM theory with $L=N_f$
fundamental hypermultiplets is a quiver gauge theory with gauge
group $U(N)^L$.

Under mirror symmetry, the Higgs and Coulomb branches of ADHM
theory and the necklace theory are interchanged. Therefore, the
Coulomb-branch operators discussed above in ADHM theory map to
Higgs-branch operators in the necklace theory. This provides a
complementary description of protected ADHM monopole correlators
in terms of winding operators in the Higgs sector of the necklace
theory, which we describe in detail below.

Because winding operators are ordinary gauge-invariant polynomials in the elementary fields of the necklace theory, this description lets us study correlators without needing to deal with monopole-bubbling contributions on the ADHM side. Specifically, operators made out of a product of bi-fundemanetal scalars that ``winds" around a circular quiver are mirror dual to bubbling-completed monopole operators. We call them {\it winding operators} in what follows.
The main result of this section is a proposed map between Coulomb-branch operators in the ADHM theory and Higgs-branch operators in the necklace theory. We propose a general map between bubbling-completed monopole operators and winding operators in Subsection \ref{subsec:column-product-proposal} based on analytical computations at low charge in Subsection \ref{subsec:necklace-winding-correlators} and numerical checks for minuscule-charge monopoles in Subsection \ref{subsec:minuscule-schur}. We then discuss a mirror map for neutral Coulomb-branch insertions in the ADHM theory in Subsection \ref{subsec:neutral-necklace}, which we test by computing selected correlators both analytically and numerically at finite $(N,L)$.


\subsection{Review of the necklace theory}
\label{subsec:necklace-review}

The mirror dual of the three-dimensional \(\mathcal N=4\) \(U(N)\) ADHM theory with one adjoint hypermultiplet and \(N_f=L\) fundamental hypermultiplets is an \(\mathcal N=4\) necklace quiver gauge theory~\cite{Hanany:1996ie,Kapustin:2010xq} with gauge group
\[
G=\prod_{I=1}^{L}U(N)^{(I)}.
\]
The quiver consists of \(L\) nodes arranged in a circle, each representing a \(U(N)\) gauge factor and carrying an \(\mathcal N=4\) vector multiplet. Adjacent nodes are connected by  bifundamental hypermultiplets, denoted by \((Q_I,\tilde Q_I)\), with node indices understood modulo \(L\). In \(\mathcal N=2\) notation, the chiral multiplets \(Q_I\) and \(\tilde Q_I\) transform in the representations \((\mathbf N_I,\overline{\mathbf N}_{I+1})\) and \((\overline{\mathbf N}_I,\mathbf N_{I+1})\), respectively. 

In addition, a single fundamental hypermultiplet is attached to node \(1\) which we denote by $q,\tilde q$. Under the $SU(2)_H$ of the necklace theory, the bifundamental fields $(Q_I,\tilde Q_I)$
 and the fundamental $q,\tilde q$ transform as doublets. On the other hand, the scalars in the vector multiplet trasform as a spin 1 vector under $SU(2)_C$.

Mirror symmetry exchanges the $SU(2)_H$ and $SU(2)_C$ isometries, mapping Higgs-branch operators of the necklace theory to Coulomb-branch operators of the ADHM theory, and vice versa.

In particular, monopole operators on the Coulomb branch of the ADHM theory map schematically to the following gauge-invariant operators in the dual necklace theory:\footnote{Conversely, traces of the ADHM adjoint Higgs fields map to necklace monopole operators with aligned GNO charges at every node.}
\begin{align}
 V_+^m&\longleftrightarrow\Tr(\widetilde Q_L\cdots\widetilde Q_1)^m,
 \label{eq:positive-monopole-winding-map}\\
 V_-^m&\longleftrightarrow\Tr(Q_1\cdots Q_L)^m.
 \label{eq:negative-monopole-winding-map}
\end{align}
The normalization of this map is fixed below.  We refer to products of the form $\widetilde Q_L\cdots\widetilde Q_1$ and $Q_1\cdots Q_L$ as winding operators. They transform in the adjoint representation of the $U(N)$ gauge group at the first node.

The map written above is only a schematic, in the subsections below, we will explain in detail our proposal for a precise map between a monopole operator, with a given choice of GNO charges in ADHM theory and the corresponding gauge invariant operator formed from the winding operators in the necklace theory.

\subsection{Winding correlators and the mirror operator map}
\label{subsec:necklace-winding-correlators}

In this subsection, we review the protected Higgs branch correlators in the necklace theory which under the mirror duality, map to the Coloumb branch correlators of the ADHM theory that we discusssed in Subsection \ref{subsec:adhm-localization}. Just like in the ADHM correlators, the insertions in the necklace theory also lie on the great circle on $S^3$ or equivalently on a line in $\mathbb R^3$. 

Let us denote the winding operators defined above as 
\begin{equation}
    W_{-}(\varphi)=Q_1(\varphi)\cdots Q_L(\varphi),
    \qquad
    W_{+}(\varphi)=\widetilde Q_L(\varphi)\cdots\widetilde Q_1(\varphi),
    \label{eq:necklace-winding-matrices}
\end{equation}
where the products are based at node $1$. The subscripts $\pm$ are chosen so that they reflect the sign of monopole charges. We define the following basis of the gauge-invariant operators constructed from the winding operators: 
\begin{equation}
    p_k(W_{\pm})=\Tr W_{\pm}^k,
    \qquad
    p_\lambda(W_{\pm})
    =\prod_{k\geq1}p_k(W_{\pm})^{m_k},
    \qquad
    \lambda=(1^{m_1}2^{m_2}\cdots),
    \label{eq:necklace-power-sums}
\end{equation}
 $p_k$ is a single trace winding $k$ times, whereas $p_\lambda$
describes several winding cycles.  

\paragraph{Finite-$(N,L)$ Higgs-branch localization.}

Just like for the ADHM correlators that we discussed in \ref{subsec:adhm-localization}, the correlators in mirror dual theory also reduce to the constant eigenvalues of the localizing vector multiplet scalar (formed from linear combination of vector multiplet scalars as in \eqref{eq:adhm-twisted-dressing-scalar}) and the localization integral measure gets the contribution from the one-loop determinants of the matter fields. 

At the superconformal fixed point where masses and FI paramters vanish, the $S^3$ measure, taking into account the one-loop determinants from the matter fields it is, up to an
overall constant~\cite{Kapustin:2009kz,Dedushenko:2016jxl},
\begin{equation}
    d\mu_{\rm neck}(\boldsymbol\sigma)
    =\frac{1}{(N!)^L}
      \prod_{I=1}^{L}\prod_{a=1}^{N}d\sigma_{I,a}\,
      \frac{
       \displaystyle
       \prod_{I=1}^{L}\prod_{a<b}
       4\sinh^2\!\bigl(\pi\sigma_{I,ab}\bigr)}{
       \displaystyle
       \prod_{I=1}^{L}\prod_{a,b}
       2\cosh\!\bigl[\pi(\sigma_{I,a}-\sigma_{I+1,b})\bigr]
       \prod_{a=1}^{N}2\cosh(\pi\sigma_{1,a})},
\label{eq:necklace-S3-measure}
\end{equation}
where $\sigma_{I,ab}=\sigma_{I,a}-\sigma_{I,b}$ and
$\sigma_{L+1,a}=\sigma_{1,a}$. $\sigma_{I,a}$ refers to the $a$th diagonal value of the vector multiplet scalar in the $I$ th node. 

For the given values of $\boldsymbol{\sigma}$, the
one-dimensional Higgs-sector theory is Gaussian~\cite{Dedushenko:2016jxl}.
The propagator for the bifundamental fields $Q,\tilde Q$ takes the following form on the great circle as a function of the angle $\varphi$ which is the coordinate on the great circle:\footnote{For simplicity, we set the radius $r$ of $S^3$ in \cite{Dedushenko:2016jxl} to be unity.}
\begin{equation}
\begin{split}
 &\left\langle
 (Q_I)^{a}{}_{b}(\varphi_1)
 (\widetilde Q_I)^{c}{}_{d}(\varphi_2)
 \right\rangle_{\boldsymbol\sigma}
=
 \delta^{a}{}_{d}\delta^{c}{}_{b}\,
 G_{\sigma_{I,a}-\sigma_{I+1,b}}(\varphi_{12}),
 \\
 &G_{\omega}(\varphi)
 =-\frac{\text{sgn}\varphi+\tanh(\pi\omega)}{8\pi }
   e^{-\omega\varphi},
 \qquad -\pi<\varphi<\pi.
\end{split}
\label{eq:necklace-1d-propagator}
\end{equation}
At the topological insertion points, taken to be separated by $\varphi_{12}=\pi$, this propagator reduces, up to the conventional overall normalization, to the kernel
\begin{align}
    \mathsf K(x)\equiv -4\pi G_x(\pi)
=\frac{1}{2\cosh(\pi x)}.\label{eq:propagatornecklace}
\end{align}
Thus, at fixed necklace eigenvalues, the weight associated with a bifundamental contraction across link $I$ is
$G_I(a,b)=\mathsf K\left(\sigma_{I,a}-\sigma_{I+1,b}\right).$
All correlators of winding operators are consequently given by the exact
finite-dimensional integral
\begin{equation}
 \left\langle\prod_{s=1}^{n}\mathcal O_s(\varphi_s)\right\rangle_{\rm neck}
 =\frac{1}{Z_{\rm neck}}
  \int d\mu_{\rm neck}(\boldsymbol\sigma)\,
  \left\langle\prod_{s=1}^{n}\mathcal O_s(\varphi_s)
  \right\rangle_{\boldsymbol\sigma}^{\rm Wick} .
\label{eq:necklace-general-Higgs-correlator}
\end{equation}
Equation~\eqref{eq:necklace-general-Higgs-correlator} is the starting point for arbitrary $N$, $L$, and winding representation. 

For comparison, let us remind ourselves of the results for the ADHM theory; the zero-flux ADHM partition-function measure is~\cite{Mezei:2013gqa,Hatsuda:2014vsa}
\begin{equation}
    d\mu_{\rm ADHM}^{(0)}(\boldsymbol\sigma)
    =\frac{1}{N!}\prod_{a=1}^{N}d\sigma_a\,
      \frac{\displaystyle
       \prod_{a<b}4\sinh^2(\pi\sigma_{ab})}{\displaystyle
       \prod_{a,b}2\cosh(\pi\sigma_{ab})
       \prod_{a=1}^{N}[2\cosh(\pi\sigma_a)]^L}.
    \label{eq:ADHM-zero-flux-measure}
\end{equation}
Coulomb insertions are represented by shift operators
\cite{Dedushenko:2017avn,Dedushenko:2018icp}; their correlators take the gluing form
\begin{equation}
 \left\langle\mathcal O_1\cdots\mathcal O_n\right\rangle_{\rm ADHM}
 =\frac{1}{N!Z_{\rm ADHM}}
   \sum_{\boldsymbol B\in\mathbb Z^N}
   \int d^N\sigma\,
   \mu(\boldsymbol\sigma,\boldsymbol B)\,
   \Psi_0(\boldsymbol\sigma,\boldsymbol B)
   \widehat{\mathcal O}_1\cdots\widehat{\mathcal O}_n
   \Psi_0(\boldsymbol\sigma,\boldsymbol B).
 \label{eq:ADHM-general-Coulomb-correlator}
\end{equation}
The shift operators act both on $\boldsymbol\sigma$ and on the magnetic flux
$\boldsymbol B$; their lower-charge terms give the monopole-bubbling
completion~\cite{Dedushenko:2018icp}.

One can explicitly show the equality of the partition functions~\cite{Kapustin:2010xq}. With operator insertions, however, the Wick contractions multi-particle kernels. Therefore the equality of partition functions does
not by itself prove the equality of the correlators
\eqref{eq:necklace-general-Higgs-correlator} and
\eqref{eq:ADHM-general-Coulomb-correlator}.

\subsubsection{Mirror map for charge one and two}
\label{subsubsec:necklace-low-charge-mirror-map}
\paragraph{Charge-one operator.}
The mirror map is most naturally formulated first in the product basis.
In the conventions of the one-dimensional topological theories, the
minimal winding operator is identified with the minimal monopole according
to
\begin{equation}
 \frac{1}{(4\pi)^{L/2}}
 \mathcal{M}_{\pm(1,0^{N-1})}
 \quad\longleftrightarrow\quad
 p_1(W_{\pm})=\Tr W_{\pm} .
 \label{eq:minimal-monopole-winding-map}
\end{equation}
\paragraph{First charge-two operator.}Let us next consider the charge-two operators. Both in ADHM theory and the necklace theory, there are two charge-two operators; $\mathcal{M}_{\pm (2,0^{N-1})}$ and $\mathcal{M}_{\pm (1.1,0^{N-2})}$ for ADHM, and $p_1(W_{\pm})^2$ and $p_2(W_{\pm})$ for the necklace theory.

Among them, one combination can be easily identified by taking a product of the charge-one operators. The product in the one-dimensional topological theory is the
point-split, ordered star product, which is preserved by the mirror
map \cite{Dedushenko:2017avn,Dedushenko:2018icp}.
Explicitly, 
\begin{align}\label{starorder}
    (A\star B)(\varphi):=\lim_{\varepsilon\to0^+}A(\varphi+\varepsilon)\,B(\varphi),
\end{align}
In the localized necklace theory, \(W\) contains fields of only
one orientation, while Wick propagators pair fields of opposite
orientations \cite{Dedushenko:2016jxl}.
Consequently,
\(p_1(W_{\pm})\star p_1(W_{\pm})=p_1(W_{\pm})^2\);
the same-orientation product in an abelian necklace is worked out
explicitly in \cite{Dedushenko:2017avn}.

The same product on the ADHM side produces the bubbling-completed monopole operator with charge $\pm (2,0^{N-1})$;
\begin{equation}
 \widehat{\mathcal{M}}_{\pm (2,0^{N-1})}
 :=
 \mathcal{M}_{\pm (1,0^{N-1})}
 \star
 \mathcal{M}_{\pm (1,0^{N-1})}\,.
 \label{eq:completed-charge-two-definition}
\end{equation}
 The assumed minimal map \eqref{eq:minimal-monopole-winding-map} then
implies
\begin{equation}
 \frac{1}{(4\pi)^L}
 \widehat{\mathcal{M}}_{\pm(2,0^{N-1})}
 \quad\longleftrightarrow\quad
 p_1(W_{\pm})^2 .
 \label{eq:charge-two-product-map}
\end{equation}
Note that $\widehat{\mathcal{M}}_{\pm(2,0^{N-1})}$ includes monopole bubbling: composing
the minimal monopole operators produces, in addition to the
leading \((2,0^{N-1})\) flux sector, a screened contribution with the
\((1,1,0^{N-2})\) flux. This relation was already given in \cite{Dedushenko:2018icp} for \(U(2)\) with one adjoint and arbitrary
\(N_f=L\).



\paragraph{Second charge-two operator.}
The second charge-two operator requires more care.  Naively, one might
expect
\begin{equation}
 \frac{2}{(4\pi)^L}
 \mathcal{M}_{\pm(1,1,0^{N-2})}
 \stackrel{?}{\longleftrightarrow}
 p_2(W_{\pm})=\Tr W_{\pm}^2.
 \label{eq:naive-charge-two-single-cycle-map}
\end{equation}
However, by computing the correlator explicitly using
localization at finite $N$, one finds that this naive identification does not
hold in general.  Instead, one has to take an appropriate linear combination of 
\(p_2(W_{\pm})\) and \(p_1(W_{\pm})^2\) to reproduce $\mathcal{M}_{\pm(1,1,0^{N-2})}$.  

A convenient way to determine the linear combination is to compute the Gram matrices, i.e.~matrices whose elements are two-point functions, on both sides and compare them. To simplify the notation, let us denote the basis of charge-two operators in the necklace theory by
\begin{equation}
 A_\pm:=p_1(W_{\pm})^2=(\operatorname{Tr}W_{\pm})^2,
 \qquad
 B_\pm:=p_2(W_{\pm})=\operatorname{Tr}W_{\pm}^2,
 \label{eq:charge-two-Higgs-basis}
\end{equation}
and the basis in the ADHM theory by
\begin{equation}
 \mathcal M_{2,\pm}
 :=
 \widehat{\mathcal M}_{\pm(2,0^{N-1})},
 \qquad
 \mathcal M_{11,\pm}
 :=
 \mathcal M_{\pm(1,1,0^{N-2})},
 \label{eq:charge-two-Coulomb-basis}
\end{equation}
with the completion defined in Eq.~\eqref{eq:completed-charge-two-definition}.
Compatibility with the minimal mirror map and the protected star product
fixes the image of the first basis vector,
\begin{equation}
 A_\pm
 =
 \frac{1}{(4\pi)^L}\mathcal M_{2,\pm}.
 \label{eq:charge-two-product-map-short}
\end{equation}
For the second operator we write a general ansatz
\begin{equation}
 B_\pm
 =
 \frac{1}{(4\pi)^L}
 \left(
  c_{N,L}\mathcal M_{2,\pm}
  +
  f_{N,L}\mathcal M_{11,\pm}
 \right).
 \label{eq:charge-two-general-triangular-map}
\end{equation}

To determine \(c_{N,L}\) and \(f_{N,L}\), we compare the Gram matrices in
the two bases.  Define
\begin{equation}
 G_{\rm H}
 =
 \begin{pmatrix}
  H_{AA}&H_{AB}\\
  H_{AB}&H_{BB}
 \end{pmatrix},
 \qquad
 G_{\rm C}
 =
 \begin{pmatrix}
  C_{11}&C_{12}\\
  C_{12}&C_{22}
 \end{pmatrix},
 \label{eq:charge-two-Gram-matrices}
\end{equation}
with
\begin{align}
 H_{AA}&=\langle A_+A_-\rangle,
 &
 H_{AB}&=\langle A_+B_-\rangle,
 &
 H_{BB}&=\langle B_+B_-\rangle,
 \\
 C_{11}&=\langle\mathcal M_{2,+}\mathcal M_{2,-}\rangle,
 &
 C_{12}&=\langle\mathcal M_{2,+}\mathcal M_{11,-}\rangle,
 &
 C_{22}&=\langle\mathcal M_{11,+}\mathcal M_{11,-}\rangle.
\end{align}
It is convenient to introduce
\begin{equation}
 d_L=(4\pi)^{-L},
 \qquad
 R_{N,L}
 =
 \begin{pmatrix}
  1&0\\
  c_{N,L}&f_{N,L}
 \end{pmatrix}.
\end{equation}
The operator map then implies
\begin{equation}
 G_{\rm H}
 =
 d_L^2
 R_{N,L}
 G_{\rm C}
 R_{N,L}^{T}.
 \label{eq:charge-two-Gram-matching}
\end{equation}
which gives
\begin{equation}
 f_{N,L}^2
 =
 \frac{
  H_{BB}-H_{AB}^2/H_{AA}
 }{
  d_L^2
  \left(
   C_{22}-C_{12}^2/C_{11}
  \right)
 }.
 \label{eq:charge-two-f-from-Gram}
\end{equation}
Thus the Gram matrices determine the magnitude of \(f_{N,L}\) directly.
Once a sign for \(f_{N,L}\) is chosen, the remaining coefficient follows from
\begin{equation}
 c_{N,L}
 =
 \frac{
  H_{AB}/d_L^2
  -
  f_{N,L}C_{12}
 }{
  C_{11}
 }.
 \label{eq:charge-two-c-from-Gram}
\end{equation}
There is, however, a discrete ambiguity which is invisible to the Gram
matrix.  Equation~\eqref{eq:charge-two-f-from-Gram} determines only
\(f_{N,L}^2\), and the two choices of sign are accompanied by different
values of \(c_{N,L}\).  More explicitly, if
\((c_{N,L},f_{N,L})\) solves
Eq.~\eqref{eq:charge-two-Gram-matching}, then
\begin{equation}
 \left(
  c_{N,L}
  +
  2f_{N,L}\frac{C_{12}}{C_{11}},
  -f_{N,L}
 \right)
 \label{eq:charge-two-reflection}
\end{equation}
gives the same full Gram matrix.  
In the next subsection we will use the proposed correspondence between minuscule monopoles and antisymmetrized Schur polynomials to select the
appropriate sign.

\paragraph{Example: $\mathbf{N=2}$, $\mathbf{L=1}$.} For a concrete example, take \(N=2\) and \(L=1\).  With the
unnormalized trace and Weyl-orbit conventions above, localization gives
\begin{equation}
 G_{\rm H}
 =\frac{1}{96\pi^2r^2}
 \begin{pmatrix}12&6\\6&7\end{pmatrix},
 \qquad
 G_{\rm C}
 =\frac{1}{r^2}
 \begin{pmatrix}2&\frac12\\[2pt]\frac12&\frac7{24}\end{pmatrix}.
 \label{eq:charge-two-N2-L1-Gram-example}
\end{equation}
Substitution into Eqs.~\eqref{eq:charge-two-f-from-Gram} and
\eqref{eq:charge-two-c-from-Gram} gives
\(f_{2,1}^2=4\) and \(c_{2,1}=\frac12-\frac14f_{2,1}\).
Thus both \((c_{2,1},f_{2,1})=(0,2)\) and \((1,-2)\) reproduce
the full Gram matrix. For other values of $N,L$, the Gram-matrix equations likewise admit the solution $(c_{N,L},f_{N,L})=(1,-2)$, while the alternative solution generally depends on $N,L$.

As we discuss in the next subsection, the correct choice turns out to be \((c_{2,1},f_{2,1})=(1,-2)\). With this choice, we obtain the following relation
\begin{equation}
\frac{\mathcal{M}_{2,\pm}}{4\pi}=A_{\pm}=({\rm Tr}W_{\pm})^2\,,\qquad \frac{\mathcal{M}_{1,1,\pm}}{4\pi}=\frac{1}{2}(A_{\pm}-B_{\pm})=\frac{1}{2}\left(({\rm Tr}W_{\pm})^2-{\rm Tr}W_{\pm}^2\right)\,.
 \label{eq:charge-two-N2-L1-Schur-branch}
\end{equation}

\subsection{Mirror map for minuscule monopole operators}
\label{subsec:minuscule-schur}
It turns out that a rather simple mirror map exists for  monopole operators with minuscule charges $\pm (1^{K},0^{N-K})$. We propose
\begin{equation}
 \frac{1}{(4\pi)^{LK/2}}\mathcal{M}_{\pm (1^{K},0^{N-K})}\ \longleftrightarrow\
 e_K(W_{\pm})=\operatorname{Tr}_{\wedge^K\mathbf N}W_{\pm}\,,
 \label{eq:minuscule-schur-map}
\end{equation}
where $e_K(W)$ is the so-called {\it Schur polynomial operator}, which can be defined by taking a trace of $W_{\pm}$ in a totally antisymmetric representation of size $K$.
Below, we support the correspondence through the  numerical checks.

The normalized fixed-flux ADHM integral for minuscule monopole operators is
\begin{equation}
\begin{split}
 \mathcal A_{K;N,L}
 ={}&\frac{1}{K!(N-K)!\,\mathcal Z_{N,L}^{(0)}}
 \int_{\mathbb R^K}\!d^K\sigma
 \int_{\mathbb R^{N-K}}\!d^{N-K}\rho\;
 \prod_{\alpha=1}^K h_1(\sigma_\alpha)^L
 \prod_{i=1}^{N-K}h_0(\rho_i)^L\\[-2pt]
 &\quad\times
 \prod_{\alpha<\beta}\mathcal F_0(\sigma_\alpha-\sigma_\beta)
 \prod_{i<j}\mathcal F_0(\rho_i-\rho_j)
 \prod_{\alpha,i}\mathcal F_1(\sigma_\alpha-\rho_i),\\
 \mathcal Z_{N,L}^{(0)}
 ={}&\frac1{N!}\int_{\mathbb R^N}\!d^Ns\;
 \prod_a h_0(s_a)^L
 \prod_{a<b}\mathcal F_0(s_a-s_b).
\end{split}
\label{eq:minuscule-adhm-integral}
\end{equation}
where $\sigma=(\sigma_1,\ldots,\sigma_K)$,
$\rho=(\rho_1,\ldots,\rho_{N-K})$ and
\begin{equation}
\begin{aligned}
 h_0(s)&=\frac{1}{2\cosh\pi s},
 &h_1(s)&=\frac{s}{2\sinh\pi s},\\
 \mathcal F_0(u)&=\tanh^2\pi u,
 &\mathcal F_1(u)&=\frac{u^2}{u^2+1/4}\coth^2\pi u.
\end{aligned}
\label{eq:schur-adhm-kernels}
\end{equation}
where $1/[K!(N-K)!]$ is the Weyl-stabilizer
factor. 
On the other hand, the corresponding quantity in the necklace theory is
\begin{equation}
 \mathcal H_{K;N,L}
 =\bigl\langle e_K(W_{+})e_K(W_{-})\bigr\rangle_{\rm neck},
 \label{eq:minuscule-necklace-integral}
\end{equation}
where the expectation value $\langle \bullet\rangle_{\rm neck}$ is normalized, i.e.~it is divided by the zero-insertion necklace partition
function.

To compare the two expressions, we first use the Cauchy-Binet formula to
expand the Schur polynomial $e_K(W_{-})$ in
\eqref{eq:minuscule-necklace-integral} as a sum over products of
subdeterminants (cf.~\cite{Balasubramanian:2001nh,Corley:2001zk,Giovannoni:2011pn}):
\begin{equation}
e_K(W_{-})=\sum_{\substack{|{\bf A_1}|=\cdots=|{\bf A_L}|=K\\{\bf A_{L+1}}={\bf A_1}}}
 \prod_{I=1}^L\det(Q_I^{[{\bf A_I}, {\bf A_{I+1}}]})\,.
\end{equation}
Here each $\mathbf A_I$ is a $K$-element subset of
$\{1,\ldots,N\}$ and the sum $\sum$ is over the $\binom{N}{K}$ possible choices of such subsets while
$Q_I^{[\mathbf A_I,\mathbf A_{I+1}]}$ denotes the corresponding
$K\times K$ submatrix. The condition
$\mathbf A_{L+1}=\mathbf A_1$ implements the cyclic structure
of the necklace quiver. Performing the Wick contraction, for each node we obtain
\begin{equation}
 \left\langle
 \det(Q_I^{[{\bf A},{\bf B}]})\det(\widetilde Q_I^{[{\bf B},{\bf A}]})
 \right\rangle_x
 =\operatorname{perm}(G_I^{[{\bf A},{\bf B}]}),
 \label{eq:minor-permanent}
\end{equation}
where ${\rm perm}$ is the permanent of a matrix while $G_{I}$ is a matrix of propagators \eqref{eq:propagatornecklace} whose elements are given by
\begin{align}
G_I(a,b)=\mathsf K(\sigma_{I,a}-\sigma_{I+1,b}) \quad \text{with} \quad \mathsf K (x)=\frac{1}{2\cosh\pi x}\,.
\end{align}
Thus, defining $P_I({\bf A},{\bf B})=\operatorname{perm}(G_I^{[{\bf A},{\bf B}]})$, we obtain the following expression for the two-point function of the necklace theory,
\begin{equation}
 \mathcal H_{K;N,L}
 =\left\langle\operatorname{Tr}_{\wedge^K\mathbf N}(P_1P_2\cdots P_L)
 \right\rangle_{\boldsymbol\sigma,\rm neck}\,,
 \label{eq:schur-transfer}
\end{equation}
where the trace is over all possible $K$-element subsets of $\{1,\ldots, N\}$ and $\langle \bullet \rangle_{\boldsymbol\sigma,\rm neck}$ means the average over integrals of the eigenvalues $\sigma$'s.

The expression \eqref{eq:schur-transfer} is particularly suited for the numerical computation and Tables~\ref{tab:schur-e2} and \ref{tab:schur-e3} give the numerical
comparisons of Eqs.~\eqref{eq:minuscule-adhm-integral} and
\eqref{eq:minuscule-necklace-integral}.  On the necklace side we
evaluate ${\rm Tr}(P_1\cdots P_L)$ in
Eq.~\eqref{eq:schur-transfer} and average it in the normalized vacuum
measure.  On the ADHM side we integrate the $K$ charged variables $\sigma$ and
evaluate the neutral-sea integral independently, by recasting it as
the Fredholm-determinant, as in the Fermi-gas
formulation of the ADHM partition function~\cite{Hatsuda:2014vsa}. As shown in the tables, the two results agree within the numerical precision. 

\begin{table}[t]
\centering
\begin{tabular}{c c c c}
\hline
$N$ & $L$ & $\mathcal H_{2;N,L}$ & $\mathcal A_{2;N,L}$\\
\hline
$2$ & $1$ & $7/24$ (exact) & $7/24$ (exact)\\
$8$ & $1$ & $7.878690$ & $7.878659$\\
$20$ & $1$ & $52.008586$ & $52.008542$\\
$30$ & $1$ & $117.697317$ & $117.697266$\\
$20$ & $2$ & $49.83\pm0.17$ & $49.66$\\
$25$ & $5$ & $62.00\pm1.28$ & $61.947$\\
\hline
\end{tabular}
\caption{Normalized $e_2(W_{+})e_2(W_{-})$ correlators versus
minuscule ADHM charge $(1,1,0^{N-2})$. }
\label{tab:schur-e2}
\end{table}

\begin{table}[t]
\centering
\begin{tabular}{c c c c}
\hline
$N$ & $L$ & $\mathcal H_{3;N,L}$ & $\mathcal A_{3;N,L}$\\
\hline
$3$ & $1$ & $0.191407$ & $0.191407$\\
$3$ & $3$ & $0.000448\pm0.000006$ & $0.0004536$\\
$8$ & $2$ & $2.268\pm0.011$ & $2.2661$\\
$12$ & $3$ & $5.05\pm0.10$ & $5.1351$\\
$16$ & $4$ & $11.41\pm0.30$ & $11.743$\\
$20$ & $1$ & $186.52\pm0.07$ & $186.48$\\
$25$ & $5$ & $204\pm7$ & $207.6$\\
\hline
\end{tabular}
\caption{Normalized $e_3(W_{+})e_3(W_{-})$ correlators versus
minuscule ADHM charge $(1,1,1,0^{N-3})$.  For $N=25,L=5$,
$N/L^2=1$; that row also tests the scaling regime of interest.}
\label{tab:schur-e3}
\end{table}

\paragraph{An exact identity for K=N.} As we explain below, for the special case of $K=N$, one can  analytically establish the mirror map between $e_K(W)$ and the minuscule monopole.
Note that this is not the regime directly relevant for the BMN triality where we take $N\to\infty$ with $K$ fixed.

In this case, $e_N(W)=\det W=\prod_I\det Q_I$, and there are no charged--neutral eigenvalues.  Define
\begin{equation}
 \mathsf K_0(x)=\mathsf K(x)=\frac1{2\cosh\pi x},
 \qquad
 \mathsf K_1(x)=\mathsf K(x)^2.
\end{equation}
Without operator insertions, the Cauchy identity allows us to rewrite the measure as a product of determinants of  $\mathsf K_0$ on each link
times $\prod_a h_0(x_{1,a})$ at the first node \cite{Marino:2011eh,Gaiotto:2020vqj}. With the  insertions of $e_N(W_{\pm})$, the Wick contraction gives permanents of $\mathsf K_0$ as discussed above. The key observation is that the product of the determinant and the permanent can be rewritten as follows using {\it Borchardt's identity}~\cite{Ishikawa:2005Borchardt},
\begin{equation}
 \det[\mathsf K_0(x_a-y_b)]\,
 \operatorname{perm}[\mathsf K_0(x_a-y_b)]
 =\det[\mathsf K_1(x_a-y_b)].
 \label{eq:full-marking-Borchardt}
\end{equation}
Thus the numerator and denominator of the normalized correlator, $\langle e_N(W_{+})e_N(W_{-})\rangle_{\rm neck}
=\mathsf Z_1^{(N,L)}/\mathsf Z_0^{(N,L)}$,
can be recast into the same form, with $\mathsf K_1$ and $\mathsf K_0$ respectively:
\begin{equation}
\begin{aligned}
 \mathsf Z_q^{(N,L)}
 := {}&\frac1{(N!)^L}\int\prod_{I=1}^L d^Nx_I\;
 \prod_{a=1}^N h_0(x_{1,a})\\
 &\times\prod_{I=1}^L
 \det\!\left[\mathsf K_q(x_{I,a}-x_{I+1,b})\right]_{a,b=1}^N,
 \qquad q=0,1,
\end{aligned}
 \label{eq:full-marking-necklace-Zq}
\end{equation}
where common, insertion-independent measure factors have been
suppressed. 

We can then follow the standard argument in the fermi-gas approach \cite{Marino:2011eh} to integrate out  nodes $2,\ldots,L$ successively with Andreief's
determinant identity~\cite{Forrester:2018Andreief} to get
\begin{equation}
 \mathsf Z_q^{(N,L)}
 =\frac1{N!}\int_{\mathbb R^N}d^Nx\;
 \prod_{a=1}^Nh_0(x_a)
 \det\!\left[\mathsf K_q^{*L}(x_a-x_b)\right]_{a,b=1}^N\,,
 \label{eq:full-marking-Andr-I}
\end{equation}
where $\ast$ denotes the convolution. As usual, the expression involving the convolution can be simplified by the Fourier transform. That is in fact the rewriting used by Kapustin, Willett and Yaakov \cite{Kapustin:2010xq} to demonstrate the mirror symmetry. Following their analysis and performing the Fourier transformation, we obtain
\begin{equation}
 \mathsf Z_q^{(N,L)}
 =\frac1{N!}\int_{\mathbb R^N}d^Ns\;
 \prod_{a=1}^N h_q(s_a)^L
 \det\!\left[h_0(s_a-s_b)\right]_{a,b=1}^N\,.
 \label{eq:full-marking-Andr-II}
\end{equation}
Here we used $\widehat{\mathsf K_0}=h_0$,
$\widehat{\mathsf K_1}=h_1$, and $\widehat h_0=h_0$, where $\widehat{f}$ is the Fourier transform 
$\widehat f(s)=\int_{\mathbb R}dx\,e^{2\pi isx}f(x)$.
Finally, the Cauchy determinant identity allows us to rewrite the determinant as
\begin{equation}
 \det\!\left[h_0(s_a-s_b)\right]_{a,b=1}^N
 =2^{-N}\prod_{a<b}\tanh^2\!\bigl(\pi(s_a-s_b)\bigr)
 =2^{-N}\prod_{a<b}\mathcal F_0(s_a-s_b).
 \label{eq:full-marking-spectral-Cauchy}
\end{equation}
The factors $2^{-N}/N!$ cancel in the ratio. Hence we arrve at
\begin{equation}
 \left\langle\det W_{+}\det W_{-}\right\rangle_{\rm neck}
 =\frac{\displaystyle\int d^Ns\;
   \prod_a h_1(s_a)^L\prod_{a<b}\mathcal F_0(s_a-s_b)}
  {\displaystyle\int d^Ns\;
   \prod_a h_0(s_a)^L\prod_{a<b}\mathcal F_0(s_a-s_b)}
 .
 \label{eq:full-marking-exact-equality}
\end{equation}
This is an exact identity at finite $N$ and $L(=N_f)$. This agrees precisely with the corresponding quantity on the ADHM side $\mathcal A_{N;N,L}$, \eqref{eq:minuscule-adhm-integral}.

\subsection{Mirror map for general monopole operators}
\label{subsec:column-product-proposal}
The mirror map for the minuscule monopole operators \eqref{eq:minuscule-schur-map} with the star-product structure \eqref{starorder} naturally leads to the following proposal for a map between general monopole operators in the ADHM theory and the winding Higgs-branch operators in the necklace theory: Consider a bubbling-completed monopole operator $\widehat{\mathcal{M}}_{\pm \lambda}$ corresponding to a partition $\lambda=(\lambda_1 \ge \lambda_2 \ge \cdots)$, with $\lambda|=K$. Let $\lambda^{T}=(c_1, c_2, \ldots)$ be its transpose, where $c_j$ denotes the height of the $j$-th column of the original partition $\lambda$. We propose that the corresponding operator in the necklace theory is given by the following product of the Schur-polynomial operators:
\begin{equation}
 \frac{1}{(4\pi)^{LK/2}}\widehat{\mathcal{M}}_{\pm \lambda}
 \ \longleftrightarrow\
 e_{\lambda^{T}}(W_{\pm}):=\prod_j e_{c_j}(W_{\pm}).
 \label{eq:column-fusion-conjecture}
\end{equation}  
For clarity, the first examples are
\begin{equation}
\begin{array}{c|c|c}
 \lambda&\lambda'&e_{\lambda^{T}}(W)\\ \hline
 (K)&(1^K)&e_1(W)^K=(\operatorname{Tr}W)^K\\
 (1^K)&(K)&e_K(W)\\
 (2,1)&(2,1)&e_2(W)e_1(W)\\
 (2,2)&(2,2)&e_2(W)^2\\
 (3,1)&(2,1,1)&e_2(W)e_1(W)^2.
\end{array}
\label{eq:column-fusion-examples}
\end{equation}
The combinatorics makes the charge label natural.  Each $e_{c_j}$
chooses $c_j$ distinct color lines; choices from different columns
can overlap.  


\paragraph{A simple analytic test, charge $(2,1)$ for N=2, L=1.} As a sanity check of our proposal, let us consider
$N=2$, $L=1$, and $\lambda=(2,1)$. In this case, the candidate operator in the necklace theory is
$e_2(Q)e_1(Q)=\det Q\,\operatorname{Tr}Q$.  Direct Gaussian
contraction of the three $Q$'s with the three $\widetilde Q$'s gives
\begin{equation}
 \left\langle\det Q\,\operatorname{Tr}Q\,
  \det\widetilde Q\,\operatorname{Tr}\widetilde Q
 \right\rangle_x
 =\frac12+\frac14 \frac{1}{\cosh^2(\pi (\sigma_1-\sigma_2))}.
 \label{eq:full-support-21-wick}
\end{equation}
Performing the integral of $\sigma$ using
$\langle 1/\cosh^{2}(\pi (\sigma_1-\sigma_2))\rangle_{\boldsymbol\sigma}=1/6$ for $N=2$ and $L=1$,
the normalized necklace correlator is therefore
\begin{equation}
 \mathcal H_{(2,1);2,1}=\frac12+\frac1{24}=\frac{13}{24}.
 \label{eq:full-support-21-necklace}
\end{equation}
Here the first subscript denotes the magnetic partition, not the
number of singly charged lines in
Eq.~\eqref{eq:minuscule-adhm-integral}.

Let us now reproduce this from the ADHM side. The ADHM computation uses
$h_2(s)=(s^2+\tfrac14)h_0(s)$ and $\mathcal F_1$ from
Eq.~\eqref{eq:schur-adhm-kernels}.  Apart from the Weyl factor, the correlator is given by a ratio of two integrals
\begin{equation}
 I_{21}=\int_{\mathbb R^2}ds\,dt\;
 h_2(s)h_1(t)\mathcal F_1(s-t),\qquad
 I_{00}=\int_{\mathbb R^2}ds\,dt\;
 h_0(s)h_0(t)\mathcal F_0(s-t).
\end{equation}
The integral can be performed using the identity
\begin{equation}
 \int_{\mathbb R}dt\;h_2(t+u)h_1(t)
 =\frac{h_4(u)}{12},
 \qquad
 h_4(u)=(u^2+\tfrac14)(u^2+\tfrac94)h_0(u).
 \label{eq:full-support-21-convolution}
\end{equation}
Consequently,
\begin{align}
 I_{21}
 &=\frac1{24}\int_{\mathbb R}du\;
 \frac{u^2(u^2+\tfrac94)\cosh\pi u}{\sinh^2\pi u}
 =\frac{13}{96\pi},\\
 I_{00}&=\frac1{2\pi}.
 \label{eq:full-support-21-integrals}
\end{align}
The first integral follows by integrating by parts and using
$\int_0^\infty du\,u/\sinh\pi u=1/4$ and
$\int_0^\infty du\,u^3/\sinh\pi u=1/8$.
The charge $(2,1)$ has Weyl stabilizer $1!1!$, whereas the vacuum
has stabilizer $2!$.  Including this factor gives the exact match
\begin{equation}
 \mathcal A_{(2,1);2,1}=\frac{2I_{21}}{I_{00}}
 =\frac{13}{24}
 =\mathcal H_{(2,1);2,1}.
 \label{eq:full-support-21-exact}
\end{equation}


\subsection{Mirror map for neutral operator insertions}
\label{subsec:neutral-necklace}

Having identified the mirror duals of monopole operators, we now turn to
neutral Coulomb-branch operators in ADHM theory, which correspond to
insertions of $\operatorname{Tr}(\phi^n)$ in the BMN model. We first
construct the mirror map exactly to all orders for $N=L=1$, then propose
its higher-rank generalization. We also discuss a possible
extension to $L>1$. Appendix~\ref{num test} tests the proposed mirror map by comparing necklace and ADHM correlators numerically.

As explained in Subsection~\ref{subsec:adhm-bmn-operator-dictionary} (around equation 
\eqref{crrmat}),
BMN trace insertions correspond to Bernoulli polynomials of the ADHM
scalar $\Phi$ defined in \eqref{eq:adhm-twisted-dressing-scalar}. To construct the mirror duals of traces of this operator, it is convenient to
first identify the necklace operators $\mathcal C_j$ corresponding
to the simpler traces of powers of $\Phi$,
\begin{equation}
    \operatorname{Tr}(i\Phi)^j
    \quad\longleftrightarrow\quad
    \mathcal C_j,
    \qquad
    N\mathbf 1
    \quad\longleftrightarrow\quad
    \mathcal C_0.
    \label{eq:necklace-neutral-generators}
\end{equation}
The mirror image of the Bernoulli-polynomial insertion
\begin{equation}
    P_n(\Phi)
    =
    \frac{2^n}{n+1}
    \operatorname{Tr}
    B_{n+1}\!\left(\frac12+i\Phi\right)
\end{equation}
then follows by expanding the Bernoulli polynomials in powers of $\Phi$ and using the linearity of trace, and using the definition \eqref{eq:necklace-neutral-generators}. We define the mirror dual of Bernoulli polynomial operators as  $\mathcal P_n$:\footnote{Here we used the expansion of Bernoulli polynomials:
$B_{n+1}(\frac12+x)
=\sum_{j=0}^{n+1}\binom{n+1}{j}
B_{n+1-j}(\frac12)x^j$.}
\begin{equation}
    \mathcal P_n
    :=
    \frac{2^n}{n+1}
    \sum_{j=0}^{n+1}
    \binom{n+1}{j}
    B_{n+1-j}\!\left(\frac12\right)\mathcal C_j.
    \label{eq:necklace-neutral-commutator}
\end{equation}
The first few operators are, up to additive constants,
\begin{equation}
    \mathcal P_0=\mathcal C_1,
    \qquad
    \mathcal P_1=\mathcal C_2,
    \qquad
    \mathcal P_2
    =
    \frac43\mathcal C_3-\frac13\mathcal C_1.
    \label{eq:necklace-first-Bernoulli-generators}
\end{equation}

Recall that we need to take the star commutator of the Bernoulli polynomials with the monopole operator to match with the insertions in BMN (see the discussion around \eqref{eq:dressedmonopole}). 
Under the mirror map \eqref{eq:minuscule-schur-map}, dressing the positive monopole gives
$[\mathcal P_n,e_K(W_+)]_\star$. In the charged-pair correlator,
we can equivalently place the dressing on the negative winding.
Indeed, cyclicity gives
\[
  \big\langle [\mathcal P_n,e_K(W_+)]_\star \star e_K(W_-)\big\rangle
  =
  \big\langle e_K(W_+) \star [e_K(W_-),\mathcal P_n]_\star\big\rangle .
\]
We use the latter form below and define
\begin{equation}
  \mathcal V^{\mathrm{neck}}_{K,n}
  := [e_K(W_-),\mathcal P_n]_\star .
\end{equation}


\paragraph{The ordered product on the circle.}
Let us first start with the simple case $N=L=1$. In this case we have one link field $q$ and its conjugate $\widetilde q$. On the localization locus, the field $q$ has the following one dimensional action \cite{Dedushenko:2016jxl} (we set the radius of the circle $r=1$)
\begin{align}
    -4\pi \int d\varphi\,\widetilde q(\partial_\varphi+\sigma)q.
\end{align}
This fixes the canonical commutator as
\begin{equation}\label{eq:neutral-link-jump}
    [q,\widetilde q]_\star=-\kappa,\qquad \kappa=\frac1{4\pi }.
\end{equation}
Since the propagator $G_\sigma(\varphi)=\langle q(\varphi)\widetilde q(0)\rangle_\sigma$ is the inverse of $-4\pi (\partial_\varphi+\sigma)$, 
\begin{align}
 (\partial_\varphi+\sigma)G_\sigma(\varphi)
 =-\kappa\,\delta(\varphi).
\end{align}
Integrate this equation across the insertion gives $G_\sigma(0^+)-G_\sigma(0^-)=-\kappa$.
These two limits are precisely the contractions in $q\star\widetilde q$
and $\widetilde q\star q$. All other Wick contractions agree when the
two insertions exchange order, provided the other insertions stay
separated.

\paragraph{Weyl ordering.}
It is useful to define a local bilinear rather than the star product when its two fields coincide. We adopt symmetric point splitting, under which the ordinary monomial $q\widetilde q$ represents the Weyl-ordered symbol. Equivalently, it corresponds to the average of the two star-product orderings:
\begin{align}
 q\widetilde q\ \longleftrightarrow\ 
 \tfrac12\big(q\star\widetilde q+\widetilde q\star q\big).
\end{align}
Their difference is fixed by
\eqref{eq:neutral-link-jump}, so
\begin{equation}
 q\star\widetilde q=q\widetilde q-\tfrac\kappa2,\qquad
 \widetilde q\star q=q\widetilde q+\tfrac\kappa2 .
 \label{eq:neutral-two-orders}
\end{equation}
For a monomial of higher degree, Weyl ordering averages over all
orders of its elementary fields. 

Consider the neutral bilinear $z=-q\widetilde q/\kappa$. To insert
$z^2$, we average over the orders of all four elementary fields.
To insert $z\star z$, we first define each bilinear by symmetric
point splitting. We then place one bilinear later than the other
and bring them together in that order. The difference is finite.
In $(q\widetilde q)\star(q\widetilde q)$. the double contraction gives
$(-\kappa/2)(+\kappa/2)$. Dividing by $\kappa^2$ gives
\begin{align}
 z\star z=z^2-\tfrac14 .
\end{align}

\paragraph{The exact map at $N=L=1$.}
Here \(W_-=q\) is the winding of charge \(-1\). The elementary commutator
gives $[q,z]_\star=q$, with the sign fixed by
$z=-q\widetilde q/\kappa$. With the scalar normalization
$i\Phi\leftrightarrow z$, mirror symmetry maps a product of $j$
scalar insertions to the ordered product of $j$ bilinears,
\begin{equation}
 \mathcal C_j=z^{\star j},\qquad
 \mathcal C_1=z,\qquad \mathcal C_2=z^2-\tfrac14,\qquad
 \mathcal C_3=z^3-\tfrac54z .
 \label{eq:rank-one-neutral-star-powers}
\end{equation}
For example, the last expression follows by star multiplying
$z^2-1/4$ by $z$ once more. For a polynomial $f(z)$ the contraction
rules are
\begin{equation}
 z\star f(z)=zf(z)-\tfrac14\bigl(f'(z)+zf''(z)\bigr),
 \qquad [q,f(z)]_\star=q f'(z).
 \label{eq:neutral-rank-one-calculus}
\end{equation}

In the chosen orientation the dressed winding is
$\mathcal V^{\mathrm{neck}}_{1,n}=[q,\mathcal P_n]_\star$.
For example, \eqref{eq:necklace-first-Bernoulli-generators} gives
$\mathcal P_2=\tfrac43z^3-2z$. Equation~\eqref{eq:neutral-rank-one-calculus}
then yields
\begin{equation}
 \mathcal V^{\mathrm{neck}}_{1,0}=q,\qquad
 \mathcal V^{\mathrm{neck}}_{1,1}=2zq,\qquad
 \mathcal V^{\mathrm{neck}}_{1,2}=(4z^2-2)q .
 \label{eq:rank-one-neutral-first}
\end{equation}
Writing the commutator in the reverse order changes the sign of all
three expressions.

We can sum these dressings at all orders. Introduce a formal variable
$\tau$, so that the $n$th derivative at $\tau=0$ selects the
order-$n$ insertion. The generating function of the Bernoulli
polynomials gives
\begin{align}
 \sum_{n\ge0}\frac{\tau^n}{n!}\mathcal P_n
 =\frac{E(\tau,z)}{2\sinh\tau}-\frac1{2\tau},
 \qquad E(\tau,z)=\exp_\star(2\tau z).
\end{align}
The star exponential is the series with coefficients $z^{\star j}$.
It obeys $\partial_\tau E=2z\star E$ and $E(0,z)=1$. Substituting
\eqref{eq:neutral-rank-one-calculus} gives
$E(\tau,z)=\operatorname{sech}\tau\,\exp(2z\tanh\tau)$.
Taking the commutator with $q$ therefore gives
\begin{equation}
 \sum_{n\ge0}\frac{\tau^n}{n!}\mathcal V^{\mathrm{neck}}_{1,n}
 =q\,\operatorname{sech}^2\tau\,\exp(2z\tanh\tau).
 \label{eq:rank-one-neutral-generating}
\end{equation}

\paragraph{Matrix-valued link fields at $L=1$.}
We now take $N>1$ and keep $L=1$, so that the winding matrix is
$W^a_b=Q^a_b$ \footnote{Since we have only one node, $Q$ is adjoint under the gauge group $U(N)$.}. The link fields obey $
 [Q^a_{b},\widetilde Q^c_{d}]_\star
 =-\kappa\,\delta^a_{d}\delta^c_{b}. $
Define the adjoint ``meson" matrix and its normalized form by
\begin{equation}
 Z=Q\widetilde Q,\qquad \mathcal Z=-\frac{Z}{\kappa}.
\end{equation}
The ordinary product in $Z$ is Weyl ordered. Thus each term
$Q^a_{e}\widetilde Q^e_{b}$ in $Z^a_{b}$ represents the average of the
two star orderings of its elementary fields. Using the above commutation relation for $Q^a_{b}$, we get
\begin{align}
 [Z^a_{b},Z^c_{d}]_\star
 =\kappa\bigl(\delta_{b}^c Z^a_{d}-\delta^a_{d}Z^c_{b}\bigr).
\end{align}
The minus sign in $\mathcal Z$ agrees with our
rank-one convention $\mathcal Z=z$.

We seek the necklace operator $\mathcal C_n$ that is mirror to
$\operatorname{Tr}[(i\Phi)^n]$ in the ADHM theory. At rank one,
the mirror image is the star power $z^{\star n}$. At higher rank,
the ``meson" $\mathcal Z$ is a matrix. The natural classical expression
is $\operatorname{Tr}(\mathcal Z^n)$, but the meson entries need not
commute under the star product. We must therefore specify their
order. To do so, start with a polynomial $P(X)$ in the commuting
entries of an auxiliary matrix $X$. One prescription is to replace
each $X^a{}_b$ by $\mathcal Z^a{}_b$ and average each monomial over
all star orderings of its meson factors.

The mirror map must also preserve products. On the ADHM side, two
consecutive neutral insertions on the same arc act by ordinary
multiplication. Their mirror image must therefore equal the star
product of their individual mirror images. For example, quantizing
$(\operatorname{Tr}X^2)^2$ directly must give the same result as
quantizing $\operatorname{Tr}X^2$ and taking its star square.
Simple averaging does not in general satisfy this condition.
The so-called {\it Duflo map} $D$ corrects the polynomial before averaging so that
$D(P_1P_2)=D(P_1)\star D(P_2)$ for any two invariant polynomials
$P_1$ and $P_2$~\cite{Duflo:1977BiInvariant,Chervov:2012hmx}.

Motivated by the rank-one map, we conjecture that
$\mathcal C_n=D(\operatorname{Tr}X^n)$. We now describe the two
steps of this prescription.
\begin{itemize}
\item \textbf{First, correct the polynomial.}
The Duflo prescription uses a scalar function $j^{1/2}(X)$ to specify
the coefficients of the correction. This function is defined by a
determinant. To construct it, fix $X$ and consider the linear map
that sends a matrix $Y$ to $\operatorname{ad}_X(Y)=[X,Y]$.
Each power of $\operatorname{ad}_X$ takes one more commutator with
$X$. Thus, for $f(t)=\sinh(t/2)/(t/2)$, the Taylor series gives
\begin{align}
 f(\operatorname{ad}_X)(Y)
 =Y+\frac{1}{24}[X,[X,Y]]
   +\frac{1}{1920}[X,[X,[X,[X,Y]]]]+\cdots .
\end{align}
This map acts linearly on the $N^2$ entries of $Y$, so it is
represented by an $N^2\times N^2$ matrix. The square root of its
determinant defines
\begin{align}
 j^{1/2}(X)=\sqrt{\det_{\mathfrak{gl}_N}f(\operatorname{ad}_X)},
 \qquad j^{1/2}(0)=1 .
\end{align}

We can evaluate this determinant for diagonal $X$. If the diagonal
entries are $x_a$, then $[X,Y]^a_{b}=(x_a-x_b)Y^a_{b}$. Hence
$f(\operatorname{ad}_X)$ multiplies each entry $Y^a_{b}$ by
$f(x_a-x_b)$. The diagonal entries contribute $f(0)=1$, and the
entries $Y^a_{b}$ and $Y^b_{a}$ give equal factors because $f$ is even.
Taking the square root leaves one factor per pair,
\begin{align}
 j^{1/2}(X)=\prod_{a<b}f(x_a-x_b)
 =1+\frac{N\operatorname{Tr}X^2-(\operatorname{Tr}X)^2}{24}
   +O(X^4),
\end{align}
where we used
$\sum_{a<b}(x_a-x_b)^2=N\operatorname{Tr}X^2-(\operatorname{Tr}X)^2$.

To act on a polynomial, replace each $X^a_{b}$ in this expansion by
$(\partial_X)^a_{b}=\partial/\partial X^b_{a}$. For
$P(X)=\operatorname{Tr}X^3$, all terms with four or more derivatives
vanish. The two remaining derivative terms give
\begin{align}
 \operatorname{Tr}(\partial_X^2)P=6N\operatorname{Tr}X,
 \qquad
 (\operatorname{Tr}\partial_X)^2P=6\operatorname{Tr}X .
\end{align}
The corrected cubic trace is therefore
\begin{align}
 j^{1/2}(\partial_X)\operatorname{Tr}X^3
 =\operatorname{Tr}X^3+\frac{N^2-1}{4}\operatorname{Tr}X .
\end{align}

\item\textbf{Second, turn the corrected polynomials into operators.}
Replace each $X^a_{b}$ by $\mathcal Z^a_{b}$. For each monomial of
degree $m$, average over the $m!$ star orderings of its
entries. We denote this operation by $\operatorname{Sym}_\star$.
For example,
\begin{align}
 X^a_{b}X^c_{d}\ \longmapsto\
 \frac12\bigl(\mathcal Z^a_{b}\star\mathcal Z^c_{d}
             +\mathcal Z^c_{d}\star\mathcal Z^a_{b}\bigr).
\end{align}
\end{itemize}
The two steps together give our candidate neutral generator,
\begin{equation}
 \mathcal C_n^{\mathrm{cand}}
 =\left.\operatorname{Sym}_\star\!
   \left[j^{1/2}(\partial_X)\operatorname{Tr}X^n\right]
   \right|_{X=\mathcal Z}.
 \label{eq:neutral-Duflo-conjecture}
\end{equation}
At $N=1$ there are no pairs $a<b$, so $j^{1/2}=1$. The prescription
then gives exactly $\mathcal C_n=z^{\star n}$.

To evaluate \eqref{eq:neutral-Duflo-conjecture} in a correlator,
we express it as a Weyl symbol in the elementary fields
$Q,\widetilde Q$. The distinction matters already for the cubic
trace. The operation $\operatorname{Sym}_\star$ averages over the
orders of $\mathcal{Z}$'s. For instance for $\mathcal{X}^3$, the average as a Weyl
symbol in the six elementary fields gives
\begin{align}
 \left.\operatorname{Sym}_\star[\operatorname{Tr}X^3]
 \right|_{X=\mathcal Z}
 =\operatorname{Tr}\mathcal Z^3
  -\frac{4N^2+1}{4}\operatorname{Tr}\mathcal Z .
\end{align}
To this we add the term
$+(N^2-1)\operatorname{Tr}\mathcal Z/4$ from the first step.
The first three generators, expressed as elementary-field Weyl
symbols and omitting additive constants, are therefore
\begin{equation}
 \begin{aligned}
 \mathcal C_1^{\mathrm{cand}}&=\operatorname{Tr}\mathcal Z,\qquad
 \mathcal C_2^{\mathrm{cand}}=\operatorname{Tr}\mathcal Z^2,\\
 \mathcal C_3^{\mathrm{cand}}&=\operatorname{Tr}\mathcal Z^3
       -\frac{3N^2+2}{4}\operatorname{Tr}\mathcal Z .
 \end{aligned}
 \label{eq:higher-rank-neutral-generators-candidate}
\end{equation}
Additive constants drop out of the dressing commutators. At $N=1$,
the cubic expression gives $z^3-5z/4$, in agreement with
\eqref{eq:rank-one-neutral-star-powers}.

We now compute the quadratic dressing of the minuscule winding
$e_K(Q)$. This dressing corresponds to the BMN insertion
$\operatorname{Tr}(\phi^2)$. By
\eqref{eq:necklace-first-Bernoulli-generators}, its neutral
generator is
\[
 \mathcal P_2^{\mathrm{cand}}
 =\frac43\mathcal C_3^{\mathrm{cand}}
  -\frac13\mathcal C_1^{\mathrm{cand}}
 =\frac43\operatorname{Tr}\mathcal Z^3
  -(N^2+1)\operatorname{Tr}\mathcal Z .
\]
Consider first a single contraction between $e_K(Q)$ and the cubic
term. Each of the three mesons contains a $\widetilde Q$ that can
contract with a $Q$ in the winding. Summing over the three choices
gives $4/\kappa^2$ times the sum obtained by replacing one of the
$K$ factors of $Q$ in $e_K(Q)$ by $Z^2Q$, where
$Z=Q\widetilde Q$. We can write this sum as
\[
 \left.\frac{d}{du}\,e_K\bigl((1+uZ^2)Q\bigr)\right|_{u=0}.
\]
The derivative holds $Z$ fixed and selects the term linear in $u$.
For example, at $K=1$ it gives $\operatorname{Tr}(Z^2Q)$.

The linear term in $\mathcal P_2^{\mathrm{cand}}$ can contract with
any of the $K$ factors of $Q$, giving $-K(N^2+1)e_K(Q)$.
Terms with two contractions cancel when we subtract the two star
orderings. The cubic term also permits three contractions with three
distinct factors of $Q$. Summing these contractions gives
$K(K-1)(K-2)e_K(Q)/3$.
Combining the terms, we obtain
\begin{equation}
 \begin{aligned}
 \mathcal V_{K,2}^{\mathrm{neck,cand}}
 &=[e_K(Q),\mathcal P_2^{\mathrm{cand}}]_\star\\
 &=\frac4{\kappa^2}
   \left[
    \left.\frac{d}{du}\,e_K\bigl((1+uZ^2)Q\bigr)\right|_{u=0}
   \right]_{\mathrm{Weyl}}
   -\left[K(N^2+1)-\frac{K(K-1)(K-2)}3\right]e_K(Q).
 \end{aligned}
 \label{eq:corrected-neutral-n2}
\end{equation}

In Appendix~\ref{num test}, we test the candidate operators at finite
rank for $L=1$ by numerically comparing necklace and ADHM correlators.
We compare quadratic moments in minuscule sectors and in the
bubbling-completed non-minuscule $(2,0)$ sector. We also compare the
quartic moment for $N=2$ and $K=1$.

\paragraph{Generalization to $L>1$.}
Let $Q_I$ be the bifundamental on link $I$, with conjugate
$\widetilde Q_I$, where $I=1,\dots,L$ and the node labels are
understood modulo $L$. The link fields obey
\begin{equation}
 \bigl[(Q_I)^a{}_{b},(\widetilde Q_J)^c{}_{d}\bigr]_\star
 =-\kappa\,\delta_{IJ}\delta^a{}_{d}\delta^c{}_{b},
 \qquad W=Q_1\cdots Q_L .
\end{equation}
Here $Q_I$ transforms as $Q_I\mapsto g_I Q_I g_{I+1}^{-1}$, so
$a$ belongs to node $I$ and $b$ to node $I+1$. The conjugate field
carries the opposite representations. The ``meson"
\begin{equation}
 \mathcal Z_I:=-\frac{Q_I\widetilde Q_I}{\kappa}
\end{equation}
therefore transforms in the adjoint of node $I$. Since it contains
only fields on link $I$, its entries obey the same commutation
relations as at $L=1$. With
$\mathcal C_0^{(I),\mathrm{cand}}=N\mathbf 1$, the natural
generalization of \eqref{eq:neutral-Duflo-conjecture} is
\begin{equation}
 \begin{aligned}
 \mathcal C_n^{(I),\mathrm{cand}}
 &=\left.\operatorname{Sym}_\star\!\left[
   j_{\mathfrak{gl}_N}^{1/2}(\partial_X)\operatorname{Tr}X^n
   \right]\right|_{X=\mathcal Z_I},\\
 \mathcal P_n^{(I),\mathrm{cand}}
 &=\frac{2^n}{n+1}\sum_{j=0}^{n+1}\binom{n+1}{j}
   B_{n+1-j}\!\left(\tfrac12\right)\mathcal C_j^{(I),\mathrm{cand}},\\
 \mathcal V_{K,n}^{(I),\mathrm{neck,cand}}
 &=[e_K(W),\mathcal P_n^{(I),\mathrm{cand}}]_\star .
 \end{aligned}
 \label{eq:general-L-candidate}
\end{equation}
Although the candidate operators carry a link label $I$, we expect
the normalized charged-pair correlators considered here to be
independent of the choice of insertion link. For $N=K=1$, this
independence follows analytically for arbitrary $L$ and all marked
moments. At higher rank, the evidence comes from the numerical
comparisons between links in Appendix~\ref{num test}.

For the quadratic dressing, we can use the $L=1$ result directly.
The neutral generator on link $I$ is obtained by replacing
$\mathcal Z$ with $\mathcal Z_I$ in $\mathcal P_2^{\mathrm{cand}}$.
Only $Q_I$ in the winding can contract with $\widetilde Q_I$ in
this generator. To write the result, start the winding at node $I$,
\[
 W_{(I)}=Q_I\cdots Q_L Q_1\cdots Q_{I-1},
 \qquad e_K(W_{(I)})=e_K(W).
\]
Set $Z_I=Q_I\widetilde Q_I$. Replacing $Q_I$ by $(1+uZ_I^2)Q_I$
then replaces $W_{(I)}$ by $(1+uZ_I^2)W_{(I)}$. All other links
retain their order. Equation~\eqref{eq:corrected-neutral-n2}
therefore gives
\begin{equation}
 \begin{aligned}
 \mathcal V_{K,2}^{(I),\mathrm{neck,cand}}
 &=\frac4{\kappa^2}
   \left[
    \left.\frac{d}{du}\,
    e_K\bigl((1+uZ_I^2)W_{(I)}\bigr)\right|_{u=0}
   \right]_{\mathrm{Weyl}}\\
 &\quad-\left[K(N^2+1)-\frac{K(K-1)(K-2)}3\right]e_K(W).
 \end{aligned}
 \label{eq:general-L-n2}
\end{equation}
For $N=K=1$, this gives $(4z_I^2-2)w$, where $z_I$ is the
rank-one meson on link $I$ and $w$ is the winding around the necklace.

In Appendix~\ref{num test}, we compare the resulting quadratic moments
with the ADHM moments for necklaces with two and three links and find
numerical agreement in the cases tested. These comparisons support
the proposed mirror map. Establishing it as an operator identity at
higher rank requires additional correlators that can distinguish
possible operator mixing.

\section{IIB pp-wave and mABJM}\label{sec:iib-ppwave-mabjm}

In this section, we develop the type IIB analogue of the triality introduced
in the Introduction, relating a large-charge sector of a four-dimensional
orbifold quiver theory, the DLCQ of type IIB string theory on the maximally
supersymmetric pp-wave, and mass-deformed ABJM theory on
$\mathbb R_t\times T^2$ at Chern--Simons level $k=1$.
The construction parallels the BMN triality:
a large-orbifold Penrose limit produces a compact lightlike
direction, and the rescaled orbifold charge
$m=J_{\rm orb}/K$ determines the rank of the proposed
Matrix Theory description. It is important to note that the role of ABJM theory
is different in the two trialities. Here, mass-deformed ABJM theory on $T^2$
provides the finite-rank Matrix Theory description, rather than the holographic
gauge-theory side as in the BMN triality.

Another difference from the BMN triality is the presence of D3-branes
wrapping torsion three-cycles of $S^5/\mathbb Z_K$.
The corresponding determinant/baryonic operators carry charge $N$
and survive the triple scaling limit (in this case, $N/K$ is fixed as we will see below).
Such wrapped D-branes are absent in the 
pp-wave geometry, motivating a restriction to a sector with vanishing
torsion wrapping charge.
Within this sector, finite-rank corrections disappear
coefficientwise at fixed longitudinal momentum and excitation
charges, allowing the quiver index to be replaced by the
projected graviton index.

The relation between $\mathcal N=2$ orbifold quiver theories and the
DLCQ of type IIB string theory was already studied in \cite{Mukhi:2002ck}.
Here we focus instead on the $\mathcal N=1$ orbifold, whose advantage is
that its $\mathbb Z_K$ action on $S^5$ is free, so the bulk geometry
has no orbifold fixed locus.
Despite this difference, both constructions lead to the same pp-wave
geometry and parameter dictionary. 

Below, we establish the parameter dictionary for the $\mathcal N=1$
orbifold and show that its limiting index reproduces the vacuum
degeneracies of level-one mass-deformed ABJM theory, providing evidence
for the proposed triality.
A fuller identification of the surviving Hilbert space and
its dynamics is left for future work. It would also be interesting to extend our index analysis to the $\mathcal{N}=2$ case, taking into account the twisted sector\footnote{In the $\mathcal{N}=2$ case, the orbifold of $S^5$
has a fixed circle supporting localized twisted-sector modes.
These modes are expected to decouple in the Penrose limit
taken along an orbifolded circle away from the fixed locus.
Since they carry zero orbifold charge, we expect that their index contribution
can be removed by dividing by the full neutral factor,
analogously to the zero-monopole normalization in the ADHM
calculation.} localized at the fixed locus.

\subsection{The $\mathcal{N}=1$ $\mathbb{Z}_K$ orbifold of
$\mathcal{N}=4$ SYM}

Consider $\mathcal{N}=4$ SYM with gauge group $SU(NK)$ and
adjoint chiral multiplets $X,Y,Z$.
We define the $\mathbb Z_K$ orbifold by imposing
\begin{equation}
    (X,Y,Z)\sim 
    \gamma (\omega X ,\omega Y,\omega^{-2}Z)\gamma^{-1},
    \label{eq:IIB-orbifold}
\end{equation}
where $\gamma$ and $\omega$ are given by
\begin{align}
\gamma =
\begin{pmatrix}
\mathbf{1}_N & 0 & \cdots & 0 \\
0 & \omega \mathbf{1}_N & \cdots & 0 \\
\vdots & \vdots & \ddots & \vdots \\
0 & 0 & \cdots & \omega^{K-1}\mathbf{1}_N
\end{pmatrix},
\qquad
\omega = e^{2\pi i/K}\,,
\end{align}
with odd $K\geq3$.
The resulting quiver theory preserves
$\mathcal N=1$ supersymmetry. In the holographic dual, the projection \eqref{eq:IIB-orbifold} induces
the $\mathbb Z_K$ identification
$(X,Y,Z)\sim(\omega X,\omega Y,\omega^{-2}Z)$ on the complex embedding
coordinates of $S^5$. For odd $K$, this action is free, so the orbifold
$S^5/\mathbb Z_K$ has no fixed loci and hence no twisted-sector modes
localized at orbifold singularities.

The resulting theory\footnote{The theory arises on the worldvolume of $N$ D3-branes
probing $\mathbb R^6/\mathbb Z_K$. Equivalently, in the covering
space there are $NK$ D3-branes, including the orbifold images.
Projecting the open-string spectrum gives the
four-dimensional $\mathcal N=1$ quiver theory.} \cite{Kachru:1998ys,Lawrence:1998ja}
has  gauge group
\begin{equation}
    G=\prod_{a\in\mathbb Z_K}SU(N)_a.
\end{equation}
together with the chiral
multiplets transforming as
\begin{equation}
    X_a,Y_a\in
    (\mathbf N_a,\overline{\mathbf N}_{a+1}),
    \qquad
    Z_a\in
    (\mathbf N_a,\overline{\mathbf N}_{a-2}),
\end{equation}
with superpotential
\begin{equation}
    W=h\sum_{a\in\mathbb Z_K}
    \operatorname{Tr}
    \left[
        (X_aY_{a+1}-Y_aX_{a+1})Z_{a+2}
    \right].
\end{equation}
All node labels are understood modulo $K$.

Let $Q_X,Q_Y,Q_Z$ generate phase rotations of the
three complex coordinates, with unit charge assigned
to the corresponding scalar. The charge corresponding to the orbifold generator is then given by
\begin{equation}
    J_{\rm orb}=Q_X+Q_Y-2Q_Z.
    \label{eq:IIB-orbifold-charge}
\end{equation}
As in the necklace theory discussed in
Section~\ref{sec:necklace-bmn-correlators}, the operators relevant
for our limit are gauge-invariant traces associated with closed
paths around the quiver. These ``winding'' operators have orbifold
charges quantized in units of $K$,
\begin{equation}
    J_{\rm orb}\in K\mathbb Z.
\end{equation}
However, since each node has gauge group ${\rm SU}(N)$, the theory also
admits baryonic operators, such as $\det X_a$ and $\det Y_a$.
Unlike winding operators, their charges need not be multiples
of $K$.
\subsection{Parameter matching and the scaling limit}

At the orbifold point, the holographic dual is
type IIB string theory on $AdS_5\times S^5/\mathbb{Z}_K$~\cite{Kachru:1998ys,Lawrence:1998ja}.
The common string-frame radius satisfies
\begin{equation}
    R^4=4\pi g_sNK\,\alpha'^2.
\end{equation}
There are $N$ units of five-form flux through the quotient
and $NK$ units through the covering space. $g_s$ is the bulk string coupling, and is related to the gauge coupling of each node of the quiver by\footnote{
Under the orbifold projection, the \(U(NK)\) gauge field of the covering-space theory is projected onto its block-diagonal components, giving rise to the \(K\) gauge-group factors of the quiver. The normalization of the orbifold-projected action differs from that of the covering-space theory by a factor of \(1/K\), which can be absorbed into a redefinition of the gauge coupling. Consequently, at the orbifold point, where all node couplings are equal, one finds \(g_{\mathrm{YM}}^2=4\pi K g_s\).}
\begin{align}\label{eq:YMcouplingrel}
g_{\mathrm{YM}}^2=4\pi K g_s .
\end{align}

\paragraph{The Penrose limit and DLCQ.}

We choose the null geodesic at the center of $AdS_5$
moving along the $X$ circle,
\begin{equation}
    X=Re^{i\phi},\qquad Y=Z=0,\qquad t=\phi,
\end{equation}
where $t$ is dimensionless global AdS time.
Introduce
\begin{equation}
    x^+=\frac{t+\phi}{2\mu},
    \qquad
    x^-=\mu R^2(t-\phi).
\end{equation}
The Penrose limit gives
\begin{equation}
    ds^2=-2dx^+dx^-
    -\mu^2\sum_{I=1}^{8}(r^I)^2(dx^+)^2
    +\sum_{I=1}^{8}(dr^I)^2,
\end{equation}
together with the self-dual five-form flux of the
maximally supersymmetric IIB pp-wave~\cite{Blau:2001ne,Blau:2002dy}.
Here $\mu$ is the transverse zero-mode frequency~\cite{Metsaev:2001bj}.

The orbifold identification induces
\begin{equation}
    \delta x^+=\frac{\pi}{\mu K},
    \qquad
    \delta x^-=-\frac{2\pi\mu R^2}{K},
\end{equation}
as well as transverse rotations of order $K^{-1}$.
Taking $K\to\infty$ with $R^2/K$ fixed leaves a purely
lightlike identification,
\begin{equation}
    x^-\sim x^-+2\pi R_-,
    \qquad
    R_-=\frac{\mu R^2}{K}.
\end{equation}
The corresponding limit for $\mathcal N=2$ quiver theories was studied
in \cite{Mukhi:2002ck}. The resulting parameter dictionary includes
\begin{equation}
    \frac{R_-}{\mu\alpha'}
    =
    \sqrt{4\pi g_s\frac{N}{K}}.
    \label{eq:IIB-radius-dictionary}
\end{equation}
To obtain the DLCQ of type IIB string theory on the pp-wave at finite
string coupling, both $g_s$ and $R_-/(\mu\alpha')$ must remain finite.
Together with \eqref{eq:YMcouplingrel}, this fixes the relevant gauge theory limit, 
\begin{align}
    N,K,g_{\rm YM}\to\infty,
    \qquad
    \frac{N}{K},\quad
    \frac{g_{\rm YM}^2}{K}
    \quad\text{fixed}.\label{eq:prelimlimit}
\end{align}

We must also identify the charge sector that survives this limit.
A state in the orbifold theory with conformal dimension $\Delta$ and charge
$Q_X$ has pp-wave light-cone energy and momentum
\begin{equation}
    H_{\rm lc}=\mu(\Delta-Q_X),
    \qquad
    P^+=\frac{\Delta+Q_X}{2\mu R^2}.
\end{equation}
We keep $\Delta-Q_X,Q_Y,Q_Z$ finite. Then
\begin{equation}
    R_-P^+
    =\frac{J_{\rm orb}}{K}
    +\frac{\Delta-Q_X}{2K}
    -\frac{Q_Y-2Q_Z}{K}
    \longrightarrow m.
\end{equation}
Thus the light-cone momentum is determined by the rescaled orbifold
charge, $m=J_{\rm orb}/K$. Combining this with
\eqref{eq:prelimlimit}, we arrive at the full scaling limit
\begin{equation}
    N,K,g_{\rm YM},J_{\rm orb}\to\infty,
    \qquad
    \frac{N}{K},\ \frac{g_{\rm YM}^2}{K},\ 
   m= \frac{J_{\rm orb}}{K},\qquad \text{fixed}.
 \label{eq:IIB-scaling}
\end{equation}
The integer $m$ counts units of longitudinal momentum along the
compact lightlike direction, $P^+=m/R_-$.

\paragraph{Relation to mABJM.}
The Matrix Theory description for the DLCQ of type IIB superstring in the pp-wave background is the mass deformation of $m$ M2-branes on a spatial torus~\cite{Gomis:2008cv,Gomis:2008vc}\footnote{This follows from the usual matrix-string-theory logic. An M2-brane wrapping the M-theory circle reduces to a fundamental string in Type IIA. Thus, \(m\) M2-branes wrapping the appropriate two-torus correspond, after reduction, to fundamental strings carrying a total winding number \(m\) around the remaining circle. Performing T-duality along this circle exchanges winding and momentum, yielding a fundamental string carrying \(m\) units of momentum.}. Its field-theoretic realization is mass-deformed ABJM
theory with gauge group $U(m)_1\times U(m)_{-1}$
\cite{Gomis:2008vc}, defined on the torus with periodic fermions
along both cycles. Here the rank $m$ corresponds to the units of light-cone
momentum.

Identifying the field-theory time with $x^+$ and defining
$\mu_{\rm ABJM}$ as the free center-of-mass scalar mass gives
\begin{equation}
    \mu_{\rm ABJM}=\mu,
    \qquad
    H_{\rm mABJM}=\mu(\Delta-Q_X).
\end{equation}
The complex structure of the torus is
\begin{equation}
    \tau_{T^2}=C_0+\frac{i}{g_s}.
\end{equation}
For $C_0=0$, the relation between the radii of the torus $r_{1,2}$ and the parameters of string theory is
\cite{Mukhi:2002ck}
\begin{equation}\label{r1r2}
    r_1=\frac{g_s\alpha'}{R_-},
    \qquad
    r_2=\frac{\alpha'}{R_-},
\end{equation}
with cycle lengths $2\pi r_1,2\pi r_2$.\footnote{This follows by T-dualizing the DLCQ circle, \(r_2=\alpha'/R_-\), and then lifting to M-theory, which gives the second torus radius \(r_1=R_{11}=g_s\alpha'/R_-\).}
Comparing \eqref{eq:IIB-radius-dictionary} with \eqref{r1r2} gives the parameter dictionary between the orbifold theory and mABJM:
\begin{equation}
    \mu^2A_{T^2}=\pi\frac{K}{N},
    \qquad \tau_{T^2} =\frac{4\pi K i}{g_{\rm YM}^2}\,,\qquad 
    m=\frac{J_{\rm orb}}{K}\,.
    \label{eq:IIB-mABJM-dictionary}
\end{equation}
Here $A_{T^2}$ is the area of the torus, so $\mu^2 A_{T^2}$ measures its
dimensionless size in units set by the mass deformation, while $m$
is the rank of the mABJM gauge group.

\paragraph{Baryonic sector.}
An important difference from the BMN triality is the presence of
baryonic operators in the $SU(N)^K$ quiver theory, such as
\begin{equation}
    B_a=\det X_a,
    \qquad
    \Delta=Q_X=J_{\rm orb}=N.
\end{equation}
These operators have vanishing light-cone energy and finite rescaled
orbifold charge,
\begin{equation}
    \Delta-Q_X=0,
    \qquad
    \frac{J_{\rm orb}}{K}=\frac{N}{K},
\end{equation}
and therefore survive the scaling limit.
Holographically, they correspond to D3-branes wrapping the torsion
three-cycle $\Sigma_X=\{X=0\}\subset S^5/\mathbb Z_K$, which generates
$H_3(S^5/\mathbb Z_K,\mathbb Z)=\mathbb Z_K$
\cite{Witten:1998xy,Arai:2019wgv}. However, this cycle lies outside the transverse neighborhood of
the null geodesic retained by the Penrose limit, suggesting
that these states should be excluded from the
pp-wave description.

To isolate the sector captured by the pp-wave background, we therefore restrict to a sector with vanishing baryonic
charge $Q_{\rm top}=0$ in the quiver theory, which corresponds to vanishing torsion
wrapping charge in the bulk. After the restriction, the surviving operators are 
the winding traces such as $\operatorname{Tr}M^\ell$, where
$M:=X_0X_1\cdots X_{K-1}$ is the product around the quiver.

\paragraph{Hilbert series for the Higgs-branch chiral ring.}
In the scaling limit \eqref{eq:IIB-scaling}, the pure-$X$ chiral ring
simplifies considerably upon restricting to $Q_{\rm top}=0$.
Although topologically neutral combinations of baryonic determinants
exist, they require at least $K$ determinants and hence carry
$J_{\rm orb}\geq NK=O(K^2)$. They therefore lie beyond the scaling sector $J_{\rm orb}=mK $ with fixed $m$ since $NK=O(K^2)$ in the limit.

The surviving operators are thus multi-traces of
$M=X_0X_1\cdots X_{K-1}$. Since each $\operatorname{Tr}M^\ell$
carries charge $\ell K$, their counting reduces to partitions
of $m$. The resulting Hilbert series is
\begin{equation}\label{eq:restrictedHilbert}
\lim_{\substack{N,K\to\infty\\N/K\text{ fixed}}}
    \mathcal H^X_{Q_{\mathrm{top}}=0}(\xi)
    =
    \prod_{\ell=1}^{\infty}\frac{1}{1-\xi^\ell}
    =
    \sum_{m=0}^{\infty}p(m)\xi^m,
    \qquad \xi=x^K,
\end{equation}
where $x$ is the fugacity for $J_{\rm orb}$ and the limit is
understood coefficientwise. As we will see below, this precisely
reproduces the vacuum degeneracy of level-one
$U(m)\times U(m)$ mass-deformed ABJM theory.

\subsection{Comparison with  graviton index}
Let us now compare the index \eqref{eq:restrictedHilbert} with the gravity dual. Thanks to $N\to \infty$ and the projection $Q_{\rm top}$, non-perturbative objects such as D-branes do not contribute to the index and hence we simply consider the graviton index of AdS$_5\times S^5/\mathbb{Z}_k$. 
We use scalar fugacities $x,y,z$ and angular fugacities
$p,q$, subject to
\begin{equation}
    xyz=pq.
\end{equation}
Begin with the single-particle index~\cite{Kinney:2005ej}\footnote{We retain the conventional singleton contribution in
this expression. For the freely acting orbifold below,
its invariant part is a neutral free vector multiplet.
Removing this contribution consistently does not change
the normalized ratio considered below.}
\begin{equation}
    f_{\rm grav}
    =\frac{x}{1-x}+\frac{y}{1-y}+\frac{z}{1-z}
    -\frac{p}{1-p}-\frac{q}{1-q}.
    \label{eq:IIB-parent-index}
\end{equation}
The plethystic exponential is then defined by
\begin{equation}
    \operatorname{PE}[f]
    =
    \exp\left[
        \sum_{r=1}^{\infty}\frac1r
        f(x^r,y^r,z^r,p^r,q^r)
    \right].
\end{equation}

\paragraph{Orbifold projection.}
The $\mathbb Z_K$ projection is implemented by averaging the
single-particle graviton index over the orbifold action
\cite{Arai:2019wgv}:
\begin{equation}
    f_{\rm grav}^{\mathbb Z_K}
    =
    \frac1K\sum_{r=0}^{K-1}
    f_{\rm grav}(\omega^r x,\omega^r y,\omega^{-2r}z,p,q).
\end{equation}
Since $K$ is odd, the orbifold weights $(1,1,-2)$ are all
coprime to $K$. The projection therefore retains only powers
of $x$, $y$, and $z$ divisible by $K$, giving
\begin{equation}
    f_{\rm grav}^{\mathbb Z_K}
    =
    \frac{x^K}{1-x^K}
    +\frac{y^K}{1-y^K}
    +\frac{z^K}{1-z^K}
    -\frac{p}{1-p}-\frac{q}{1-q}\,.
\end{equation}
Taking the plethystic exponential, we obtain the orbifold
graviton index
\begin{equation}
    \mathcal I_{\rm grav}^{\mathbb Z_K}
    =
    \prod_{\ell=1}^{\infty}
    \frac{(1-p^\ell)(1-q^\ell)}
    {(1-x^{K\ell})(1-y^{K\ell})(1-z^{K\ell})}.
    \label{eq:IIB-orbifold-index}
\end{equation}

\paragraph{The neutral single-particle factor.}
As in the BMN triality \cite{Komatsu:2026ano}, the index comparison
requires removing the contribution from single-particle states neutral
under $J_{\rm orb}$. To isolate this contribution, we introduce a
fugacity $u$ for $J_{\rm orb}$ through
\begin{equation}
    (x,y,z)\longmapsto(ux,uy,u^{-2}z)\,,
\end{equation}
and project onto the $u^0$ term:
\begin{equation}
    f_{\rm grav}^{0}
    :=
    \oint \frac{du}{2\pi i u}
    f_{\rm grav}(ux,uy,u^{-2}z,p,q)
    =
    -\frac{p}{1-p}-\frac{q}{1-q}.
\end{equation}
Taking the plethystic exponential gives the neutral factor
\begin{equation}
    \mathcal I_{\rm grav}^{0}
    :=
    \operatorname{PE}[f_{\rm grav}^{0}]
    =
    \prod_{\ell=1}^{\infty}(1-p^\ell)(1-q^\ell).
\end{equation}
\paragraph{The large-$K$ limit.}Let us finally consider the large $K$ limit relevant for the pp-wave limit.
Introduce
\begin{equation}
    \xi=x^K,\qquad a=\frac{y}{x},\qquad b=x^2z,
    \qquad ab=pq.
    \label{eq:IIB-rescaled-fugacities}
\end{equation}
These variables are adapted to
$J_{\rm orb}=Q_X+Q_Y-2Q_Z$:
\begin{equation}
    x^{Q_X}y^{Q_Y}z^{Q_Z}
    =
    x^{J_{\rm orb}}a^{Q_Y}b^{Q_Z}.
\end{equation}
The normalized graviton index becomes
\begin{equation}
    \frac{\mathcal I_{\rm grav}^{\mathbb Z_K}}
         {\mathcal I_{\rm grav}^{0}}
    =
    \prod_{\ell=1}^{\infty}
    \frac{1}{
        (1-\xi^\ell)
        (1-a^{K\ell}\xi^\ell)
        (1-b^{K\ell}\xi^{-2\ell})
    }.
\end{equation}
Taking $K\to\infty$ at fixed $\xi,a,b$, in the chamber
$0<|\xi|<1$ and $|a|,|b|<1$, gives
\begin{equation}
    \lim_{K\to\infty}
    \frac{\mathcal I_{\rm grav}^{\mathbb Z_K}}
         {\mathcal I_{\rm grav}^{0}}
    =
    \prod_{\ell=1}^{\infty}\frac{1}{1-\xi^\ell}
    =
    \sum_{m=0}^{\infty}p(m)\xi^m.
    \label{eq:IIB-graviton-limit}
\end{equation}
Here $p(m)$ is the number of partitions of $m$, which agrees with the index on the gauge-theory side \eqref{eq:restrictedHilbert}.

\subsection{Comparison with mABJM}

The level-one mass-deformed ABJM theory has gauge group
$U(m)_1\times U(m)_{-1}$, with matter fields
\begin{equation}
    Z_a\in(\mathbf m,\overline{\mathbf m}),
    \qquad
    W_{\dot a}\in(\overline{\mathbf m},\mathbf m),
    \qquad a,\dot a=1,2.
\end{equation}
Its supersymmetric vacua are described by GRVV blocks~\cite{Gomis:2008vc}.
After imposing the quantum restrictions, the vacuum
degeneracy at level one is the partition number
$p(m)$~\cite{Kim:2010mr}, with generating function
\begin{equation}
    \sum_{m=0}^{\infty}p(m)\xi^m
    =
    \prod_{\ell=1}^{\infty}\frac{1}{1-\xi^\ell}.
\end{equation}
which precisely agrees with \eqref{eq:IIB-graviton-limit}. We hence established the relation analogous to the one presented in \cite{Komatsu:2026ano} for the BMN triality:
\begin{equation}
    \left.    \lim_{\substack{N,K\to\infty\\N/K=c}}
    \frac{\mathcal I_{\mathrm{quiver},\,Q_{\mathrm{top}}=0}}
         {\mathcal I^0}
    \right|_{x^K=\xi}
    =
    \sum_{m=0}^{\infty}\xi^m\,
    \mathcal I_{\mathrm{mABJM},m}.
    \label{eq:IIB-topological-sector-matching}
\end{equation}

On the quiver theory side, one can consider a more refined counting such as the Hilbert series of the chiral ring generated by both $X$ and $Y$. It would be interesting to compute the counterpart on the mass-deformed side and test the correspondece further.
 

\section{Discussion}\label{sec:discussion}
In this paper, we have extended the holographic construction of
Matrix Theory proposed in~\cite{Komatsu:2026ano} to the
three-dimensional $\mathcal N=4$ ADHM theory and tested the
correspondence beyond protected state counting. An appropriate
Penrose limit of the ADHM bulk dual, combined with a scaling of
the orbifold identification, produces the DLCQ of the maximally
supersymmetric eleven-dimensional pp-wave. This motivates a
triality between the triple-scaled large-charge sector of ADHM
theory, M-theory on this DLCQ background, and the finite-rank
BMN matrix model, with the ADHM monopole charge identified with
the BMN rank. We showed that the ADHM superconformal index,
after dividing out the zero-monopole contribution, reproduces
the grand-canonical BMN index, with positive magnetic-flux
configurations corresponding to fuzzy-sphere vacua.

Our main new evidence comes from correlation functions in topological subsectors.
Using supersymmetric localization, we matched the normalized
localization measures and operator insertions in fixed
positive screened magnetic-flux sectors to those in the
corresponding BMN vacua. For minuscule monopole charges, this
gives a direct correspondence between normalized physical ADHM
correlators and protected correlators in the trivial BMN vacuum.
For non-minuscule charges, monopole bubbling introduces several
screened sectors, and we found that a contribution to their relative weights matches the classical BMN instanton action.
We also proposed and tested a mirror description of
bubbling-completed monopole operators in terms of Higgs-branch
winding operators in the necklace theory.

Finally, we discussed a type IIB analogue relating a large-charge
limit of four-dimensional orbifold quiver theories, the DLCQ of
the maximally supersymmetric type IIB pp-wave, and level-one
mass-deformed ABJM theory on $\mathbb R_t\times T^2$.
In the sector of vanishing torsion wrapping charge, the limiting
protected count reproduces the vacuum degeneracies of the
proposed mass-deformed ABJM description.
We conclude with several observations and directions for
future study.
\vspace{15pt}

First, there are a few directions that are directly related to the analysis presented in this paper.
\paragraph{Monopole bubbling and BMN instantons.}
The sector-by-sector correspondence between ADHM theory and the BMN model that we established raises a basic question:
what does a physical monopole operator actually create in the
BMN theory? For a non-minuscule monopole, bubbling allows
the magnetic charge to be screened in several different ways.
Each screened sector can correspond to a different BMN vacuum.
The full monopole correlator therefore involves a sum over
vacua, rather than the contribution of a single one.

But what is the BMN interpretation of this sum? We need to
identify the state or boundary condition that reproduces it,
including the relative weights dictated by monopole bubbling. 

For this purpose, an important missing piece is localization of the BMN correlator, {\it including instanton contributions}. Such a calculation
has not yet been carried out, but it is essential for
understanding precisely how the sum over vacua and its
relative weights emerge on the BMN side.
Our comparison of classical instanton actions is a first
step in this direction, but it is not the whole story.
A complete matching must also account for fluctuation
measures, external-state normalizations, and fermionic
zero modes~\cite{Yee:2003ge,Lin:2006tr}. The zero modes
are particularly important because they determine which
operator insertions can give a nonzero instanton contribution.
A natural place to start is with the simplest nontrivial
screening process, where the dictionary can be worked out
explicitly. Once that is understood, we can ask how the
picture changes in the M5-brane regime discussed in 
\cite{Maldacena:2002rb,Asano:2017nxw}.
\paragraph{Understanding the ADHM--BMN operator dictionary.}
In Section~\ref{sec:adhm-bmn-correlators}, we showed that neutral
BMN trace insertions can be reproduced by taking star commutators
of ADHM monopole operators with suitable Bernoulli-polynomial
generators. The key mathematical trick is the Bernoulli
finite-difference identity. It turns the difference between
the ADHM scalar on the two arcs into a sum over the weights
of the corresponding BMN fuzzy-sphere block. Together with
the matching localization measures, this gives us an explicit
dictionary at the level of protected correlators.

But there is still a basic question: why does this dictionary
work? Matching correlators is one thing; understanding why
these dressed monopole operators at large charge actually create the
corresponding BMN excitations  is another.
To answer that, we need to understand how the operators act
on large-charge states and compare their supersymmetry
transformations, inner products, and operator products with
those in the BMN theory. Such an analysis should tell us
why the star commutator is the right operation, where the
lower-degree terms in the Bernoulli generators come from,
and whether there is additional operator mixing that our
correlator calculations have not yet detected.

\paragraph{An exact mirror operator dictionary.}
In Section~\ref{sec:necklace-bmn-correlators}, we proposed a
mirror dictionary between bubbling-completed Coulomb-branch
operators in the ADHM theory and Higgs-branch winding operators
in the mirror necklace theory.
We also proposed neutral generators whose star commutators
reproduce the ADHM dressing prescription, including
lower-degree corrections to the classical expressions.
The explicit algebraic and correlator checks support this
proposal, but do not establish the general dictionary:
agreement in selected charged-pair correlators cannot exclude
operator mixing that those pairings do not detect.
A complete derivation should fix the neutral-operator
corrections at arbitrary degree and show that the map preserves
the protected operator products, with the ordering conventions
used on the two sides.
It should also identify the sphere expectation-value
functional, so that the algebraic map reproduces normalized
sphere correlators and fixes the remaining normalization
ambiguities.
Algebraic descriptions of quantized Coulomb
branches~\cite{Kodera:2016faj} and constructions of protected
sphere correlators~\cite{Gaiotto:2020vqj} may provide useful
starting points.
This would place the charged and neutral dictionaries in a
common framework and extend the present tests to general
bubbling-completed operators.
\paragraph{Twisted holography and twisted M-theory.}
There is another way to look at the correspondence we
have found. The topological sector of ADHM theory is
an example of twisted holography~\cite{Gaiotto:2020vqj},
where the bulk description is expected to be given
by twisted M-theory \cite{Costello:2017fbo}, formulated in terms of a
five-dimensional noncommutative Chern--Simons theory.
From this point of view, the matching of topological
correlators between ADHM and BMN may be part of this bigger picture.
This raises some interesting questions. What is the
precise relation between the topological sector of
BMN Matrix Theory and twisted M-theory? What happens
if we take the Penrose limit directly in twisted
M-theory? And can we recover the topological sector
of BMN Matrix Theory from this limit?
Understanding these questions could give us a
holographic explanation for the correspondence in topological subsectors,
rather than just a matching of correlators.
\vspace{15pt}

There are also broader, more general questions concerning our proposed trialities:
\paragraph{Higher-point correlators and interactions.}
A natural next question is whether we can use monopole
correlators to learn something about interactions in Matrix
Theory. Consider, for example, a three-point function
with monopole charges $m_1$, $m_2$, and $-(m_1+m_2)$.
Could such correlators tell us something about the
interaction vertices of Matrix Theory?
To make this connection precise, we first need to
understand how the external states are defined and
normalized on both sides of the correspondence.
Related graviton amplitudes in BFSS Matrix Theory
have been studied in
~\cite{Herderschee:2023pza,Asano:2026udd,Guevara:2026gqw}, while \cite{Kovacs:2015xha} analyzed such a process in the BMN model at large rank.

\paragraph{IIB construction and mABJM.}
For the type IIB version of our triality, we have only matched
the vacuum degeneracy so far.
The next question is whether the excitations around these
vacua also agree. A natural place to start is to pick a
particular mABJM vacuum, study its low-lying excitations,
and identify the supercharge that corresponds to the
protected sector of the parent quiver theory.
It would be especially interesting to compare more refined counting on both sides. 
Another important test is to look at quantities that
know about the geometry of the torus. Using the dictionary
in Eq.~\eqref{eq:IIB-mABJM-dictionary}, we should compare
excitation energies or correlators that depend on its
area and complex structure. Work in this direction is in progress~\cite{Inprogress}. Finally, the bulk interpretation of this correspondence also requires
a direct derivation of the dynamical decoupling of the $Q_{\mathrm{top}}=0$ sector from the rest in the relevant Penrose limit.


\paragraph{Universality and the emergence of Matrix Theory.}
The trialities for ABJM \cite{Komatsu:2026ano} and ADHM suggest a kind of
universality at large charge. Much like large-charge
effective field theory \cite{Hellerman:2015nra}, where very different CFTs can
be described by the same effective theory at large
global charge, our results suggest that different
parent CFTs can give rise to the same Matrix Theory
sector in an appropriate scaling limit. This raises a
natural question: can we derive Matrix Theory directly
from the CFT, rather than simply matching protected
observables? The ultimate goal would be to recover
the BMN Hamiltonian, including its enhanced
supersymmetry, from the triple-scaled theory.

A useful starting point may be the large-angular-momentum
analysis of ABJM theory~\cite{Kovacs:2013una}, although
it remains to be seen how much of that analysis survives
in our limit with fixed monopole charge. The first step
is to identify the excitations with finite $\Delta-J/2$
around a chosen charged background and ask whether their
spectrum, leading interactions, and supersymmetry
transformations agree with those of the corresponding
BMN vacuum. But matching the spectrum is not enough.
We also need to understand why these excitations decouple
dynamically from the neutral sector and, perhaps most
importantly, how the finite-rank matrix degrees of freedom
emerge from the original CFT. That would turn the
correspondence into a genuine derivation of Matrix Theory. Work in this direction is in progress \cite{Inprogress2}.

\paragraph{Toward BFSS Matrix Theory.}
There is one final limit worth considering. BMN Matrix Theory
is a mass-deformed version of BFSS, so we should be able to
recover BFSS by removing the pp-wave mass deformation
~\cite{Banks:1996vh,Susskind:1997cw,Sen:1997we,Seiberg:1997ad}.
To do this, we take $\mu\to0$ while keeping $R_-$,
$\ell_p$, and the matrix rank $m$ fixed.
From Eq.~\eqref{eq:adhm-bmn-coupling-map},
\begin{equation}
    g_{\rm BMN}^2
    =\frac{R_-^3}{\mu^3\ell_p^6}
    =\frac{4\pi^2}{27}\frac{N}{N_f^2},
\end{equation}
we see that this requires
$\lambda_{\rm pp}=N/N_f^2\to\infty$.
In other words, the BFSS limit takes us to strong coupling
of the BMN model.

But there is a second condition if we want to recover the DLCQ of M-theory in flat space: it is not enough to
remove the pp-wave mass deformation. We must also make
sure that  the  compact direction remains lightlike. As can be seen from
Eqs.~\eqref{eq:adhm-penrose-coordinates}
and~\eqref{eq:adhm-canonical-lightcone}, the orbifold
identification also produces a shift, not just in the $x^{-}$ direction, but also in the $x^{+}$ direction:
\begin{equation}
    |\Delta x^+|
    =\frac{6\pi}{\mu N_f}
    \propto
    \frac{\ell_p^2}{R_-}
    \left(\frac{N}{N_f^5}\right)^{1/3}.
\end{equation}
For the compactification to remain null, this shift
must vanish while the $x^-$ period stays fixed at
$2\pi R_-$. This gives the additional condition
$N/N_f^5\to0$.

Putting the two conditions together, we find a scaling
window
\begin{equation}
    N_f^2\ll N\ll N_f^5
    \quad\text{for ADHM},\qquad
    k^2\ll N\ll k^5
    \quad\text{for ABJM}\,,
\end{equation}
in which one expects to recover the DLCQ of flat space M-theory.
 The lower bound
removes the pp-wave mass deformation, while the upper
bound ensures that the compactification remains lightlike.
What remains is perhaps the most interesting question:
what becomes of the states and observables we have
identified when the BMN mass gap disappears?
Understanding how they reorganize into the physics of BFSS is a fascinating future direction, potentially giving us a handle on the non-perturbative physics of M-theory in flat space.
\section*{Acknowledgement} We thank Zhengyuan Du and Shivansh Tomar for discussions and collaboration on related topics. We also thank Elisabetta Armanini, Henry Lin, Shiraz Minwalla, Silviu Pufu, Hidehiko Shimada, and Yifan Wang for discussions. The work of C.P was supported in part by the NSF grant PHY-2609862, the Simons Foundation grant 917464, the Simons Foundation International grant SFI-MPS-QCD-00014867-01 (Simons Collaboration on Confinement and QCD Strings), and by a grant of access to OpenAI models through the ChatGPT for Academic Researchers Program. 

\paragraph{Use of artificial intelligence tools.} The authors used OpenAI's ChatGPT (Astra) at various stages in the preparation of this paper. In particular, ChatGPT assisted with calculations and technical analysis in Sections~\ref{subsec:adhm-bmn-bubbling-scope} and~\ref{sec:necklace-bmn-correlators}. The authors independently reviewed and verified the resulting analysis and take full responsibility for the content and conclusions of the paper.

\appendix
\section{Matching the ADHM and BMN localization supercharges}
\label{app:susy_matching}

In this appendix, we compare the localization supercharges used for
ADHM Coulomb-branch correlators and BMN matrix quantum mechanics.

We first construct the bulk Killing spinor associated with the original
ADHM localization charge of \cite{Dedushenko:2017avn}.
Then, we compare its spinor coefficients with
those of the BMN localization charge in \cite{Asano:2012zt}.

\subsection{Notation and conventions}
\label{app:susy_notation}

We use the following notation throughout this Appendix.

The radius of the round $S^3$ on which the ADHM theory is defined is
denoted by $r$.
The same parameter is used for the radius of the $S^2$ in the canonical
cylinder conformal representative
\begin{equation}
ds_{\mathrm{cyl}}^2
=
-dt^2+r^2d\Omega_2^2
\end{equation}

The $AdS_4$ curvature radius is denoted by $L$
and the $S^7$ radius is
$R_{S^7}=2L$.
We denote dimensionless global $AdS_4$ time by $T$
and physical Lorentzian boundary-cylinder time by
$
t=rT.
$
The polar coordinates on $S^7$ will be denoted by
$
\alpha
,
\beta
,
\vartheta
$
where $\vartheta$ is the coordinate denoted by $\gamma$ in
\cite{Nishioka:2008ib}. 
The four phases of the complex coordinates on $S^7$ are
$
\xi_1
,
\xi_2
,
\xi_3
,
\xi_4.
$
The phase of the particular $z_1$ great circle used for the Penrose limit
will be denoted by $\psi$, so that on the geodesic
\begin{equation}
\xi_1=\psi
\end{equation}
Also, $\tau_{\mathrm A}$ denotes the Euclidean BMN time used in the
Asano localization calculation, while $t_{\mathrm A}$ denotes the
corresponding Lorentzian time.

We use $\gamma_A$, $A=0,\ldots,10$, for the eleven-dimensional
tangent-space gamma matrices in the conventions of
\cite{Nishioka:2008ib},
\begin{equation}
\left\{
\gamma_A,\gamma_B
\right\}
=
2\eta_{AB}
\end{equation}
with mostly-plus signature. We define
\begin{equation}
\widehat\gamma
=
\gamma_{0123}
\end{equation}
The indices $0,1,2,3$ correspond respectively to global $AdS_4$ time,
the radial direction, and the two angular directions on the boundary $S^2$.

\subsection{The ADHM localization spinor in Dedushenko-Fan-Pufu-Yacoby}
\label{app:DFPY_spinor}

The three-dimensional $\mathcal N=4$ superconformal algebra contains
Poincar\'e and conformal supercharges
$
Q_{\alpha a\dot a},S_{\alpha a\dot a}
$
where
$
\alpha=1,2
$
is a three-dimensional spinor index and
$
a,\dot a=1,2
$
are doublet indices of $SU(2)_H$ and $SU(2)_C$.

The Coulomb-branch localization supercharge used by Dedushenko, Fan, Pufu and Yacoby  \cite{Dedushenko:2017avn} is
\begin{equation}
\mathcal Q_\beta^C
=
\mathcal Q_1^C
+
\beta\mathcal Q_2^C
\end{equation}
and the localization calculation uses
$ \beta=1$.

In flat $\mathbb R^3$, the corresponding conformal Killing spinor has the
form
\begin{equation}
\xi_{\alpha a\dot a}(x)
=
q_{\alpha a\dot a}
+
\frac{x^i}{2r}
(\gamma_i^{(3)})_\alpha{}^\beta
s_{\beta a\dot a}
\end{equation}
where
$
s_{\alpha a\dot a}
=
2r\,\eta_{\alpha a\dot a}
$
in terms of the conformal-supersymmetry parameter $\eta$ of \cite{Dedushenko:2017avn}.

It is important to distinguish the lower-index Killing-spinor components
from the upper-index coefficients multiplying the generators in the
supercharge. Extracting the lower-index components directly from the
explicit ADHM Killing spinor at $\beta=1$ gives
\begin{equation}
q_{1\,a\dot a}
=
\begin{pmatrix}
1 & -1
\\
-i & i
\end{pmatrix}
,\quad
q_{2\,a\dot a}
=
\begin{pmatrix}
1 & -1
\\
i & -i
\end{pmatrix}.
\label{eq:DFPY_q}
\end{equation}

Using the standard Pauli-matrix convention for the flat-space gamma
matrices,
\begin{equation}
\gamma_i^{(3)}
=
\sigma_i
\end{equation}
one similarly obtains
\begin{equation}
s_{1\,a\dot a}
=
\begin{pmatrix}
-i & -i
\\
-1 & -1
\end{pmatrix}
,\quad
s_{2\,a\dot a}
=
\begin{pmatrix}
i & i
\\
-1 & -1
\end{pmatrix}.
\label{eq:DFPY_s}
\end{equation}
The components which will be useful later are related as follows: 
\begin{equation}
s_{\alpha a}^{(\dot 1)}
=
\left(i\sigma_3\right)_{\alpha}{}^{\beta}\,
q_{\beta a}^{(\dot 2)} .
\label{eq:DFPY_q_s_relation}
\end{equation}


\paragraph{Conformal map to the cylinder}

Write flat $\mathbb R^3$ as
\begin{equation}
x^i
=
2r e^u n^i
\end{equation}
where
$n^in^i=1.$
Then
\begin{equation}
ds_{\mathbb R^3}^2
=
4r^2e^{2u}
\left(
du^2+d\Omega_2^2
\right).
\end{equation}

After the Weyl transformation to the cylinder, and dropping a common
position-independent normalization, the spinor becomes
\begin{equation}
\xi_{\mathrm{cyl}}^{(E)}
=
e^{-u/2}q
+
e^{+u/2}
n^i\gamma_i^{(3)}s.
\label{eq:DFPY_cylinder_E}
\end{equation}
Continuing according to
$
u=iT
$
gives the Lorentzian cylinder spinor
\begin{equation}
\xi_{\mathrm{cyl}}
=
e^{-iT/2}q
+
e^{+iT/2}
n^i\gamma_i^{(3)}s.
\label{eq:DFPY_cylinder_L}
\end{equation}

\subsection{The bulk dual to the ADHM spinor }
\label{app:full_bulk_spinor}
We now study the eleven-dimensional bulk Killing spinors and take their $AdS_4$ boundary limit to identify the one that induces the boundary conformal Killing spinor associated with the ADHM localization supercharge.
We use the conventions and explicit Killing spinor of
\cite{Nishioka:2008ib}.

After setting their $S^7$ radius $R=2L$ and introducing
$r=\sinh\rho$, the metric is
\begin{equation}
ds_{11}^2
=
L^2
\left[
-\cosh^2\rho\,dT^2
+d\rho^2
+\sinh^2\rho
\left(
d\theta^2
+
\sin^2\theta\,d\phi^2
\right)
\right]
+
4L^2d\Omega_7^2.
\label{eq:bulk_metric}
\end{equation}
The corresponding $AdS_4$ orthonormal frame is
\begin{equation}\label{adscframe}
\begin{split}
&e^0= L\cosh\rho\,dT,~e^1=L\,d\rho,\\
&e^2=L\sinh\rho\,d\theta,~ e^3=
L\sinh\rho\sin\theta\,d\phi.
\end{split}
\end{equation}
The unit $S^7$ is embedded in $\mathbb C^4$ as
$
\sum_{I=1}^4|z_I|^2=1
$
with
$
z_I
=
\mu_Ie^{i\xi_I}
$
and
\begin{equation}\label{eq:S7_mu}
\begin{split}
\mu_1&=
\sin\alpha, ~\mu_2
=\cos\alpha\sin\beta,\\
\mu_3
&=\cos\alpha\cos\beta\sin\vartheta,~\mu_4=\cos\alpha\cos\beta\cos\vartheta.
\end{split}
\end{equation}

The full eleven-dimensional bulk Killing spinor is
\begin{equation}\label{eq:NT_full_spinor}
\epsilon
=
\mathcal U_{S^7}
\mathcal U_{AdS}
\epsilon_0
\end{equation}
where $\epsilon_0$ is a constant eleven-dimensional spinor. $\mathcal U_{S^7}$ and $\mathcal U_{AdS}$ are defined as \footnote{One can easily check that $[\mathcal U_{S^7},
\mathcal U_{AdS}]=0$.}
\begin{equation}\label{eq:US7}
\begin{split}
\mathcal U_{S^7}
&=
e^{\frac{\alpha}{2}\widehat\gamma\gamma_4}
e^{\frac{\beta}{2}\widehat\gamma\gamma_5}
e^{\frac{\vartheta}{2}\widehat\gamma\gamma_6}
e^{\frac{\xi_1}{2}\gamma_{47}}
e^{\frac{\xi_2}{2}\gamma_{58}}
e^{\frac{\xi_3}{2}\gamma_{69}}
e^{\frac{\xi_4}{2}\widehat\gamma\gamma_{10}},\\
\mathcal U_{AdS}
&=
e^{-\frac{\rho}{2}\gamma_1\widehat\gamma}
e^{-\frac{T}{2}\gamma_0\widehat\gamma}
e^{\frac{\theta}{2}\gamma_{12}}
e^{\frac{\phi}{2}\gamma_{23}}.
\end{split}
\end{equation}

\subsubsection{The \texorpdfstring{$AdS_4$}{AdS4} boundary limit}
\label{app:AdS_boundary_limit}

We now analyze the large-$\rho$ behavior of
\eqref{eq:NT_full_spinor}. The $S^7$ factor is finite as
$\rho\rightarrow\infty$ and is therefore a spectator in the radial
asymptotics.

Define
\begin{equation}
K
=
-\gamma_1\widehat\gamma, \qquad
J
=
-\gamma_0\widehat\gamma.
\end{equation}
Then
\begin{equation}
K^2=1, ~ J^2=-1, \qquad
\left\{
K,J
\right\}
=
0.
\end{equation}

The $AdS_4$ part of the Killing spinor becomes
\begin{equation}
\mathcal U_{AdS}
=
e^{\frac{\rho}{2}K}
e^{\frac{T}{2}J}
R(\theta,\phi),
\end{equation}
where
\begin{equation}
R(\theta,\phi)
=
e^{\frac{\theta}{2}\gamma_{12}}
e^{\frac{\phi}{2}\gamma_{23}}.
\label{eq:AdS_angular_R}
\end{equation}

\paragraph{An explicit gamma-matrix basis}

To make the projection explicit, we use the following representation of
the four-dimensional subalgebra generated by
$\gamma_0,\gamma_1,\gamma_2,\gamma_3$:
\begin{equation}\label{eq:AdS_gamma_basis}
\begin{split}
\gamma_0
&=
i\sigma_2\otimes\mathbf1,~ \gamma_1
=
\sigma_3\otimes\mathbf1,\\
\gamma_2
&=
\sigma_1\otimes\sigma_1,~\gamma_3
=
\sigma_1\otimes\sigma_2.
\end{split}
\end{equation}
The identity in the internal spinor space is suppressed.

In this basis,
\begin{equation}
\widehat\gamma
=
-i\sigma_1\otimes\sigma_3,
\end{equation}
and hence
\begin{equation}\label{KJexpl}
K
=
-\sigma_2\otimes\sigma_3,~J=
i\sigma_3\otimes\sigma_3.
\end{equation}
Furthermore,
\begin{equation}
\begin{split}
\gamma_{12}
=
i\sigma_2\otimes\sigma_1,~
\gamma_{23}
=
i\mathbf1\otimes\sigma_3.
\end{split}
\end{equation}

Let $|\pm\rangle$ denote eigenvectors of $\sigma_3$. We order the
four-component basis as
\begin{equation}
\left(
|+-\rangle
\quad
|-+\rangle
\quad
|++\rangle
\quad
|--\rangle
\right).
\label{eq:J_basis}
\end{equation}
Then
\begin{equation}
J
=
\begin{pmatrix}
-i\mathbf1_2 & 0
\\
0 & +i\mathbf1_2
\end{pmatrix}.
\label{eq:J_block}
\end{equation}
We write the two $J$ eigenspaces as \footnote{Remember that $q_0$ and $s_0$ are themselves two component spinors chosen as linear combination of first and second pair of vectors in \eqref{eq:J_basis}.}
\begin{equation}
\begin{split}
\epsilon_0^Q
&=
\begin{pmatrix}
q_0
\\
0
\end{pmatrix},~
\epsilon_0^S
=
\begin{pmatrix}
0
\\
s_0
\end{pmatrix},
\end{split}
\end{equation}
with
\begin{equation}
\begin{split}
J\epsilon_0^Q
&=
-i\epsilon_0^Q,~
J\epsilon_0^S
=
+i\epsilon_0^S.
\end{split}
\end{equation}
These are the bulk precursors of the boundary Poincar\'e and conformal
supersymmetry parameters.

Because
$
[J,R]=0
$
the angular rotation is block diagonal in the same basis,
\begin{equation}
R
=
\begin{pmatrix}
R_- & 0
\\
0 & R_+
\end{pmatrix},
\end{equation}
with
\begin{equation}
R_-
=
e^{+\frac{i\theta}{2}\sigma_2}
e^{-\frac{i\phi}{2}\sigma_3},
\label{eq:Rminus}
\end{equation}
and
\begin{equation}
R_+
=
e^{+\frac{i\theta}{2}\sigma_2}
e^{+\frac{i\phi}{2}\sigma_3}.
\label{eq:Rplus}
\end{equation}
This can be explicitly seen by using \eqref{eq:AdS_angular_R}. Here $\sigma$ matrices act on the two components of $q_0$ and $s_0$.

\paragraph{The radial projection}

Define
\begin{equation}
P_\pm
=
\frac12
\left(
1\pm K
\right).
\end{equation}
Then
\begin{equation}
e^{\frac{\rho}{2}K}
=
e^{\rho/2}P_+
+
e^{-\rho/2}P_-.
\end{equation}
Therefore the leading large-$\rho$ term of the full Killing spinor is
\begin{equation}
\epsilon
=
e^{\rho/2}
\mathcal U_{S^7}
P_+
e^{\frac{T}{2}J}
R
\epsilon_0
+
O(e^{-\rho/2}).
\label{eq:large_rho_bulk}
\end{equation}

In the $J$-adapted basis,
\begin{equation}
K
=
\begin{pmatrix}
0 & -i\sigma_1
\\
+i\sigma_1 & 0
\end{pmatrix},
\end{equation}
and hence
\begin{equation}
P_+
=
\frac12
\begin{pmatrix}
\mathbf1_2 & -i\sigma_1
\\
+i\sigma_1 & \mathbf1_2
\end{pmatrix}.
\label{eq:Pplus}
\end{equation}

Acting on the $Q$ sector gives
\begin{equation}
P_+R\epsilon_0^Q
=
\frac12
\begin{pmatrix}
R_-q_0
\\
i\sigma_1R_-q_0
\end{pmatrix}.
\label{eq:PQ}
\end{equation}
Acting on the $S$ sector gives
\begin{equation}
P_+R\epsilon_0^S
=
\frac12
\begin{pmatrix}
-i\sigma_1R_+s_0
\\
R_+s_0
\end{pmatrix}.
\label{eq:PS}
\end{equation}

\paragraph{The projected four-spinor as a boundary two-spinor}

Equations~\eqref{eq:PQ} and~\eqref{eq:PS} show that any spinor in $\operatorname{Im}P_+$ can be written as
\begin{equation}
\Psi_+(\lambda)
=
\frac{1}{\sqrt2}
\begin{pmatrix}
\lambda
\\
i\sigma_1\lambda
\end{pmatrix}
\label{eq:Pplus_embedding}
\end{equation}
where $\lambda$ is an arbitrary two-component spinor.

We can associate this two component spinor $\lambda$ with the corresponding boundary spatial $SU(2)$ spinor. 
This identification is not merely based on counting components. The
tangent gamma matrices preserve $\operatorname{Im}P_+$. Indeed,
\begin{equation}
[K,\gamma_a]=0
\qquad
a=0,2,3
\end{equation}

We therefore define induced two-dimensional matrices
$\overline\gamma_a$ by
\begin{equation}
\gamma_a\iota(\lambda)
=
\iota
\left(
\overline\gamma_a\lambda
\right),
\end{equation}
where $\iota$ is a map that takes us from $\iota:\Psi_+\to\lambda$.
A direct calculation gives
\begin{equation}
\overline\gamma_0
=
i\sigma_3, ~\overline\gamma_2
=
\sigma_1,~
\overline\gamma_3
=
-\sigma_2.
\label{eq:induced_boundary_gamma}
\end{equation}
They obey
\begin{equation}
\left\{
\overline\gamma_a,
\overline\gamma_b
\right\}
=
2\eta_{ab}.
\end{equation}
Therefore, $\operatorname{Im}P_+$ is precisely the irreducible two-component spinor representation of $Spin(1,2)$.

After Wick rotation it becomes the fundamental $\mathbf2$
of $Spin(3)\simeq SU(2)$ which is precisely the representation carried by the ADHM spinor index
$\alpha$.
Using \eqref{eq:PQ} and \eqref{eq:PS}, the independent two-component
boundary data in the spherical frame are therefore
\begin{equation}
\xi_{\mathrm{sph}}
=
e^{-iT/2}
R_-q_0
+
e^{+iT/2}
\left(
-i\sigma_1R_+s_0.
\right)
\label{eq:boundary_spherical_spinor}
\end{equation}

\subsubsection{Fixing the bulk spinor dual to the boundary localization supercharge}
\label{app:fixing_bulk_spinor}

We now determine the bulk Killing spinor corresponding to the boundary
localization supercharge. The boundary spinor obtained from the bulk is
written in a spherical spin frame, whereas the ADHM spinor is written in
a conformally Cartesian frame. We first find the local $SO(3)$ transformation that relates the spherical frame gamma matrices and the Cartesian frame gamma matrices used in ADHM, then apply the resulting spin transformation to compare the
two spinors.

After Wick rotation, $u=iT$, the Euclidean boundary metric is
\begin{equation}
ds_{\partial,E}^2
=
r^2\left(du^2+d\theta^2+\sin^2\theta\,d\phi^2\right),
\end{equation}
with orthonormal coframe
\begin{equation}
e^{\hat u}=r\,du,
\qquad
e^{\hat\theta}=r\,d\theta,
\qquad
e^{\hat\phi}=r\sin\theta\,d\phi.
\end{equation}
The boundary gamma matrices inherited from the bulk are
\begin{equation}
\bar\gamma_{\hat u}=\sigma_3,
\qquad
\bar\gamma_{\hat\theta}=\sigma_1,
\qquad
\bar\gamma_{\hat\phi}=-\sigma_2.
\end{equation}
The corresponding boundary spinor is
\begin{equation}
\xi_{\rm sph}^{(E)}
=
e^{-u/2}R_-(\theta,\phi)q_0
+
e^{u/2}\left[-i\sigma_1R_+(\theta,\phi)s_0\right],
\label{eq:bulk_boundary_spherical}
\end{equation}
where
\begin{equation}
R_-
=
e^{+\frac{i\theta}{2}\sigma_2}
e^{-\frac{i\phi}{2}\sigma_3},
\qquad
R_+
=
e^{+\frac{i\theta}{2}\sigma_2}
e^{+\frac{i\phi}{2}\sigma_3}.
\end{equation}

In \cite{Dedushenko:2016jxl}, the standard Pauli matrices are used as the local-frame gamma
matrices. To compare with this convention, we define the ordering of the bulk chosen Vielbein in similar manner, so that they can be related by an $SO(3)$ transformation. So vielbein are written in terms of $f^i$ defined below: 

\begin{equation}
\bar\gamma_{\hat u}e^{\hat u}
+
\bar\gamma_{\hat\theta}e^{\hat\theta}
+
\bar\gamma_{\hat\phi}e^{\hat\phi}
=
\sigma_i f^i,
\qquad
(f^1,f^2,f^3)
=
(e^{\hat\theta},-e^{\hat\phi},e^{\hat u}).
\end{equation}

We identify the ADHM Cartesian coordinates with the bulk boundary
coordinates through
\begin{equation}
x^i=2re^u n^i(\theta,\phi),
\qquad
n^i=
\left(
\sin\theta\cos\phi,
-\sin\theta\sin\phi,
\cos\theta
\right).
\label{eq:boundary_n_vector}
\end{equation}
This choice fixes the relative orientation of the bulk angular
coordinates and the Cartesian axes. The conformally Cartesian frame on the cylinder is
\begin{equation}
E^i
=
\frac{e^{-u}}{2}\,dx^i
=
n^i e^{\hat u}
+
n_\theta^i e^{\hat\theta}
+
n_\phi^i e^{\hat\phi},
\end{equation}
where
\begin{equation}
n_\theta^i=\partial_\theta n^i,
\qquad
n_\phi^i=\frac{1}{\sin\theta}\partial_\phi n^i.
\end{equation}
In terms of the frame $f^i$, this becomes
\begin{equation}
E^i=\Lambda^i{}_j f^j,
\qquad
\Lambda=(n_\theta,-n_\phi,n),
\qquad
\Lambda^T\Lambda=\mathbf{1},
\qquad
\det\Lambda=+1.
\end{equation}
Thus $f^i$ and $E^i$ are related by an $SO(3)$ rotation.

Its spin lift is
\begin{equation}
\mathcal S(\theta,\phi)
=
e^{+\frac{i\phi}{2}\sigma_3}
e^{-\frac{i\theta}{2}\sigma_2},
\qquad
\mathcal S\sigma_j\mathcal S^{-1}
=
\Lambda^i{}_j\sigma_i.
\label{eq:boundary_spin_rotation}
\end{equation}
Equivalently,
\begin{equation}
\begin{aligned}
\mathcal S\bar\gamma_{\hat u}\mathcal S^{-1}
&=n^i\sigma_i,
\\
\mathcal S\bar\gamma_{\hat\theta}\mathcal S^{-1}
&=n_\theta^i\sigma_i,
\\
\mathcal S\bar\gamma_{\hat\phi}\mathcal S^{-1}
&=n_\phi^i\sigma_i.
\end{aligned}
\end{equation}
These relations ensure that the transformed bulk gamma matrices
agree with the ADHM gamma matrices in the Cartesian coframe.

The spinor components therefore transform as
\begin{equation}
\xi_{\rm cart}^{(E)}
=
\mathcal S\,\xi_{\rm sph}^{(E)}.
\end{equation}
Using
\begin{equation}
\mathcal S R_-=\mathbf{1},
\qquad
\mathcal S(-i\sigma_1R_+)
=
(n^i\sigma_i)\sigma_2,
\label{eq:spin_rotation_useful}
\end{equation}
we obtain
\begin{equation}
\xi_{\rm cart}^{(E)}
=
e^{-u/2}q_0
+
e^{u/2}(n^i\sigma_i)\sigma_2s_0.
\label{eq:bulk_boundary_cartesian}
\end{equation}

The ADHM localization spinor, after Weyl rescaling to the same cylinder,
is
\begin{equation}
\xi_{\rm ADHM}^{(E)}
=
e^{-u/2}q
+
e^{u/2}(n^i\sigma_i)s.
\label{eq:DFPY_boundary_spinor_matching}
\end{equation}
Both spinors are now expressed in the same frame and gamma-matrix
convention. Comparing their two independent $u$ dependences gives
\begin{equation}
q_0=q,
\qquad
s_0=\sigma_2s.
\end{equation}
The scalar Weyl rescaling back to the ADHM stereographic frame
multiplies both spinors by the same factor and does not affect this
identification.

Thus, in the $J$-adapted basis used above, the constant bulk spinor
corresponding to the original ADHM localization supercharge is
\begin{equation}
\epsilon_{0,\rm ADHM}
=
\begin{pmatrix}
q\\[1mm]
\sigma_2s
\end{pmatrix}.
\label{eq:epsilon0_DFPY}
\end{equation}
Restoring the internal indices,
\begin{equation}
\left(\epsilon_{0,\rm ADHM}\right)_{a\dot a}
=
\begin{pmatrix}
q_{\alpha a\dot a}\\[1mm]
(\sigma_2)_\alpha{}^\beta s_{\beta a\dot a}
\end{pmatrix}.
\label{eq:epsilon0_DFPY_indices}
\end{equation}
The corresponding bulk Killing spinor is therefore
\begin{equation}
\epsilon_{\rm ADHM}^{\rm bulk}(x,y)
=
\mathcal U_{S^7}(y)\,
\mathcal U_{AdS}(x)\,
\epsilon_{0,\rm ADHM}.
\label{eq:fixed_bulk_DFPY_spinor}
\end{equation}

\subsubsection{The ADHM orbifold acting on the \texorpdfstring{$S^7$}{S7} spinor}
\label{app:ADHM_orbifold}

We now impose the ADHM orbifold condition on the $S^7$ part of
the full Killing spinor \eqref{eq:NT_full_spinor}.
In subsection~\ref{app:Penrose_spinor}, we use these constant spinor
data to construct the aligned localization charge and take its
Penrose limit.

The four mutually commuting matrices appearing with the four complex-plane
phases are
\begin{equation}
i\gamma_{47}
\qquad
i\gamma_{58}
\qquad
i\gamma_{69}
\qquad
i\widehat\gamma\gamma_{10}
\end{equation}

Following the conventions of \cite{Nishioka:2008ib}, choose a
constant spinor with definite eigenvalues
\begin{equation}
i\gamma_{47}\epsilon_0
=
\varsigma_1\epsilon_0,~i\gamma_{58}\epsilon_0
=
\varsigma_2\epsilon_0,~
i\gamma_{69}\epsilon_0
=
\varsigma_3\epsilon_0,~
i\widehat\gamma\gamma_{10}\epsilon_0
=
\varsigma_4\epsilon_0,
\label{eq:internal_weights}
\end{equation}
where
\begin{equation}
\varsigma_I=\pm1, \quad I\in(1,2,3,4).
\end{equation}
The eleven-dimensional Clifford-algebra condition implies
\begin{equation}
\varsigma_1
\varsigma_2
\varsigma_3
\varsigma_4
=
1.
\label{eq:internal_chirality}
\end{equation}
Thus, the $S^7$ Killing spinors carry the eight weights of one chiral spinor representation of $Spin(8)$.

\paragraph{The ADHM quotient}

The ADHM orbifold acts on the first $\mathbb C^2$ according to
\begin{equation}
(z_1,z_2,z_3,z_4)
\longrightarrow
\left(
\omega z_1,
\omega^{-1}z_2,
z_3,
z_4
\right)
\label{eq:ADHM_orbifold_action}
\end{equation}
where
\begin{equation}
\omega
=
e^{2\pi i/k}
\end{equation}

Since
\begin{equation}
z_I=\mu_Ie^{i\xi_I}
\end{equation}
this means
\begin{equation}
\left(\xi_1
\longrightarrow
\xi_1+\frac{2\pi}{k}\right),~ 
\left(\xi_2
\longrightarrow
\xi_2-\frac{2\pi}{k}\right)
\end{equation}
while $\xi_3$ and $\xi_4$ are unchanged.

From the explicit Killing spinor \eqref{eq:NT_full_spinor}, a weight
eigenstate therefore transforms as
\begin{equation}
\epsilon_0
\longrightarrow
\exp
\left[
\frac{i\pi}{k}
\left(
\varsigma_1-\varsigma_2
\right)
\right]
\epsilon_0
\end{equation}

For the standard spin lift and $k>2$, orbifold invariance requires
\begin{equation}
\varsigma_1
=
\varsigma_2
\label{eq:orbifold_condition}
\end{equation}

Combining \eqref{eq:orbifold_condition} with
\eqref{eq:internal_chirality} gives
\begin{equation}
\varsigma_3
=
\varsigma_4
\end{equation}

The four surviving internal weight assignments are consequently
\begin{equation}
\begin{split}
&(+,+,+,+),~
(+,+,-,-)\\
&(-,-,+,+),~
(-,-,-,-)
\end{split}
\end{equation}

These are the four internal Killing spinors corresponding to the sixteen
supercharges of the boundary $OSp(4|4)$ algebra.

We denote the four surviving internal states by
\begin{equation}
\zeta_{a\dot a}
\end{equation}
where $a$ and $\dot a$ transform as doublets of $SU(2)_H$ and $SU(2)_C$.

\subsection{Penrose limit of the bulk Killing spinor}
\label{app:Penrose_spinor}

We first evaluate the bulk spinor associated with the original ADHM
localization supercharge on the reference null geodesic. This identifies
its kinematical and dynamical components. We then choose an
$SU(2)_C$-related localization supercharge whose independent
kinematical component vanishes. All operator insertions and monopole
boundary data are transformed by the same symmetry.

The reference geodesic lies at the center of $AdS_4$ and moves along
the $z_1$ circle of $S^7$:
\begin{equation}
\rho=0,
\qquad
z_1=e^{i\psi},
\qquad
z_2=z_3=z_4=0.
\label{eq:Penrose_geodesic}
\end{equation}
In the angular coordinates used above, this means
$\alpha=\pi/2$ and $\xi_1=\psi$.

\paragraph{A regular frame at the center of $AdS_4$.}

The spatial coframe in \eqref{adscframe} is
\begin{equation}
e^1=L\,d\rho,
\qquad
e^2=L\sinh\rho\,d\theta,
\qquad
e^3=L\sinh\rho\sin\theta\,d\phi.
\end{equation}
This polar frame is singular at $\rho=0$. To evaluate the spinor at
the center, introduce
\begin{equation}
n_{\rm AdS}^i
=\left(\cos\theta,\sin\theta\cos\phi,\sin\theta\sin\phi\right),
\qquad
X^i=L\rho\,n_{\rm AdS}^i,
\end{equation}
and define
\begin{equation}
n_{{\rm AdS},\theta}^i=\partial_\theta n_{\rm AdS}^i,
\qquad
n_{{\rm AdS},\phi}^i
=\frac{1}{\sin\theta}\partial_\phi n_{\rm AdS}^i.
\end{equation}
The rotated orthonormal coframe is
\begin{equation}
E^i
=n_{\rm AdS}^i e^1
+n_{{\rm AdS},\theta}^i e^2
+n_{{\rm AdS},\phi}^i e^3
=\left(\delta^i{}_j+O(\rho^2)\right)dX^j.
\end{equation}
The last equality follows from $\sinh\rho/\rho=1+O(\rho^2)$.
Thus $E^i$ extends smoothly to the center. The two coframes obey
\begin{equation}
E^i=\Lambda_{\rm AdS}^i{}_j e^j,
\qquad
\Lambda_{\rm AdS}
=\left(n_{\rm AdS},n_{{\rm AdS},\theta},n_{{\rm AdS},\phi}\right)
\in SO(3).
\end{equation}

The angular factor in the bulk Killing spinor is
\begin{equation}
\mathcal R(\theta,\phi)
=e^{\frac{\theta}{2}\gamma_{12}}
 e^{\frac{\phi}{2}\gamma_{23}}.
\end{equation}
The spin lift of the coframe rotation is
\begin{equation}
\mathcal S_{\rm AdS}
=\mathcal R^{-1}
=e^{-\frac{\phi}{2}\gamma_{23}}
 e^{-\frac{\theta}{2}\gamma_{12}},
\qquad
\mathcal S_{\rm AdS}\gamma_j\mathcal S_{\rm AdS}^{-1}
=\Lambda_{\rm AdS}^i{}_j\gamma_i.
\end{equation}
In particular,
$\mathcal S_{\rm AdS}\gamma_1\mathcal S_{\rm AdS}^{-1}
=n_{\rm AdS}^i\gamma_i$.
The components in the regular frame are therefore
$\epsilon_F=\mathcal S_{\rm AdS}\epsilon_{\rm polar}$.
Applying this transformation to \eqref{eq:fixed_bulk_DFPY_spinor} gives
\begin{equation}
\epsilon_F
=\mathcal U_{S^7}
 e^{-\frac{\rho}{2}n_{\rm AdS}^i\gamma_i\widehat\gamma}
 e^{-\frac{T}{2}\gamma_0\widehat\gamma}
 \epsilon_{0,\rm ADHM}.
\label{eq:regular_bulk_spinor}
\end{equation}
Since $\rho n_{\rm AdS}^i=X^i/L$, this expression is smooth at the center.

\paragraph{The original spinor on the reference geodesic.}

On \eqref{eq:Penrose_geodesic}, the spinor becomes
\begin{equation}
\epsilon_\gamma(T,\psi)
=C\,e^{\frac{\psi}{2}\gamma_{47}}
 e^{\frac{T}{2}J}\epsilon_{0,\rm ADHM},
\qquad
J=-\gamma_0\widehat\gamma,
\qquad
C=e^{\frac{\pi}{4}\widehat\gamma\gamma_4}.
\label{eq:spinor_on_geodesic}
\end{equation}
At these transverse coordinates, the metric is
\begin{equation}
ds^2=-L^2dT^2+4L^2d\psi^2.
\end{equation}
Introduce light-cone coordinates
\begin{equation}
x^+=\frac12(T+2\psi),
\qquad
x^-=L^2(T-2\psi),
\end{equation}
for which $ds^2=-2dx^+dx^-$ on this two-dimensional subspace.
The reference geodesic has $x^-=0$, so
\begin{equation}
T=x^+,
\qquad
\psi=\frac{x^+}{2}.
\label{eq:Tpsi_xplus}
\end{equation}
This choice of $x^+$ gives $\mu=3$ in the standard pp-wave
normalization. In \eqref{eq:adhm-ppwave-intermediate}, it corresponds
to $U=x^+$ and $2V=x^-$.

Decompose the original constant spinor into its $Q$ and $S$ sectors
and its two internal $SU(2)_C$ components:
\begin{equation}
\epsilon_{0,\rm ADHM}
=\epsilon^0_{Q,\dot1}+\epsilon^0_{Q,\dot2}
 +\epsilon^0_{S,\dot1}+\epsilon^0_{S,\dot2}.
\end{equation}
The $Q$ and $S$ sectors have
\begin{equation}
J\epsilon^0_{Q,\dot a}=-i\epsilon^0_{Q,\dot a},
\qquad
J\epsilon^0_{S,\dot a}=+i\epsilon^0_{S,\dot a}.
\end{equation}
Within the orbifold-invariant internal spinor space, our dotted-index
convention is
\begin{equation}
\gamma_{47}\epsilon^0_{\dot1}=-i\epsilon^0_{\dot1},
\qquad
\gamma_{47}\epsilon^0_{\dot2}=+i\epsilon^0_{\dot2}.
\end{equation}
It follows that
\begin{align}
\epsilon_\gamma(x^+)
=C\Big(
&e^{-3ix^+/4}\epsilon^0_{Q,\dot1}
+e^{-ix^+/4}\epsilon^0_{Q,\dot2}
\nonumber\\
&+e^{+ix^+/4}\epsilon^0_{S,\dot1}
+e^{+3ix^+/4}\epsilon^0_{S,\dot2}
\Big).
\label{eq:four_penrose_frequencies}
\end{align}

\paragraph{The large Lorentz boost.}

The Penrose limit requires a boost to a finite null frame.
We use the same spinor-component convention as for the spatial rotation:
\begin{equation}
E^a=\Lambda^a{}_b e^b,
\qquad
S\gamma_bS^{-1}=\Lambda^a{}_b\gamma_a,
\qquad
\epsilon_E=S\epsilon_e.
\end{equation}
The Lorentzian gamma matrices have lower frame indices, with
$\gamma_0^2=-1$.
On the reference geodesic, the original coframe is
\begin{equation}
e^0=L\,dT,
\qquad
e^7=2L\,d\psi.
\end{equation}
The inverse coordinate transformation is
\begin{equation}
T=x^++\frac{x^-}{2L^2},
\qquad
\psi=\frac{x^+}{2}-\frac{x^-}{4L^2}.
\end{equation}
Hence
\begin{equation}
e^0=L\,dx^++\frac{dx^-}{2L},
\qquad
e^7=L\,dx^+-\frac{dx^-}{2L},
\end{equation}
and the null coframe satisfies
\begin{equation}
e^\pm=\frac{e^0\pm e^7}{\sqrt2},
\qquad
e^+=\sqrt2L\,dx^+,
\qquad
e^-=\frac{dx^-}{\sqrt2L}.
\end{equation}
For $L=\Omega L_0$ and $\Omega\to\infty$, with $L_0$ fixed, a finite
coframe is
\begin{equation}
E^+=\Omega^{-1}e^+,
\qquad
E^-=\Omega e^-.
\end{equation}

Define
\begin{equation}
\mathcal A=\gamma_0\gamma_7,
\qquad
\Pi_\pm=\frac12(1\pm\mathcal A),
\qquad
\mathcal A^2=1.
\label{eq:lightcone_projectors}
\end{equation}
For $\gamma_\pm=(\gamma_0\pm\gamma_7)/\sqrt2$, one has
\begin{equation}
[\mathcal A,\gamma_+]=2\gamma_+,
\qquad
[\mathcal A,\gamma_-]=-2\gamma_-.
\end{equation}
The spin lift of the boost is therefore
\begin{equation}
S_\Omega
=e^{-\frac12(\log\Omega)\mathcal A}
=\Omega^{-1/2}\Pi_++\Omega^{1/2}\Pi_-.
\label{eq:large_boost}
\end{equation}
Indeed,
\begin{equation}
S_\Omega\gamma_+S_\Omega^{-1}=\Omega^{-1}\gamma_+,
\qquad
S_\Omega\gamma_-S_\Omega^{-1}=\Omega\gamma_-.
\label{eq:boost_null_gamma}
\end{equation}
Thus
\begin{equation}
\epsilon_\Omega
=S_\Omega\epsilon_\gamma
=\Omega^{-1/2}\Pi_+\epsilon_\gamma
+\Omega^{1/2}\Pi_-\epsilon_\gamma.
\end{equation}

\paragraph{Kinematical and dynamical components.}

The identity
\begin{equation}
C^{-1}\mathcal A C=J\gamma_{47}
\label{eq:A_conjugation}
\end{equation}
determines the two projections. Its eigenvalues are
\begin{align}
J\gamma_{47}\epsilon^0_{Q,\dot2}&=+\epsilon^0_{Q,\dot2},
&J\gamma_{47}\epsilon^0_{S,\dot1}&=+\epsilon^0_{S,\dot1},\\
J\gamma_{47}\epsilon^0_{Q,\dot1}&=-\epsilon^0_{Q,\dot1},
&J\gamma_{47}\epsilon^0_{S,\dot2}&=-\epsilon^0_{S,\dot2}.
\end{align}
Consequently,
\begin{align}
\epsilon_{\rm dyn}(x^+)
&\equiv\Pi_+\epsilon_\gamma(x^+)
=C\left(e^{-ix^+/4}\epsilon^0_{Q,\dot2}
+e^{+ix^+/4}\epsilon^0_{S,\dot1}\right),\\
\epsilon_{\rm kin}(x^+)
&\equiv\Pi_-\epsilon_\gamma(x^+)
=C\left(e^{-3ix^+/4}\epsilon^0_{Q,\dot1}
+e^{+3ix^+/4}\epsilon^0_{S,\dot2}\right).
\end{align}
The first pair has frequency $\mu/12=1/4$ and belongs to the
dynamical sector. The second pair has frequency $\mu/4=3/4$ and
belongs to the kinematical sector.
For the original ADHM charge, both projections are nonzero:
\begin{equation}
S_\Omega\epsilon_\gamma
=\Omega^{1/2}\epsilon_{\rm kin}
+\Omega^{-1/2}\epsilon_{\rm dyn}.
\end{equation}
These different boost weights are the standard relative weights of
the two supersymmetry sectors in the Penrose
limit~\cite{Hatsuda:2002kx,Blau:2002mw}.

\paragraph{An equivalent choice with no independent kinematical component.}

The explicit coefficients in \eqref{eq:DFPY_q} and \eqref{eq:DFPY_s}
obey
\begin{equation}
q_{\alpha a\dot2}=-q_{\alpha a\dot1},
\qquad
s_{\alpha a\dot2}=s_{\alpha a\dot1}.
\label{eq:ADHM_dotted_relations}
\end{equation}
These are relations between numerical coefficients. The full spinors
$\epsilon^0_{Q,\dot1}$ and $\epsilon^0_{Q,\dot2}$ also contain distinct
internal basis spinors $\zeta^0_{a\dot1}$ and $\zeta^0_{a\dot2}$, so they
do not cancel in their sum. Instead, \eqref{eq:ADHM_dotted_relations}
shows that the $Q$ and $S$ coefficients occupy the two orthogonal
directions $(1,-1)$ and $(1,1)$ in the dotted-index space.

Choose the $SU(2)_C$ transformation whose action on the lower-index
coefficient doublets is
\begin{equation}
\mathcal R_C=\frac1{\sqrt2}
\begin{pmatrix}1&1\\-1&1\end{pmatrix}.
\label{eq:ADHM_alignment_rotation}
\end{equation}
For convenience, include an overall normalization $1/\sqrt2$ of the
supercharge and define
\begin{equation}
\begin{pmatrix}\widetilde q_{\alpha a\dot1}\\
                \widetilde q_{\alpha a\dot2}\end{pmatrix}
=\frac1{\sqrt2}\mathcal R_C
 \begin{pmatrix}q_{\alpha a\dot1}\\q_{\alpha a\dot2}\end{pmatrix},
\qquad
\begin{pmatrix}\widetilde s_{\alpha a\dot1}\\
                \widetilde s_{\alpha a\dot2}\end{pmatrix}
=\frac1{\sqrt2}\mathcal R_C
 \begin{pmatrix}s_{\alpha a\dot1}\\s_{\alpha a\dot2}\end{pmatrix}.
\label{eq:ADHM_aligned_coefficients_def}
\end{equation}
Then
\begin{equation}
\widetilde q_{\alpha a\dot1}=0,
\qquad
\widetilde q_{\alpha a\dot2}=q_{\alpha a\dot2},
\qquad
\widetilde s_{\alpha a\dot1}=s_{\alpha a\dot1},
\qquad
\widetilde s_{\alpha a\dot2}=0.
\label{eq:ADHM_aligned_coefficients}
\end{equation}
Both the rotation and the common normalization are invertible. They
preserve the rank of the coefficient tensors. The zero components
follow from the special relations \eqref{eq:ADHM_dotted_relations};
the same construction need not work for a general supercharge.
The rotation acts within the orbifold-invariant $SU(2)_C$ doublet
and therefore preserves the ADHM orbifold condition.

Let $U_C$ denote the corresponding symmetry operator. The new charge is
\begin{equation}
\widetilde{\mathcal Q}
=\frac1{\sqrt2}U_C\mathcal Q^C_{\beta=1}U_C^{-1}.
\label{eq:ADHM_aligned_charge}
\end{equation}
At the undeformed superconformal fixed point, $SU(2)_C$ is an exact
symmetry. We transform all insertions by
$\widetilde{\mathcal O}=U_C\mathcal O U_C^{-1}$ and transform the
monopole boundary conditions and their counterterms accordingly.
The transformation of monopole polarizations and boundary conditions
is described in equations (2.48)--(2.50) of
\cite{Dedushenko:2017avn}. We also transform the localization deformation
and its auxiliary-field and background data. In this way the
supersymmetric path integral is carried to an equivalent one.
In particular,
\begin{equation}
\left\langle\prod_i\widetilde{\mathcal O}_i\right\rangle
=\left\langle\prod_i\mathcal O_i\right\rangle,
\label{eq:ADHM_rotated_correlators}
\end{equation}
where all insertions defining the correlator are included in the
product. If boundary states are used, they are transformed as well.

We keep the reference geodesic and the local frames fixed. The bulk
spinor associated with the new charge is obtained by replacing its
constant data with
\begin{equation}
\left(\widetilde\epsilon_0\right)_{a\dot a}
=\begin{pmatrix}
\widetilde q_{\alpha a\dot a}\\[1mm]
(\sigma_2)_\alpha{}^\beta\widetilde s_{\beta a\dot a}
\end{pmatrix}.
\label{eq:ADHM_aligned_bulk_constant}
\end{equation}
With the normalization in \eqref{eq:ADHM_aligned_coefficients_def},
this constant spinor is simply
\begin{equation}
\widetilde\epsilon_0
=\epsilon^0_{Q,\dot2}+\epsilon^0_{S,\dot1}.
\end{equation}
Its restriction to the geodesic is
\begin{equation}
\widetilde\epsilon_\gamma(x^+)
=C\left(e^{-ix^+/4}\epsilon^0_{Q,\dot2}
+e^{+ix^+/4}\epsilon^0_{S,\dot1}\right),
\qquad
\Pi_-\widetilde\epsilon_\gamma=0,
\qquad
\Pi_+\widetilde\epsilon_\gamma=\widetilde\epsilon_\gamma.
\label{eq:ADHM_aligned_lightcone_projection}
\end{equation}
The boost therefore gives
\begin{equation}
S_\Omega\widetilde\epsilon_\gamma
=\Omega^{-1/2}\widetilde\epsilon_\gamma.
\end{equation}
After the usual dynamical normalization, the whole aligned charge
has a finite limit:
\begin{equation}
\epsilon^{\rm PW}_{\rm aligned}(x^+)
=\lim_{\Omega\to\infty}
 \Omega^{1/2}S_\Omega\widetilde\epsilon_\gamma(x^+)
=\widetilde\epsilon_\gamma(x^+).
\label{eq:ADHM_aligned_Penrose_limit}
\end{equation}
No additional light-cone projection is required in this limit.
The absence of a kinematical component here refers to the independent
Killing-spinor data on the reference geodesic. The transverse
dependence of the dynamical Killing spinor is fixed by the Killing
spinor equation.

\paragraph{Spinor data for the BMN comparison.}

The limiting spinor of the aligned charge takes the form
\begin{equation}
\epsilon^{\rm PW}_{\rm aligned}(x^+)
=e^{-ix^+/4}\chi_-+e^{+ix^+/4}\chi_+,
\label{eq:DFPY_PW_spinor}
\end{equation}
where
\begin{equation}
\chi_-=C\epsilon^0_{Q,\dot2},
\qquad
\chi_+=C\epsilon^0_{S,\dot1}.
\label{eq:chi_pm_def}
\end{equation}
The normalization of the aligned charge was chosen so that these
coefficients agree with the dynamical coefficients of the original
spinor. Explicitly,
\begin{equation}
\chi_-
=C\sum_{a=1}^2
\begin{pmatrix}
q_{1a\dot2}\\q_{2a\dot2}\\0\\0
\end{pmatrix}
\otimes\zeta^0_{a\dot2},
\label{chiform1}
\end{equation}
and
\begin{equation}
\chi_+
=C\sum_{a=1}^2
\begin{pmatrix}
0\\0\\(\sigma_2s)_{1a\dot1}\\(\sigma_2s)_{2a\dot1}
\end{pmatrix}
\otimes\zeta^0_{a\dot1}.
\label{chiform2}
\end{equation}
The nonzero coefficient matrices are
\begin{equation}
q^{(\dot2)}_{\alpha a}
=\begin{pmatrix}-1&i\\-1&-i\end{pmatrix}_{\alpha a},
\qquad
(\sigma_2s^{(\dot1)})_{\alpha a}
=\begin{pmatrix}1&i\\1&-i\end{pmatrix}_{\alpha a}.
\label{eq:explicit_surviving_coefficients}
\end{equation}

To determine their spatial $SU(2)$ transformation, use the regular
Cartesian frame introduced above. The rotation
$\epsilon_F=\mathcal R^{-1}\epsilon_{\rm polar}$ removes the angular
dependence of the polar spin frame. Spatial rotations then act
through the constant Cartesian-frame gamma matrices.
With rotations represented by $e^{-i\omega^i\mathsf S_i}$, define
\begin{equation}
\mathsf S_1=-\frac{i}{2}\gamma_{23},
\qquad
\mathsf S_2=-\frac{i}{2}\gamma_{31},
\qquad
\mathsf S_3=-\frac{i}{2}\gamma_{12}.
\label{eq:bulk_spatial_spin_generators}
\end{equation}
In the $J$-adapted basis of \eqref{eq:AdS_gamma_basis}, these generators
are block diagonal:
\begin{equation}
\mathsf S_i
=\begin{pmatrix}\mathsf S_i^Q&0\\0&\mathsf S_i^S\end{pmatrix},
\qquad
\begin{aligned}
(\mathsf S_1^Q,\mathsf S_2^Q,\mathsf S_3^Q)
&=\frac12(-\sigma_3,-\sigma_1,\sigma_2),\\
(\mathsf S_1^S,\mathsf S_2^S,\mathsf S_3^S)
&=\frac12(\sigma_3,\sigma_1,\sigma_2).
\end{aligned}
\label{eq:bulk_spatial_spin_blocks}
\end{equation}
Each block satisfies
$[\mathsf S_i^{Q,S},\mathsf S_j^{Q,S}]
=i\epsilon_{ijk}\mathsf S_k^{Q,S}$.
Both blocks therefore transform as spatial spin-$1/2$ doublets.
Their different $J$ eigenvalues distinguish the $Q$ and $S$ sectors.

Choosing $\mathsf S_1$ as the Cartan generator gives
\begin{equation}
\mathsf S_1=\frac12\operatorname{diag}(-1,+1,+1,-1).
\end{equation}
The upper block has weights $-1/2$ and $+1/2$.
The lower block has the opposite ordering.
These weights refer to rotations about the Cartesian $X^1$ axis.

The two spatial doublet bases are related by
\begin{equation}
\sigma_2\mathsf S_i^S\sigma_2=\mathsf S_i^Q.
\label{eq:bulk_spatial_doublet_intertwiner}
\end{equation}
Thus, if $s_0=\sigma_2s$ is the lower-block coefficient, then
$\sigma_2s_0=s$ transforms with the same spatial generator matrices
as $q_0=q$. The same statement holds for the rotated coefficients,
since $\mathcal R_C$ acts on the dotted index and commutes with the
spatial matrices. We therefore use $q^{(\dot2)}$ and $s^{(\dot1)}$
when comparing the two spatial doublets with the BMN spinor.

\subsection{The common unbroken subgroup}
\label{app:common_subgroup}

The three transverse directions inherited from the spatial part of
$AdS_4$ form the $SO(3)$ factor of BMN. The convention in \cite{Asano:2012zt} labels these
directions by
\begin{equation}
2,3,4
\end{equation}
so we identify
\begin{equation}
SO(3)_{\mathrm{space}}^{\mathrm{ADHM}}
=
SO(3)_{234}^{\mathrm{BMN}}
\label{eq:SO3_match}
\end{equation}

The six transverse directions inherited from $S^7$ are the real directions
contained in
\begin{equation}
z_2
\qquad
z_3
\qquad
z_4
\end{equation}

We choose the BMN axes so that
\begin{equation}
z_2
\longleftrightarrow
(Y^9,Y^{10})
\end{equation}
while
\begin{equation}
(z_3,z_4)
\longleftrightarrow
(Y^5,Y^6,Y^7,Y^8)
\end{equation}

Thus
\begin{equation}
SO(6)
\supset
SO(4)_{5678}
\times
SO(2)_{9\,10}
\end{equation}

We identify
\begin{equation}
SO(4)_{5678}
\simeq
SU(2)_H
\times
SU(2)_F
\end{equation}
while
\begin{equation}
SO(2)_{9\,10}
\simeq
U(1)_C\subset SU(2)_C
\end{equation}
on the inherited supersymmetry parameters.

The latter identification follows directly from the orbifold condition
\begin{equation}
\varsigma_1=\varsigma_2
\end{equation}
because the charge under the transverse $z_2$ plane is then equal to the
$SU(2)_C$ weight carried by the dotted index.

Under
\begin{equation}
SU(2)_H
\times
SU(2)_F
\times
U(1)_C
\end{equation}
the relevant chiral spinor of $Spin(6)$ decomposes as
\begin{equation}
\mathbf4
\longrightarrow
(\mathbf2,\mathbf1)_{+1/2}
\oplus
(\mathbf1,\mathbf2)_{-1/2}
\end{equation}

The supersymmetries inherited from the ADHM theory lie in the
$SU(2)_H$-doublet and $SU(2)_F$-singlet sector. Therefore
\begin{equation}
\epsilon^{\rm PW}_{\rm aligned}
\in
(\mathbf2_{\mathrm{space}},
 \mathbf2_H,
 \mathbf1_F)_{+1/2}
\oplus
(\mathbf2_{\mathrm{space}},
 \mathbf2_H,
 \mathbf1_F)_{-1/2}
\label{eq:common_rep}
\end{equation}

\subsection{Comparison with the BMN localization spinor in Asano-Ishiki-Okada-Shimasaki}
\label{app:comparison_pwmm_spinor}

We now compare the Penrose-limit Killing spinor of the aligned charge,
\begin{equation}
\epsilon^{\rm PW}_{\rm aligned}(x^+)
=
e^{-ix^+/4}\chi_-
+
e^{+ix^+/4}\chi_+ ,
\label{eq:DFPY_PW_recalled}
\end{equation}
with the localization Killing spinor used by Asano, Ishiki, Okada and Shimasaki in \cite{Asano:2012zt} (see equation 3.2 in \cite{Asano:2012zt})
\begin{equation}
\epsilon_{\rm A}(\tau_{\rm A})
=
e^{\frac{-\tau_{\rm A}}{2}\Gamma^{09}}
e^{-\frac{\pi}{4}\Gamma^{49}}
\chi,
\qquad
\chi=
\begin{pmatrix}
\eta_1\\
0\\
0\\
0
\end{pmatrix}.
\label{eq:Asano_spinor}
\end{equation}
Using the Gamma matrix basis given in appendix A of \cite{Asano:2012zt}, we can easily see that for this spinor\footnote{We have to remember that as per the conventions in \cite{Asano:2012zt}, $\Gamma^{MN}=\tilde\Gamma^{[M}\Gamma^{N]}$, and in the conventions of the same paper, $\Gamma^0=-\tilde\Gamma^0=\mathbf{1}_{16\times 16}$.},
\begin{equation}
\Gamma^{09}\chi=-\chi\implies \Gamma^{9,10}\chi=-i\chi,
\qquad
\Gamma^{5678}\chi=+\chi .
\label{eq:Asano_projections}
\end{equation}

Above we have used the fact that $X^0=i X_{10}$ as reviewed in subsection \ref{subsec:asano-supercharge}.
The first condition fixes one of the two $U(1)_C$ weights, while the
second selects one chirality of
\begin{equation}
Spin(4)_{5678}\simeq SU(2)_H\times SU(2)_F .
\end{equation}
We choose the convention such that positive chirality transforms as
$({\bf 2}_H,{\bf 1}_F)$, therefore the spinor in \eqref{eq:Asano_spinor} transforms as
\begin{equation}
\epsilon_A \in {\bf 2}_{\rm space}\otimes{\bf 2}_H .
\label{eq:eta1_rep}
\end{equation}

We first make explicit the two time-dependent components of
\eqref{eq:Asano_spinor}. Since
\begin{equation}
(\Gamma^{49})^2=-1,
\end{equation}
we have
\begin{equation}
e^{-\frac{\pi}{4}\Gamma^{49}}
=
\frac{1}{\sqrt2}
\left(1-\Gamma^{49}\right).
\end{equation}
Furthermore,
\begin{equation}
\{\Gamma^{09},\Gamma^{49}\}=0 .
\end{equation}
Using $\Gamma^{09}\chi=-\chi$, it follows that
\begin{equation}
\epsilon_{\rm A}(\tau_{\rm A})
=
\frac{1}{\sqrt2}
\left[
e^{\tau_{\rm A}/2}\chi
+
e^{-\tau_{\rm A}/2}\chi'
\right],
\qquad
\chi'\equiv-\Gamma^{49}\chi .
\label{eq:Asano_two_components}
\end{equation}

In Lorentzian worldine time, $t_A=-i\tau_A$, we get

\begin{equation}
\epsilon_{\rm A}(\tau_{\rm A})
=
\frac{1}{\sqrt2}
\left[
e^{i t_{\rm A}/2}\chi
+
e^{-it_{\rm A}/2}\chi'
\right],
\qquad
\chi'\equiv-\Gamma^{49}\chi .
\label{eq:Asano_spin_lor}
\end{equation}

The conventions in \cite{Asano:2017nxw} are chosen so that $\mu=6$ in their notation, however in our coordinates, $\mu=3$, so the time coordinates should be identified as $2t_A= x^+$. We therefore get the following BMN localization spinor: 
\begin{equation}
\epsilon_{\rm A}(\tau_{\rm A})
=
\frac{1}{\sqrt2}
\left[
e^{i x^+/4}\chi
+
e^{-ix^+/4}\chi'
\right],
\qquad
\chi'\equiv-\Gamma^{49}\chi .
\label{eq:Asano_spin_lor2}
\end{equation}

Comparing it with \eqref{eq:DFPY_PW_spinor}, we need to make the comparison as : $\chi_-\leftrightarrow \chi', \chi_+\Leftrightarrow \chi$. 

The $SU(2)$ components of $\chi_{\pm}$ are related by:
\begin{equation}
    (\chi_-)_\alpha=(-i\sigma_3)_{\alpha}^{\beta}(\chi_+)_\beta
\end{equation}

On the other hand, the $SU(2)_F$ components are the same, and $U(1)_C$ charges are reversed. We want to check if the same relation holds between $\chi$ and $\chi'$. The relation between $\chi$ and $\chi'$ is given in \eqref{eq:Asano_two_components}. Since $\Gamma^{49}$ does not act on the planes $5678$, the $SU(2)_H$ components are the same in both $\chi$ and $\chi'$. 

Similarly, since $U(1)_C$ is generated by $\Gamma^{9,10}$, it is easy to see that $\Gamma^9$, changes $U(1)_C$ charge. In the explicit gamma matrix basis of \cite{Asano:2012zt}, it is easy to check that 
\begin{equation}
\chi'
=
-\Gamma^{49}\chi
=
\begin{pmatrix}
\mathbf{0}_4\\[1mm]
\mathbf{0}_4\\[1mm]
-J_4\eta_1\\[1mm]
\mathbf{0}_4
\end{pmatrix},
\qquad
\eta_1=
\begin{pmatrix}
1\\0\\0\\0
\end{pmatrix}.
\end{equation}
Here $J_4=-\left.\Gamma^{23}\right|_{\rm block}$ on the first
and third four-component blocks. The generator $\Gamma^{23}$
describes rotations about the $X^4$ axis. With the standard
normalization of the Hermitian angular-momentum generators,
\begin{equation}
S_4
=
-\frac{i}{2}\left.\Gamma^{23}\right|_{\rm block}
=
\frac{i}{2}J_4.
\end{equation}
In the spatial spin-$\frac12$ basis, choosing\footnote{The reason for this choice is to compare with the choice of rotation matrices in the bulk calculation in \eqref{eq:bulk_spatial_spin_generators}.}
\begin{equation}
S_4=\frac{-\sigma_3}{2}\otimes\mathbf{1}_H,
\qquad
J_4=i\sigma_3\otimes\mathbf{1}_H.
\end{equation}
Therefore, denoting the nonzero block coefficients of $\chi$
and $\chi'$ in this basis by $v_{\alpha a}$ and $v'_{\alpha a}$,
respectively, we obtain
\begin{equation}
v'_{\alpha a}
=
-i(\sigma_3)_\alpha{}^\beta v_{\beta a}.
\end{equation}
which is consistent with the identification $\chi_-\leftrightarrow \chi', \chi_+\Leftrightarrow \chi$ since $SU(2)$ components of $\chi_{\pm}$ obey the same relation. This also identifies the axis $X^4$ in \cite{Asano:2012zt} with $X^1$ in the bulk cartesian frame. 

The aligned ADHM localization charge therefore has a purely
dynamical Penrose limit. With the time and frame identifications
specified above, its two frequency components have the same quantum
numbers under the common symmetry subgroup and the same relative
polarization as the BMN localization spinor of \cite{Asano:2012zt}.
The comparison applies to the whole aligned charge.
The original ADHM correlators are related to this choice by the
simultaneous $SU(2)_C$ transformation of the charge, the operators,
and the monopole boundary data.

\section{Generic positive-flux sectors and the BMN measure}\label{app:generic-positive-flux-measure}

In this appendix, we work with fixed screened GNO charges in ADHM theory. We take the triple-scaling limit of the measure appearing in the associated Coulomb branch correlators and compare it with the measure appearing in the protected BMN correlators in the corresponding fuzzy sphere vacua.

As explained in \cite{Dedushenko:2018icp}, for non-miniscule GNO charges, the screened charges can differ from those of the inserted local monopole operator. Throughout this appendix, we compare the measures at fixed screened charges, without including the associated bubbling coefficients that weight these sectors in the full Coulomb branch correlators.

As discussed in subsection~\ref{subsec:adhm-bmn-fixed-backgrounds}, the ADHM correlators localize to an integral over the eigenvalues of an $N\times N$ Hermitian matrix. On the other hand, correlators in a fuzzy sphere vacuum of the $U(m)$ BMN model localize to an integral over the eigenvalues of an $m\times m$ Hermitian matrix.

As we will see, the integration variables in the BMN correlators correspond to the eigenvalues associated with nonzero flux in ADHM theory. We integrate out the zero-flux eigenvalues in the triple-scaling limit discussed in the main text and compare the resulting effective measure for the remaining eigenvalues with the measure appearing in the BMN correlators.

\subsection{Associating localization integration variables}

Consider the following choice of screened GNO charges
\begin{equation}
  \vec b
 =
 \left(
  n_1^{N_1},n_2^{N_2},\ldots,n_{\Lambda}^{N_{\Lambda}},
  0^{N-\nu}
 \right),
 \qquad
 n_s\in\mathbb Z_{>0},
 \qquad
 \nu\equiv\sum_{s=1}^{\Lambda}N_s .
 \label{eq:app-positive-flux-partition}
\end{equation}
Here the convention $n_s^{N_s}$ means that the distinct charges $n_s$ occurs with multiplicities
$N_s$. As per the dictionary explained in the main text, the corresponding BMN gauge group rank is
\begin{equation}
 m=\sum_{s=1}^{\Lambda}n_sN_s,
 \label{eq:app-bmn-matrix-size}
\end{equation}
and the corresponding BMN fuzzy sphere vacuum is
\begin{equation}
 L_i(\vec b)
 =
 \bigoplus_{s=1}^{\Lambda}
 \left(
  L_i^{[n_s]}\otimes\mathbf 1_{N_s}
 \right).
 \label{eq:app-general-bmn-vacuum}
\end{equation}

On BMN side, localization around this vacuum restricts the commuting localizing matrix $M$ to
\begin{equation}
 M
 =
 \bigoplus_{s=1}^{\Lambda}
 \left(
  \mathbf 1_{n_s}\otimes M_s
 \right),
 \qquad
 \operatorname{Spec}(M_s)=\{q_{s,a}\}_{a=1}^{N_s}.
 \label{eq:app-general-bmn-M}
\end{equation}

On ADHM side, the localization restricts the path integral to constant values of the eigenvalues of the adjoint scalar $\sigma$. 

Let us denote the eigenvalues carrying positive flux
by $\sigma_{s,a}$, while $ \rho_I,
 I\in(1,\ldots,N-\nu)$
denote the eigenvalues in the zero-flux sea.  We propose the following identification of the integration variables 
\begin{equation}
 \begin{gathered}
 q_{s,a}=2\sigma_{s,a},\\
 n_s=\text{dimension of the corresponding BMN irreducible block}.
 \end{gathered}
 \label{eq:app-charge-block-dictionary}
\end{equation}

\subsection{ADHM one-loop determinants in the magnetic background}

We now compute the one-loop measure for the flux configuration \(\vec b\) in \eqref{eq:app-positive-flux-partition}. We set the sphere radius to one and take the real masses and FI parameter to vanish.

For each weight $\vec w$ of the hypermultiplet representation,
Eq.~(2.38) of \cite{Dedushenko:2018icp} gives a factor
$h_{|\vec w\cdot\vec b|}(\vec w\cdot\vec\sigma)$, where
\begin{equation}
    h_n(x)\equiv \frac{1}{2\pi}
    \left|\Gamma\!\left(\frac{n+1}{2}+ix\right)\right|^2,
    \qquad n=0,1,2,\ldots .
    \label{eq:app-hn-definition}
\end{equation}
The weights of the fundamental representation of $U(N)$ are the unit
vectors $e_i$. Thus $e_i\cdot\vec b=b_i$ and
$e_i\cdot\vec\sigma=\sigma_i$. One fundamental hypermultiplet  one-loop determinant gives
\begin{equation}
    Z_f(\vec\sigma,\vec b)
    =\prod_{i=1}^{N}\frac{1}{2\pi}
      \left|\Gamma\!\left(\frac{1+|b_i|}{2}+i\sigma_i\right)\right|^2
    =\prod_{i=1}^{N}h_{|b_i|}(\sigma_i).
\end{equation}
The $N_f$ fundamental hypermultiplets each give the same factor, so
their product is
\begin{equation}
    \bigl[Z_f(\vec\sigma,\vec b)\bigr]^{N_f}
    =\prod_{i=1}^{N}\bigl[h_{|b_i|}(\sigma_i)\bigr]^{N_f}.
\end{equation}

For the adjoint and vector multiplets, write
$x_{ij}=\sigma_i-\sigma_j$ and $n_{ij}=|b_i-b_j|$.
The $N$ diagonal entries of an adjoint matrix have zero weight, so
each contributes $h_0(0)=1/2$. For each $i<j$, the two off-diagonal
entries have weights $e_i-e_j$ and $e_j-e_i$. Since $h_n(-x)=h_n(x)$,
these two weights give the same factor. The adjoint hypermultiplet
therefore contributes
\begin{equation}
    Z_{\mathrm{adj}}(\vec\sigma,\vec b)
    =2^{-N}\prod_{i<j}h_{n_{ij}}(x_{ij})^2.
\end{equation}

For the vector multiplet, Eq.~(2.39) of \cite{Dedushenko:2018icp}
has one numerator factor for each positive root $e_i-e_j$ with $i<j$.
Both roots $\pm(e_i-e_j)$ enter the denominator. Combining them gives
\begin{equation}
    Z_{\mathrm{vec}}(\vec\sigma,\vec b)
    =\prod_{i<j}
      \frac{(2\pi)^2\bigl(x_{ij}^2+n_{ij}^2/4\bigr)}
           {\left|\Gamma\!\left(1+\frac{n_{ij}}{2}
                               +ix_{ij}\right)\right|^4}.
\end{equation}

The terms appearing in the products in $Z_{adj}$ and $Z_{vec}$ can be combined to a single function below:
\begin{equation}
 \mathcal F_n(x)
 =
 \left(x^2+\frac{n^2}{4}\right)
 \left[
  \frac{
   \Gamma\!\left(\frac{n+1}{2}+ix\right)
   \Gamma\!\left(\frac{n+1}{2}-ix\right)
  }{
   \Gamma\!\left(1+\frac n2+ix\right)
   \Gamma\!\left(1+\frac n2-ix\right)
  }
 \right]^2,
 \label{eq:app-Fn-definition}
\end{equation}
so that the determinant contribution from adjoint and vector multiplets can be written as:
\begin{equation}
    Z_{adj}Z_{vec}=2^{-N}\prod_{i<j} \mathcal{F}_{|b_i-b_j|}(\sigma_i-\sigma_j)
\end{equation}
It is also convenient to introduce the following functions: 
\begin{equation}
 U_n(x)\equiv-\log h_n(x),
 \qquad
 L_n(x)\equiv\log\mathcal F_n(x).
 \label{eq:app-Un-Ln-definition}
\end{equation}
Up to constant factors, the contribution of the one-loop determinants to the path integral in a given screened GNO sector is 
\begin{align}
 Z_{\bm m}
 \propto{}&
 \frac{1}{(N-\nu)!\prod_sN_s!}
 \int
 \prod_{I=1}^{N-\nu}\dd\rho_I
 \prod_{s,a}\dd\sigma_{s,a}
 \nonumber\\
 &\times
 \prod_Ih_0(\rho_I)^{N_f}
 \prod_{I<J}\mathcal F_0(\rho_I-\rho_J)
 \nonumber\\
 &\times
 \prod_{s,a}h_{n_s}(\sigma_{s,a})^{N_f}
 \prod_{I,s,a}
 \mathcal F_{n_s}(\sigma_{s,a}-\rho_I)
 \nonumber\\
 &\times
 \prod_{(s,a)<(t,b)}
 \mathcal F_{|n_s-n_t|}
 \left(\sigma_{s,a}-\sigma_{t,b}\right).
 \label{eq:app-exact-fixed-flux-integral}
\end{align}

We now explain the above decomposition in detail. 
The factor $(N-\nu)!\prod_sN_s!$ is the order of the Weyl group preserving
the magnetic background. The eigenvalues $\vec\sigma$ are split into two groups: $\sigma_{s,a}$ and $\rho_I$. $\rho_I$ correspond to eigenvalues carrying zero flux, and $\sigma_{s,a}$ denotes $a$th eigenvalue with flux $n_s$. 

The one-loop determinants of the fundamental hypermultiplets split into zero-flux contribution $h_0(\rho_I)$ and nonzero flux contributions $h_{n_s}(\sigma_{s,a})$. Similarly the combined adjoint and vector multiplet contributions split into zero-zero, zero-nonzero, and nonzero-nonzero sectors $\mathcal{F}_0(\rho_{I}-\rho_J)$, $\mathcal{F}_{n_s}(\sigma_{s,a}-\rho_I)$ and $\mathcal{F}_{|n_s-n_t|}(\sigma_{s,a}-\sigma_{t,b})$ respectively.  

To compare with the measure appearing in BMN correlators computed in \cite{Asano:2017nxw}, we need to integrate out the zero flux eigenvalues $\rho_I$. We define the effective action for the remaining eigenvalues $\sigma_{s,a}$ as 
\begin{align}
&e^{S_{\eff}(\bm\sigma)}
 ={}
 \prod_{s,a}e^{-N_fU_{n_s}(\sigma_{s,a})}
 \prod_{(s,a)<(t,b)}
 e^{L_{|n_s-n_t|}(\sigma_{s,a}-\sigma_{t,b})}
 \nonumber\\
 &\times
 \left\langle
  \exp\!\left[
   \sum_{I,s,a}
   L_{n_s}(\sigma_{s,a}-\rho_I)+\sum_{I<J} L_0(\rho_I-\rho_J) - N_f \sum_I U_0(\rho_I)
  \right]
 \right\rangle_{0,N-\nu}.
 \label{eq:app-exact-charged-effective-weight}
\end{align}
The brackets $\langle~~\rangle_{0,N-\nu}$ mean integrating out over the $N-\nu$ eigenvalues $\rho_I$.

It is convenient to move to the momentum space.
We use the convention
\begin{equation}
 \widehat f(k)
 =
 \int_{-\infty}^{\infty}\dd x\,e^{-ikx}f(x).
 \label{eq:app-fourier-convention}
\end{equation}
Terms supported only at $k=0$ correspond to additive constants or
normalizations independent of the eigenvalues and will be omitted.  For
$k\neq0$, the following are the Fourier transforms of the functions $U_n$ and $L_n$.
\begin{equation}
 \widehat U_n(k)
 =
 -\frac{\pi}{|k|\sinh(|k|/2)}e^{-n|k|/2},
 ~~
 \widehat L_n(k)
 =
 -\frac{2\pi}{|k|}
 e^{-n|k|/2}
 \tanh\frac{|k|}{4}. 
 \label{eq:app-general-fourier-transforms}
\end{equation}
In particular, note that
\begin{equation}
 \widehat U_n(k)=e^{-n|k|/2}\widehat U_0(k),
 ~~
 \widehat L_n(k)=e^{-n|k|/2}\widehat L_0(k).
 \label{eq:app-smearing-identities}
\end{equation}

For completeness, these identities can also be obtained recursively.  The
Gamma-function recursion relation implies
\begin{align}
 U_{n+2}(x)-U_n(x)
 &=-\log\!\left(x^2+\frac{(n+1)^2}{4}\right),
 \label{eq:app-Un-recursion}
 \\
 \frac{\mathcal F_{n+2}(x)}{\mathcal F_n(x)}
 &=
 \frac{
  \left(x^2+\frac{(n+1)^2}{4}\right)^2
 }{
  \left(x^2+\frac{n^2}{4}\right)
  \left(x^2+\frac{(n+2)^2}{4}\right)
 }.
 \label{eq:app-Fn-recursion}
\end{align}
Using
\begin{equation}
 \int_{-\infty}^{\infty}\dd x\,
 e^{-ikx}\log(x^2+a^2)
 =
 -\frac{2\pi}{|k|}e^{-a|k|},
 \qquad k\neq0,
 \label{eq:app-log-fourier-transform}
\end{equation}
together with the $n=0,1$ cases proves
\eqref{eq:app-general-fourier-transforms} for every nonnegative integer $n$.

\subsection{The saddle point solution}

We use the large-$N$ saddle-point approximation for this integration by treating the discrete eigenvalues $\rho_I$ as continuously distributed over the real line with eigenvalue density $\varrho(x)$. 

Let us define the unnormalized eigenvalue density as
\begin{equation}
 \varrho(x)=\sum_{I=1}^{N-\nu}\delta(x-\rho_I).
 \label{eq:app-sea-density}
\end{equation}
The effective action in terms of $\rho(x)$ is then given by
\begin{align}
 S_0[\varrho]
 =
 &N_f\int\dd x\,\varrho(x)U_0(x)
 -\frac12
 \int\dd x\,\dd y\,
 \varrho(x)\varrho(y)L_0(x-y)-\sum_{s,a}\int\dd x \varrho(x)L_{n_s}(\sigma_{s,a}-x)\nonumber\\
 &-\mu\left(\int \dd x\varrho(x)-(N-\nu)\right)
 \label{eq:app-sea-collective-action}
\end{align}
We have also added a Lagrange multiplier term at the end to make sure that the total number of eigenvalues is fixed. 
Since we also take the large $N_f$ limit, the third term is subleading compared to the first and can be ignored for the leading eigenvalue distribution. 

Varying with respect to $\varrho$, the saddle point equation  is given by
\begin{equation}
 N_fU_0(x)
 -\int_{-\infty}^{\infty}\dd y\,\varrho_0(y)L_0(x-y)
 =\mu.
 \label{eq:app-sea-saddle-equation}
\end{equation}
Here we assume that the eigenvalue density has a finite support over the whole real line. We will see that this leads to a problem below. 

Going to the Fourier space, for nonzero momentum $k$, we can easily see that 
\begin{equation}
 \widehat\varrho_0(k)
 =N_f\frac{\widehat U_0(k)}{\widehat L_0(k)}
 =\frac{N_f}{4\sinh^2(|k|/4)},
 \qquad k\neq0.
 \label{eq:app-sea-density-fourier}
\end{equation}

Performing the inverse Fourier transform, and including the undetermined constant part, we get
\begin{equation}
 \varrho_0(x)
 =
 2N_f\left[A-x\coth(2\pi x)\right]
 \label{eq:app-sea-bulk-density}
\end{equation}
and hence
\begin{equation}
 \varrho_0(x)\simeq2N_f(A-|x|),
 \qquad
 1\ll|x|\ll A.
 \label{eq:app-triangular-sea}
\end{equation}
This would make the eigenvalue density negative for \(|x|>A\), violating its non-negativity. 

Therefore, we demand that the eigenvalue density be truncated in finite support $[-A,A]$. 

Let us first determine the constant $A$ assuming that we simply use the solution in \eqref{eq:app-sea-bulk-density} and truncate it in the range $[-A,A]$, using the normalization constraint on $\varrho$, we deduce that 
\begin{equation}
 N-\nu
 =2N_fA^2+O(N_f),
 \label{eq:app-sea-normalization}
\end{equation}
and therefore
\begin{equation}
 A^2
 =\frac{\lambda_{\pp}N_f}{2}[1+o(1)].
 \label{eq:app-sea-endpoint}
\end{equation}

Thus, in the double scaling limit, where $N_f$ is large, and $\lambda_{\pp}=N/N_f^2$ is fixed, we see that the endpoints of the support are very far from the origin. There is, however, a subtlety we should point out. Simply restricting the solution we found above by using Fourier transforms to a finite range does not satisfy the saddle point equation for points inside the range. This is because \eqref{eq:app-sea-saddle-equation} was satisfied for points inside the range, assuming the density was extended to the full real line. 

However, this is not a major obstacle. We can find a correction to the above solution which make sure that the new corrected solution satifies the saddle point equation everywhere in the range $[-A,A]$. Let us assume that the correct solution is 
\begin{equation}
\varrho_A(x)=
\begin{cases}
\varrho_{0}(x)+\eta_A(x), & |x|<A,\\
0, & |x|>A.
\end{cases}
\end{equation}
where 
\begin{equation}
    \varrho_0(x)= 2N_f\left[B-x\coth(2\pi x)\right].
\end{equation}

Since we know that $\varrho_0$ satisfies the saddle point equation on the whole real line, it is easy to see that demanding that $\varrho_A$ satisfies the saddle point equation in the given range implies that 

\begin{equation}
\int_{-A}^{A}\eta_A(y)L_0(x-y)\,dy
=
\int_{|y|>A}\rho_{0}(y)L_0(x-y)\,dy,
\qquad |x|<A.
\end{equation}

Using the simplified $L_0(y)=\log \tanh^2(\pi y)$, it is easy to see that, when $x$ is deep inside the range, the RHS goes to zero, therefore, $\eta$ should also go to zero. Therefore, we see that deep inside the range, the solution is not much altered from $\varrho$ as we should expect. Also since, $L_0$ goes to zero exponentially on order scale of its argument, so should $\eta_A$ from the endpoint. 

This means that $\eta_A$ only affect the solution at order one distance from the endpoint $A$, therefore it is also reasonable that the constant $B$ in the solutions $\varrho_0$ is different from $A$ by an order one number, which can be computed, but we will not need it. Therefore, for all practical purposes, we can simply truncate the original solution $\varrho_0$ we found earlier, with endpoint $A$ written in \eqref{eq:app-sea-endpoint}. We will also not need the explicit form of the corrected solution $\eta_A$ for leading order analysis. 

\subsection{Effective action of flux eigenvalues}\label{appsub:2body}

Now we use the leading order eigenvalue density derived in the previous subsection to compute the effective action of the flux eigenvalues $\sigma_{s,a}$. We will first compute the interactions among the eigenvalues, and then the single eigenvalue potential.

The interaction among different eigenvalues arise from the terms $L_{|n_s-n_t|}$ and the induced potential from terms $L_{n_s}(\sigma_{s,a}-\rho_I)$ in \eqref{eq:app-exact-charged-effective-weight}. The latter term induces two body interactions once we integrate out $\rho_I$. 

In \eqref{eq:app-sea-collective-action}, we expand the density as $\varrho=\varrho_0+ \delta \varrho$, where $\varrho_0$ is the saddle point density we found previously. Because $\varrho_0$ satisfies saddle point equation, there is no term linear in $\delta\varrho$ from the first two terms, however the second terms contributes a quadratic term. The third term however contribute to the linear term since we did not include it in saddle point equation. Therefore the effective quadratic action for the fluctuations is of the form
\begin{equation}
    S[\delta\varrho]= -\frac{1}{2}\int dx dy \delta\varrho(x)\delta\varrho(y)L_0(x-y)-\sum_{s,a}\int dx\delta\varrho(x)L_{n_s}(\sigma_{s,a}-x).
\end{equation}

In Fourier space, using the Fourier transforms in \eqref{eq:app-general-fourier-transforms}\footnote{Here we will ignore the fact that the leading density profile has a finite support. The fact that the saddle point equation is only satisfied inside the interval can give rise to a linear term in $\delta\varrho$ which acts as an additional source. However, this source only grows linearly, 
but for $A$ taken to be large, the propagator $L_0(x-y)^{-1}$ decays exponentially, and therefore, does not affect $\delta\varrho$ to the leading order. However, as we will see below for one body potential, the finite support will be important.}, 
\begin{align}
S[\delta\varrho]
&=
\int\frac{dk}{2\pi}\,
\frac{2\pi}{|k|}\tanh\!\left(\frac{|k|}{4}\right)
\left[
\frac12\widehat{\delta\varrho}(k)
\widehat{\delta\varrho}(-k)
+
\widehat{\delta\varrho}(k)
\sum_{s,a}e^{-n_s|k|/2}e^{ik\sigma_{s,a}}
\right].
\end{align}

Integrating out $\delta\varrho$ by completing the square, we get the following contribution to the action of $\sigma_{s,a}$ in Fourier space: 
\begin{equation}
\begin{split}
    &\int \frac{dk}{2\pi}\sum_{(s,a)\neq (s',a')}\left[-\frac{\pi}{|k|}\tanh\left(\frac{|k|}{4}\right)e^{-(n_s+n_{s'})|k|/2}e^{ik(\sigma_{s,a}-\sigma_{s',a'})}\right]\\
    &=\frac{1}{2}\sum_{(s,a)\neq (s',a')}L_{n_s+n_s'}(\sigma_{s,a}-\sigma_{s',a'}).
\end{split}
\end{equation}
\footnote{The contribution from diagonal terms $(s,a)=(s',a')$ just gives a constant in $\sigma_{s,a}$.}

Combining this with the contribution to the two body interaction from $L_{n_s-n_t}$, we get
\begin{equation}
    S_{int}(\sigma_{s,a})=\sum_{(s,a)<(t,b)}\left(L_{|n_s-n_t|}(\sigma_{s,a}-\sigma_{t,b})-L_{n_s+n_t}(\sigma_{s,a}-\sigma_{t,b})\right).
\end{equation}
Exponentiating the above action and using the explicit forms of functions $L_n$ defined in \eqref{eq:app-Fn-definition} and \eqref{eq:app-Un-Ln-definition}, we get the following contribution to the measure(we simply use $i,j$ for the collective indices $(s,a)$ below):
\begin{equation}\label{intprod}
    \prod_{i<j}\frac{\mathcal F_{|n_i-n_j|}(\sigma_i-\sigma_j)}{\mathcal F_{|n_i+n_j|}(\sigma_{i}-\sigma_{j})}=\prod_{i<j}\prod_{l=0}^{\min(n_i,n_j)-1}\frac{\left[(\sigma_i-\sigma_j)^2+\left(\frac{|n_i-n_j|}{2}+l\right)^2\right]\left[(\sigma_i-\sigma_j)^2+\left(\frac{|n_i-n_j|}{2}+l+1\right)^2\right]}{\left[(\sigma_i-\sigma_j)^2+\left(\frac{|n_i-n_j|}{2}+l+1/2\right)^2\right]^2}
\end{equation}

\paragraph{Comparison with the BMN eigenvalue interaction}

To compare the contribution from eigenvalue integral above with the same in BMN localization, consider the one-loop contribution written in \eqref{eq:pwmm-fixed-vacuum-measure}.
For two BMN blocks of dimensions $n_s$ and $n_t$, the contribution to the one-loop determinant from the pair of eigenvalues $q_{s,a}$ and $q_{t,b}$  from equation \eqref{eq:pwmm-fixed-vacuum-measure} is given by

\begin{equation}
\prod_{J=|n_s-n_t|/2}^{(n_s+n_t)/2-1}\frac{\left((2J+2)^2+(q_{s,a}-q_{t,b})^2\right)\left((2J)^2+(q_{s,a}-q_{t,b})^2\right)}{\left((2J+1)^2+(q_{s,a}-q_{t,b})^2\right)^2}.
\end{equation}
This is the same product that appears in \eqref{intprod} with the replacement $l=J-|n_s-n_t|/2$ and $q_{s,a}=2\sigma_{s,a}$. 
Therefore we see that the pair interaction of the ADHM flux eigenvalues matches that of the BMN eigenvalues. 

\subsection{One-body potential flux eigenvalues}

Now let us obtain the one-body potential for flux eigenvalues in ADHM. This gets contribution from two terms: the term proportional to $U_n$ in \eqref{eq:app-exact-charged-effective-weight}, and the interaction term between zero flux and nonzero flux eigenvalues, proportional to $L_{n}$ in \eqref{eq:app-exact-charged-effective-weight}. In the continuum limit of zero flux eigenvalues, these add to 
\begin{equation}\label{sigmact}
\begin{split}
    S_{\sigma_{s,a}}&=\sum_{s,a} \left(-N_f U_{n_s}(\sigma_{s,a})+ \int_{-A}^A dx \varrho_0(x)L_{n_s}(\sigma_{s,a}-x)\right)\\
    &=\sum_{s,a} \left(-N_f U_{n_s}(\sigma_{s,a})+ \int_{-\infty}^\infty dx \varrho_0(x)L_{n_s}(\sigma_{s,a}-x)-\left(\int_{-\infty}^{-A}+\int_A^{\infty}\right)dx \varrho_0(x)L_{n_s}(\sigma_{s,a}-x)\right),
\end{split}    
\end{equation}
where we use the leading order contribution of the eigenvalue density we obtained above in \eqref{eq:app-sea-bulk-density}. 

In Fourier space, the first two terms can be written as 
\begin{equation}
    -N_f \hat U_{n_s}(k)+ \hat\varrho_0(-k)\hat L_{n_s}(k)=e^{-n_s|k|/2}\left( -N_f \hat U_{0}(k)+ \hat\varrho_0(-k)\hat L_{0}(k)\right)=0,
\end{equation}
where we used the Fourier space identities \eqref{eq:app-smearing-identities} and saddle point equation \eqref{eq:app-sea-saddle-equation} in Fourier space. Note that to do this, we need to extend the truncated solution \eqref{eq:app-sea-bulk-density} to the whole real line, therefore we subtracted off the contribution from rest of the real line in \eqref{sigmact}. Therefore, the only contribution to the single body potential comes from the regions outside the finite interval. 

Suppose we are interested in $\sigma_{s,a}$ that is deep inside the interval. Then only large distance behaviour of $L_{n}$ can be used to extract the leading contribution, since the support over the integral starts from $A$ which is large. It is not difficult to see that for large $x$, 
\begin{equation}
    L_n(x)\approx \frac{-n}{4x^2}.
\end{equation}

Also, $\varrho$ behaves linearly at large $x$. Therefore the integrals outside the interval can be approximated as
\begin{equation}
\begin{split}
  V(\sigma)&= \int_A^\infty dx ~2N_f(A-x)\frac{-n_s}{4(\sigma_{s,a}-x)^2} + \int_{-\infty}^{-A} dx ~2N_f(A+x)\frac{-n_s}{4(\sigma_{s,a}-x)^2}\\
  &=\frac{N_f n_s}{2}\int_A^\infty dx (x-A)\left(\frac{1}{(\sigma_{s,a}-x)^2}+\frac{1}{(\sigma_{s,a}+x)^2}\right).
\end{split}
\end{equation}

The above integrals have logarithmic independent of $\sigma$, which can be removed as follows:
\begin{equation}
\begin{split}
    V(\sigma)-V(0)&=\frac{N_f n_s}{2}\int_A^\infty dx (x-A)\left(\frac{1}{(\sigma_{s,a}-x)^2}+\frac{1}{(\sigma_{s,a}+x)^2}-\frac{2}{x^2}\right)\\
    &=-\frac{n_s N_f}{2}\log\left(1-\frac{\sigma_{s,a}^2}{A^2}\right)\approx \frac{n_sN_f}{2A^2}\sigma_{s,a}^2=\frac{n_sq_{s,a}^2}{4\lambda_{\pp}} .
\end{split}
\end{equation}
above we use the identification between the BMN eigenvalues and ADHM flux eigenvalues that we deduced above. Comparing the potential to the potential that appears in BMN measure in \eqref{eq:pwmm-fixed-vacuum-measure} we can deduce that
\begin{equation}
    \lambda_{\pp}=\frac{N}{N_f^2}=\frac{g_{loc}^2}{8}.
\end{equation}

\paragraph{Summary} 

To summarize, in the double scaling limit, we identify the measure appearing in BMN localization integral with the measure that appears in ADHM correlator in the given screened GNO charge background. 

Comparing the measures we made the following identifications:
\begin{itemize}
    \item The eigenvalues appearing in BMN intergral and the flux eigenvalues in ADHM are related by $q_{s,a}=2\sigma_{s,a}$.
    \item The parameters $N,N_f$ of ADHM are related to the BMN dimensionless coupling by: $\frac{N}{N_f^2}=\frac{g_{loc}^2}{8}$.
\end{itemize}

\subsection{Relative normalization from the zero-flux eigenvalues}
\label{app:neutral-sea-normalization}

The one-body potential calculation above extracts only the $\sigma$-dependent part of the potential. However, comparing the relative weights of the screened sectors in a correlator involving a non-minuscule monopole operator (see subsection~\ref{subsec:adhm-bmn-bubbling-scope}) also requires us to retain the constant term $V_n(0)$ that was subtracted above and account for the change in the free energy of the zero-flux sea. In this subsection, we determine these contributions to the relative weights of the screened sectors.

\paragraph{The source potential and the change in sea rank.}
For a partition $p=(n_1^{N_1},\ldots,n_\Lambda^{N_\Lambda})$,
write
\begin{equation}
 \nu_p=\sum_sN_s,
 \qquad
 m=\sum_s n_sN_s.
 \label{eq:app-sea-partition-data}
\end{equation}
Thus the zero-flux eigenvalues has rank $N-\nu_p$, whereas $m$ is the
total magnetic charge. Let $Z_{\mathrm{sea}}(M)$ be the neutral
partition function evaluated at its continuum saddle, and define
the effective potential of a charge-$n$ source at the origin by
\begin{equation}
 V_n(0)=N_fU_n(0)
       -\int_{-A}^{A}dy\,\varrho_A(y)L_n(y).
 \label{eq:app-sea-source-potential}
\end{equation}
We use the functions $U_n,L_n$ and the corrected finite-support
density $\varrho_A$ introduced above. All source potentials can
be evaluated in a common neutral background of rank $N$:
changing that rank by a fixed number changes these potentials
only by $o(1)$ in the limit under consideration. The neutral-sea
factor is therefore
\begin{equation}
 \mathscr A_p^{\mathrm{sea}}
 =Z_{\mathrm{sea}}(N-\nu_p)
   \exp\left[-\sum_sN_sV_{n_s}(0)\right]
   [1+o(1)].
 \label{eq:app-sea-factor-definition}
\end{equation}

The neutral saddle equation gives $V_0(0)=\mu$, where $\mu$
is the Lagrange multiplier fixing the sea rank. Equivalently,
the saddle free energy obeys
$\partial_M\log Z_{\mathrm{sea}}(M)=-\mu$.
Since $\mu=O(N_fA)$ and $M\simeq2N_fA^2$,
$\partial_M\mu=O(A^{-1})$. For fixed $\nu_p$, this gives
\begin{equation}
 \log\frac{Z_{\mathrm{sea}}(N-\nu_p)}
          {Z_{\mathrm{sea}}(N)}
 =\nu_p\mu+o(1).
 \label{eq:app-sea-rank-shift}
\end{equation}

To compare one charge-$n$ source with $n$ unit-charge sources,
we must also account for the $n-1$ additional eigenvalues
removed from the zero-flux eigenvalues. The relevant combination is
\begin{equation}
 D_n=V_n(0)-nV_1(0)+(n-1)\mu.
 \label{eq:app-sea-source-difference}
\end{equation}
Substituting this into
\eqref{eq:app-sea-factor-definition} gives
\begin{equation}
 \log\frac{\mathscr A_p^{\mathrm{sea}}}
          {Z_{\mathrm{sea}}(N)}
 =m\bigl(\mu-V_1(0)\bigr)-\sum_sN_sD_{n_s}+o(1).
 \label{eq:app-sea-common-charge-cancellation}
\end{equation}
The first term depends only on the total charge $m$.
It cancels when comparing partitions of the same charge.
This cancellation is why the change in sea rank must be
included along with the source potentials.

\paragraph{The finite-support contribution.}
We now evaluate $D_n$. The Fourier identities for $U_n$ and
$L_n$ imply the whole-line identity
\begin{align}
 &N_f\bigl[U_n(0)-nU_1(0)+(n-1)U_0(0)\bigr]
 \nonumber\\
 &\hspace{15mm}=
 \int_{\mathbb R}dy\,\varrho_0(y)
       \bigl[L_n(y)-nL_1(y)+(n-1)L_0(y)\bigr],
 \label{eq:app-sea-whole-line-cancellation}
\end{align}
where $\varrho_0$ is the formal whole-line profile used in
the preceding saddle calculation. No additive constant is
discarded in this identity. The combination of $U_n$'s tends
to zero at infinity, while the kernel combination has zero
integral. Its convolution with $\varrho_0$ is convergent and
also tends to zero at infinity. These properties fix the
constant that would remain undetermined if the individual
Fourier transforms were used separately.

Replacing the whole-line profile by the physical finite-support
density therefore gives
\begin{align}
 D_n={}&
 \int_{|y|>A}dy\,\varrho_0(y)
       \bigl[L_n(y)-nL_1(y)+(n-1)L_0(y)\bigr]
 \nonumber\\
 &-\int_{-A}^{A}dy\,\eta_A(y)
       \bigl[L_n(y)-nL_1(y)+(n-1)L_0(y)\bigr],
 \label{eq:app-sea-tail-and-edge}
\end{align}
where $\eta_A$ is the endpoint correction described above.

The large-distance expansion of the function $L_n$ is
\begin{equation}
 L_n(y)=-\frac{n}{4y^2}
        +\frac{n(2n^2-1)}{32y^4}+O(y^{-6}).
\end{equation}
Thus the $y^{-2}$ terms cancel, leaving
\begin{equation}
 L_n(y)-nL_1(y)+(n-1)L_0(y)
 =\frac{n(n^2-1)}{16y^4}+O(y^{-6}).
 \label{eq:app-sea-cubic-kernel-tail}
\end{equation}
This cancellation also makes the endpoint correction
subleading. In the endpoint estimate used above,
$\eta_A$ has size $O(N_f)$ in layers of width $O(1)$
near $\pm A$. Its contribution to
\eqref{eq:app-sea-tail-and-edge} is therefore
$O(N_f/A^4)=o(1)$.

Outside the support,
$\varrho_0(y)=2N_f(A-|y|)+O(N_f)$, up to exponentially
small terms. The $O(N_f)$ term allows for the $O(1)$
shift of the bulk constant discussed above. It contributes
only $O(N_f/A^3)=o(1)$. The leading tail integral is
therefore elementary:
\begin{align}
 D_n
 &=\frac{N_f n(n^2-1)}4
       \int_A^\infty dy\,\frac{A-y}{y^4}+o(1)
 \nonumber\\
 &=-\frac{N_f n(n^2-1)}{24A^2}+o(1)
  =-\frac{n(n^2-1)}{12\lambda_{\mathrm{pp}}}+o(1),
 \label{eq:app-sea-source-difference-result}
\end{align}
where we used
$A^2=\lambda_{\mathrm{pp}}N_f/2\,[1+o(1)]$.
The inverse-coupling coefficient thus comes from the finite
support of the zero-flux eigenvalues, while its cubic charge dependence
comes from the first surviving term in the kernel difference.

\paragraph{Relative weights.}
Define
\begin{equation}
 C(p)=\sum_sN_s n_s(n_s^2-1).
\end{equation}
Equations~\eqref{eq:app-sea-common-charge-cancellation}
and~\eqref{eq:app-sea-source-difference-result} give
\begin{equation}
 \log\frac{\mathscr A_p^{\mathrm{sea}}}
          {Z_{\mathrm{sea}}(N)}
 =m\bigl(\mu-V_1(0)\bigr)
   +\frac{C(p)}{12\lambda_{\mathrm{pp}}}+o(1).
\end{equation}
Hence, for two fixed partitions $B,v$ with the same total
magnetic charge,
\begin{equation}
 \log\frac{\mathscr A_v^{\mathrm{sea}}}
          {\mathscr A_B^{\mathrm{sea}}}
 =-\frac{C(B)-C(v)}{12\lambda_{\mathrm{pp}}}+o(1)
 =-2S_{\mathrm{cl}}+o(1),
 \qquad
 S_{\mathrm{cl}}
 =\frac{C(B)-C(v)}{3g_{\mathrm{loc}}^2}
 \label{eq:app-sea-relative-normalization}
\end{equation}
with $g_{\mathrm{loc}}^2=8\lambda_{\mathrm{pp}}$.
For $B=(2^k)$ and $v=(2^{k-1},1,1)$, the Casimir
difference is six, so the neutral factor is
$\exp[-1/(2\lambda_{\mathrm{pp}})]$ at this order.
Writing $t=kg_{\mathrm{loc}}^2$, its exponent is $-4k/t$.
Here the neutral-sea limit is taken with fixed $k$; the
large-$k$ charged saddle is a subsequent limit.

This establishes the neutral-saddle factor used in the
bubbling discussion. It does not include the remaining
normalization factors or the charged integral, and hence
does not determine the full relative bubbling weight.

\section{Numerical tests of the mirror map for neutral insertions}
\label{num test}
In this Appendix, we test the mirror map for neutral insertions proposed in Subsection \ref{subsec:neutral-necklace} by comparing charged-pair correlators in the necklace and ADHM theories at finite rank.

\paragraph{Finite-rank tests.}
For $N=K=3$, $e_3(Q)=\det Q$ and
$\mathcal A_3=\det Q\operatorname{Tr}Z^2$. The commutator therefore
predicts
\begin{equation}
    \mathcal V^{\mathrm{neck,cand}}_{3,2}
    =4[\det Q\operatorname{Tr}\mathcal Z^2]_{\mathrm{Weyl}}
      -28\det Q.
    \label{eq:neutral-N3K3-operator}
\end{equation}
We evaluated the ADHM moment directly using
\begin{equation}
    R_{\mathrm{ADHM}}
    =\frac{\int_{\mathbb R^3}d^3\sigma\,
      \prod_a h_1(\sigma_a)
      \prod_{a<b}\tanh^2(\pi\sigma_{ab})
      \left(-4\sum_a\sigma_a^2\right)}
    {\int_{\mathbb R^3}d^3\sigma\,
      \prod_a h_1(\sigma_a)
      \prod_{a<b}\tanh^2(\pi\sigma_{ab})}
    \simeq-8.32250232.
    \label{eq:neutral-N3K3-ADHM-moment}
\end{equation}
An independent necklace computation, retaining the eigenvalue-dependent
propagators and the coincident-point Weyl contractions, gives
\begin{equation}
    \frac{\langle\det\widetilde Q\star
      [\det Q\operatorname{Tr}\mathcal Z^2]_{\mathrm{Weyl}}\rangle}
    {\langle\det\widetilde Q\star\det Q\rangle}
    \simeq4.91937441930.
\end{equation}
Thus $4(4.91937441930)-28\simeq -8.32250232$.

We also test the same quadratic generator in the non-minuscule
$(2,0)$ sector at $N=2$, $L=1$. For the completion tested here, we take
$\mathcal H_{(2,0)}=(\operatorname{Tr}Q)^2$. This specifies the
fusion-product completion, whose highest magnetic charge is $(2,0)$.
The proposed commutator gives
\begin{equation}
    \mathcal V^{\mathrm{neck,cand}}_{(2,0),2}
    =[\mathcal H_{(2,0)},\mathcal P_2^{\mathrm{cand}}]_\star
    =8[\operatorname{Tr}Q\,
       \operatorname{Tr}(\mathcal Z^2Q)]_{\mathrm{Weyl}}
      -10(\operatorname{Tr}Q)^2.
    \label{eq:neutral-completed20-operator}
\end{equation}
On the ADHM side, we include both the unscreened $(2,0)$ contribution
and the screened $(1,1)$ contribution, with the bubbling factors at
both poles, and summed their unnormalized contributions before
normalizing (The bubbling coefficient is written in \eqref{eq:bubbling-elementary-coefficient}). 
The resulting quadratic moment agrees numerically with the necklace
computation:
\begin{equation}
    R_{\mathrm{ADHM}}^{(2,0),\mathrm{comp}}
    \simeq R_{\mathrm{neck}}^{(2,0)}\simeq-5.00000000.
    \label{eq:neutral-completed20-moment}
\end{equation}
This test therefore probes the bubbling-completed
charged operator and extends the evidence beyond minuscule sectors.

\paragraph{A higher-moment test.}
For $N=2$ and $L=1$, we take the winding in the fundamental
representation, $K=1$, so that $e_1(Q)=\operatorname{Tr}Q$.
The cubic moment of this charged pair vanishes by symmetry.
We therefore test the quartic dressing.
\begin{equation}
    \mathcal V_{1,4}^{\mathrm{neck,cand}}
    =[\operatorname{Tr}Q,\mathcal P_4^{\mathrm{cand}}]_\star,
    \qquad
    \mathcal P_4^{\mathrm{cand}}
    =\frac{16}{5}\mathcal C_5^{\mathrm{cand}}
      -\frac83\mathcal C_3^{\mathrm{cand}}
      +\frac7{15}\mathcal C_1^{\mathrm{cand}}.
\end{equation}
Without fitting any new coefficients, the Duflo prescription gives
\begin{equation}
    \frac{\langle\operatorname{Tr}\widetilde Q\star
      \mathcal V_{1,4}^{\mathrm{neck,cand}}\rangle}
    {\langle\operatorname{Tr}\widetilde Q\star
      \operatorname{Tr}Q\rangle}
    =\frac{67}{3}.
\end{equation}
Independent numerical integration of the ADHM measure with insertion
$(2i\sigma)^4$ gives the same value.

{\paragraph{Quadratic moments for necklaces with two and three links.}
For $L=2$, we tested Eq.~\eqref{eq:general-L-n2} numerically through
the quadratic marked moment. For $N=K=1$, both sides give $-0.8$. For
$N=2$, $K=1$, with the insertion at one specified link and the same
candidate coefficient without adjustment, the necklace and ADHM
computations both give $-1$.

For $L=3$, we additionally tested the quadratic marked moment in
the determinant sectors $N=K=1,2$. The necklace and ADHM computations
agree, giving respectively $-0.489657785539$ and $-1.321356841456$.
In each case, insertion at any of the three links gives the same
result, without adjusting the neutral generator.

\bibliography{biblio}
\end{document}